\documentclass{article}

\usepackage{amsmath}
\usepackage{amssymb}
\usepackage{float}
\usepackage{graphicx}
\usepackage{authblk}
\usepackage{hyperref}
\usepackage{natbib}
\usepackage{xcolor}

\newcommand{\lcdm}{$\Lambda$CDM}
\newcommand{\kms}{$\,{\rm km\,s}^{-1}{\rm\,Mpc}^{-1}$}

\title{CosmoVerse Data Challenge: \\
How Tip of the Red Giant Branch Analysis Choices Shape the Inferred Hubble Constant}

\author{
\textbf{Lead Author:} Siyang Li
\\
\textbf{Coordinators:} Eleonora Di Valentino, Jackson Levi Said
\\
\textbf{Lecturers:} Kayla Owens, Yukei Murakami, Louise Breuval, Teresa Sicignano
\\
\textbf{Data Challenge Notebook Contributors:}
Luis A. Escamilla,
Jéferson A. S. Fortunato,
Payam Ghafari,
Ayush Hazarika,
Vasiliki Karanasou,
Trisha Khan,
Yoelsy Leyva,
Mahdi Najafi,
Swayamtrupta Panda,
Masroor C. Pookkillath,
Antonio Quintana-Estellés,
Masoume Reyhani,
Nandan Roy,
Chandra Shekhar Saraf,
Rahul Shah,
Jess Worsley

\textbf{Data Challenge Participants:}
Alejandro Aguirre,
Miguel Barroso Varela,
Anupam Bhardwaj,
Elliot Dutnall,
Francesca Di Nella,
Hussain Gohar,
Paridhi Jain,
Gracelynn Jost,
Shivani Kulkarni,
Dhiraj Kuniyal,
Mariia Marinichenko,
Yash Mehta,
Vasudev Mittal,
Lily Myers,
Devaraja Mallesha Naik,
José Antonio Nájera,
Antonio Quintana-Estellés,
Pilar Ruiz-Lapuente,
Kathan Dave,
Elaina Sadler,
Gurnoor Sandhu,
Sarang Shah,
Petar Suman

\textbf{Internal Reviewers:}
Stefano Casertano,
Alexei V. Filippenko,
Richard Stiskalek,
Syed Ashraf Uddin

\textbf{Paper Review Contributors:}
Du\v{s}ko Borka,
Vesna Borka Jovanovi\'{c},
Youri Carloni,
Umut Demirbozan,
Predrag Jovanovi\'{c},
Mohammad Malekjani,
Nils A. Nilsson,
Masroor C. Pookkillath,
Saeed Pourojaghi,
Masoume Reyhani,
Jorge Venzor

\textbf{Endorsers:}
Lilia Anguelova,
Marika Asgari,
Nihan Kat{\i}rc{\i},
Mar\'{i}a P\'{e}rez Garrote,
Leandros Perivolaropoulos,
Pradyumn Kumar Sahoo,
Denitsa Staicova
}

\begin{document}

\maketitle
\newpage

\begin{abstract}
Tip of the red giant branch (TRGB)-based determinations of the Hubble constant ($H_0$), the current expansion rate of the Universe, depend on choices
made in selecting and analyzing stellar photometry and in assembling
the sample of calibrator galaxies hosting Type Ia supernovae (SNe~Ia). The CosmoVerse Data Challenge invites participants to analyze public photometric catalogs within a shared distance-ladder framework to open each step of the measurement to the community and investigate how these choices propagate into the inferred value of $H_0$. Participants independently
selected photometric catalogs, galaxies, and TRGB measurement
settings, while using a common SN~Ia data compilation and fitting
procedure to convert their distances into $H_0$. The challenge,
supported by publicly available notebooks and an associated training
school, received 26 submissions from 22 participants. The challenge was not blinded, as one of its goals was to make the full $H_0$ measurement as transparent as possible. All submissions used the same analysis notebook, many participants were trainees at the associated school, and the true distances to the galaxies are unknown; the results therefore characterize the sensitivity of the challenge submissions to analysis and sample-selection choices and do not constitute a combined determination of $H_0$ or an estimate of the systematic uncertainty in published TRGB measurements. The inferred
$H_0$ values have a median of
$77.10$~km~s$^{-1}$~Mpc$^{-1}$ and a standard deviation of
$7.38$~km~s$^{-1}$~Mpc$^{-1}$ ($5.11$~km~s$^{-1}$~Mpc$^{-1}$ for the scaled median absolute deviation, which is less sensitive to the highest values), $\sim 3.4$ ($\sim 2.4$) times
the typical quoted uncertainty of an individual submission.
Differences in the measured tip of the
anchor galaxy NGC~4258 introduce an additional coherent shift in
the distance scale. The challenge demonstrates how different choices within a common
analysis framework can produce substantially different inferred
values of $H_0$ and provides an open basis for tracing and
investigating these differences as well as increasing the transparency and accessibility of datasets and tools for the community.
\end{abstract}

\tableofcontents
\newpage

\section{Introduction}

\subsection{The Hubble Tension} \label{sec:Hubb_ten}

The Hubble constant, $H_0$, sets the present-day expansion rate of the Universe. Over the past decade, two largely independent routes to $H_0$ have converged on discrepant values. Early-Universe inferences from the cosmic microwave background (CMB), assuming the standard $\Lambda$-cold-dark-matter ($\Lambda$CDM) model, yield $H_0=67.4\pm0.5$\kms{} from Planck \citep{Planck2018, Planck:2018vyg}, or $H_0=67.19\pm0.38$\kms{} from the combined South Pole Telescope (SPT), Planck, and Atacama Cosmology Telescope (ACT) dataset
\citep[CMB-SPA;][]{SPT-3G:2025bzu}. Local measurements combined within the Local Distance Network instead yield $H_0=73.50\pm0.81$\kms{} \citep{Casertano2025}, differing from the CMB inferences by
$\sim 6.4\sigma$ ($7.1\sigma$) when the quoted uncertainties are combined in quadrature. This discrepancy is referred to as the ``Hubble Tension.''

The Local Distance Network combines a range of local distance indicators within a distance-ladder framework \citep{Casertano2025}. In the first rung of the distance ladder, geometric distances to nearby anchor galaxies, such as the distance to NGC~4258 from the orbital motion of water masers around its central black hole \citep{Reid2019}, calibrate the luminosity of standardizable candles
such as Cepheid variables or the tip of the red giant branch (TRGB). In the second rung, the same candle is measured in more distant galaxies that have hosted Type~Ia supernovae (SNe~Ia); this step calibrates the luminosity of the SNe~Ia. In the third rung, this calibration is applied to SNe~Ia sufficiently far away that their redshifts are dominated by the expansion of the Universe (the Hubble flow; $0.023 < z < 0.15$ for the SNe~Ia used in this challenge), and the relation between their distances and redshifts gives $H_0$. SNe~Ia are used in the last two rungs because they are luminous enough to be observed at these distances and, once their peak magnitudes are corrected using the shapes of their light curves, they become precise relative distance indicators \citep{Phillips1993, Hamuy1996, Riess1996}. SN~Ia samples used for this purpose include the Pantheon+ compilation \citep{Scolnic2022, Brout2022b, Riess2016, Riess2022}, the Carnegie Supernova Project (CSP) sample \citep{Hamuy2006, Freedman2019, Uddin2024, FreedmanCCHP2025}, and the Zwicky Transient Facility (ZTF) sample \citep{Rigault2025, Dhawan2022}.

The discrepancy is not easily dismissed as a single measurement error, because each side is reached by multiple routes: several CMB experiments give consistent early-Universe values, and several distance indicators give consistent local values. The tension has been approached in two complementary ways: through searches for new physics beyond $\Lambda$CDM that could reconcile the CMB-inferred value with the local one, and through renewed, independent scrutiny of the systematics in both approaches, including the local distance-ladder measurements themselves. The TRGB method, which this Data Challenge focuses on, provides a specialized, partial cross-check of the local measurement: it replaces Cepheids as the primary distance indicator that calibrates the SNe~Ia, but its dependence on the geometric anchor distances and on the SN~Ia data is essentially the same as that of Cepheids.

The TRGB method rests on a well-understood feature of the evolution of old, low-mass stars. Red giant stars grow steadily brighter as they evolve, until helium ignition abruptly ends the red giant phase at a nearly constant luminosity \citep{Iben_1983A}. The bright end of the red giant branch (RGB), the TRGB, therefore appears as a sharp drop-off in the luminosity function of an old stellar population, and because this drop-off occurs at nearly the same luminosity in all such populations, comparing the apparent magnitudes of TRGB stars in a galaxy with their intrinsic magnitudes, calibrated with a geometric distance, gives the distance modulus of that galaxy.

In practice, the tip is located by selecting red giants from a catalog of resolved-star photometry, building their luminosity function, and detecting the sharp drop in star counts at its bright end. How cleanly and precisely this edge is detected depends on the stellar population of each galaxy and on analysis choices, such as which stars are selected and how much the luminosity function is smoothed; these choices can shift the inferred distance and therefore $H_0$. Measured in the outer halos of galaxies and most commonly in the $I$ band, the TRGB is affected only weakly by several systematics that affect Cepheids: halo fields have low reddening internal to the host galaxy and little crowding or blending of the photometry, and the $I$-band tip has only a shallow, calibrated dependence on metallicity \citep{Rizzi2007, Jang2021, Newman2024a, Newman2024b}. In practice, however, the tip is often not a clean edge, and its localization relies on choices made case by case \citep{Wu2022}; in eleven Hubble Space Telescope (HST) fields around NGC~4258, for example, the measured $F814W$ tips can span a range of 0.3~mag \citep{Li2023}. The tip contrast, the ratio of the numbers of stars half a magnitude fainter and brighter than the tip, has been used to assess the quality of tip measurements \citep{Wu2022} and to standardize them \citep{Li2023, Scolnic2023}. Other approaches have also been explored, such as utilizing the variability of red giants \citep{Anderson2024, Koblischke2024} and excluding subsamples of measurements based on their edge detector responses \citep{Hoyt2023}; in the context of measuring $H_0$, the applicability of these approaches can be limited in the hosts of SNe~Ia.

Measurements of $H_0$ calibrated with the TRGB currently span a range of
values. Using HST imaging of SN~Ia host galaxies \citep{JangLee2015}, \citet{JangLee2017a} obtained $H_0 = 71.17 \pm 1.66\,(\mathrm{stat})
\pm 1.87\,(\mathrm{sys})$\kms{}. The Carnegie-Chicago Hubble Program (CCHP) found
$69.8 \pm 0.8\,(\mathrm{stat}) \pm 1.7\,(\mathrm{sys})$\kms{} \citep{Freedman2019} and
$69.8 \pm 0.6\,(\mathrm{stat}) \pm 1.6\,(\mathrm{sys})$\kms{} with an updated
calibration \citep{Freedman2021}, with further distances to NGC~5643 and
NGC~1404 \citep{Hoyt2021} and a calibration of the TRGB in the maser host
NGC~4258 \citep{Jang2021}. An independent reduction of the same HST data by
the Extragalactic Distance Database (EDD) team gave $71.5 \pm 1.8$\kms{}
\citep{Anand2022}, and the Comparative Analysis of TRGBs (CATs), which
standardizes the tip using its contrast, gave
$73.22 \pm 2.06$\kms{} \citep{Wu2022, Li2023, Scolnic2023}. With JWST, the TRGB has been
calibrated in NGC~4258 \citep{Anand2024} and measured in further SN~Ia
hosts by the SH0ES team \citep{Li2024} and by the CCHP \citep{FreedmanCCHP2025, Hoyt2026}.
Combining HST and JWST calibrators, the CCHP finds
$70.39 \pm 1.22\,(\mathrm{stat}) \pm 1.33\,(\mathrm{sys}) \pm 0.70\,(\sigma_{\rm SN})$\kms{}
\citep{FreedmanCCHP2025}, while \citet{Li2026} trace the differences
in $H_0$ to the subsample selection of SNe~Ia. Compiling all
35 SNe~Ia with HST or JWST TRGB distances, \citet{Li2026} find
(72.1--73.3) $\pm 1.8$\kms{}, depending on methodology, and can closely reproduce lower measures of $H_0$ by selecting subsamples of the full SNe~Ia set.
Other TRGB-based
determinations include a calibration of ZTF SNe~Ia
using the TRGB distance to NGC~7814, $76.94 \pm 6.4$\kms{} \citep{Dhawan2022}; the
Carnegie Supernova Project analysis, which uses the TRGB alongside Cepheid
and surface brightness fluctuation (SBF) calibrators \citep{Uddin2024}; and
the TRGB--SBF Project, which measures JWST TRGB distances to early-type
galaxies in the Fornax and Virgo clusters \citep{Anand2024Fornax,
Anand2025Virgo} to calibrate SBF distances, giving
$73.8 \pm 0.7\,(\mathrm{stat}) \pm 2.3\,(\mathrm{sys})$\kms{} \citep{Jensen2025}. The
TRGB is also one of the distance indicators combined in the Local Distance
Network \citep{Casertano2025}.
These TRGB-based analyses differ in how the tip is located and standardized, in the photometric reductions, and in the SN~Ia hosts included. For example, the EDD analysis models the luminosity function of RGB and asymptotic giant branch stars instead of using edge detection \citep{Anand2022}, CATs standardizes the tip by its contrast \citep{Scolnic2023}, and \citet{Li2026} use all publicly available data rather than a subgroup.

Besides the three-rung distance ladder described above, $H_0$ has also been measured with two-rung versions of the ladder, calibrated with Cepheids \citep{Kenworthy2022,Stiskalek2026Cepheid} or with the TRGB \citep{Stiskalek2026TRGB}, and from megamasers in galaxies in the Hubble flow, independently of the ladder \citep{Pesce2020,Stiskalek2026Maser}.

\subsection{CosmoVerse and the Hubble Tension}

The CosmoVerse network connects over 1000 cosmologists spanning observational surveys, data analysis, and fundamental physics in over 40 European countries as well as another 50 globally. The CosmoVerse community provides research exchange opportunities for early-career researchers, the promotion of research results by senior researchers, as well as a vehicle to communicate breakthroughs to the general public and science students. The main aim of CosmoVerse is to establish an interdisciplinary synergy to confront the growing challenges of tensions in recent cosmological surveys. The network does this by harmonizing approaches to European observational cosmology (Working Group~1), connecting data-analysis teams in cosmology (Working Group~2), and supporting new physics development in Europe (Working Group~3). The CosmoVerse network has come to represent a significant portion of the observational, data analysis, and theory communities in cosmology. This includes the funding of research exchanges and conference presentations, as well as holding conferences, workshops, and schools. Additionally, the community hosts biweekly seminars on leading research results, a cross-university journal club, a platform to share the release of data-analysis toolkits, and a framework to hold virtual lectures and training series. The network has a strong public-dissemination presence through in-person talks at every CosmoVerse event, public articles and profile pieces, and other efforts targeted at the education sector.

The network was established in 2022 and holds a large international conference each year. There are also workshops and graduate schools that are targeted at different aspects of the network. The Data Challenge emerged from discussions between the different working-group subcommunities, and particularly at the CosmoVerseWorkshop@Naples 2025, which gave a platform for an open discussion on the possible effects of survey systematics and the indication of new physics that has emerged in the comparison of different observational surveys. This led to several funded research exchanges -- the CosmoVerseSchool@Sofia 2026 and this Data Challenge. These efforts primarily aimed at dispelling some confusion about the appearance of cosmological tensions between different specific surveys and the possible role that systematics may play in the statistical robustness of the specific cosmic tensions that have been raised in the last decade of cosmology.

Indeed, the Data Challenge aims to shed light on the complexities of analyzing observational data and their implications for determining the Hubble constant. Its scope may expand in future years, but its initial objective is to clarify how constraints on the Hubble constant are obtained. Cosmology has entered an era in which measurements from probes of the early and late Universe have reached the precision needed to reveal statistically significant discrepancies. These tensions have challenged the status of \lcdm{} as a concordance model, prompting renewed scrutiny of its foundations and closer consideration of possible extensions involving new physics in gravity, dark matter, dark energy, or particle physics.

\begin{figure}[htbp]
    \centering
    \includegraphics[scale=0.7, trim=0 0 0 0, clip]{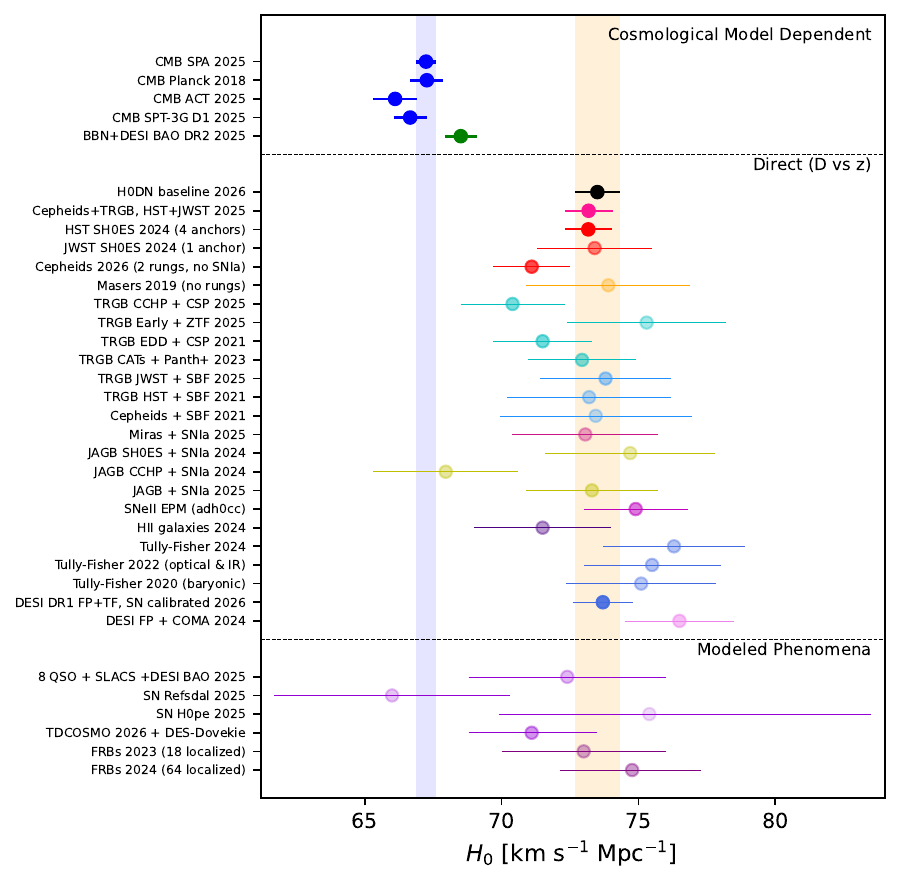}
    \caption{Whisker plot showing the most precise measurements of $H_0$, updated from~\cite{CosmoVerseNetwork:2025alb}.}
    \label{fig:Intro_H0_whisk}
\end{figure}

A big contribution by the CosmoVerse network to the discussion on cosmic tensions has been the CosmoVerse White Paper \citep{CosmoVerseNetwork:2025alb}. This brought together leading research groups making precise direct and indirect measurements of $H_0$, developers of new data-analysis methods including different applications of inference-free methods, as well as different flavors of new physics proposals. One of the main outputs of this collaboration is the comparison of these different measurements as shown in Fig.~\ref{fig:Intro_H0_whisk} (updated from \citealp{CosmoVerseNetwork:2025alb}). The benchmark constraints on $H_0$ based on the early Universe come from CMB-SPA, giving a value of $H_0 = 67.19\pm0.38$\kms{} (\citealt{SPT-3G:2025bzu}; shown in blue), while late-Universe measurements combined within the Local Distance Network give $H_0 = 73.50 \pm 0.81$\kms{} (\citealt{Casertano2025}; shown in black). These CMB and local measurements expose a tension of $\sim 7.1\sigma$.

The development of new data-analysis techniques and the expansion of mature cosmological simulations that can contribute to producing CMB and other constraints on cosmological parameters have been major outputs of the CosmoVerse network. This extends to the implementation of new machine-learning applications to reconstruct cosmological functions and distance–redshift relations, such as through ReFANN \citep{Wang:2019vxv} and LADDER \citep{Shah:2024slr}, as well as likelihood-free inference packages such as CoLFI \citep{Wang:2023vej}. The use of these methods may be important for accurately producing predictions from extended cosmological models.

The network has also contributed to addressing the question of cosmic tensions by promoting the development of new fundamental physics through various vehicles to determine
which are the most realistic directions to consider for cosmologies beyond \lcdm. In the early Universe, there is ample ground to inject new physics prior to recombination through different inflationary scenarios, the variation of fundamental constants, early dark energy, and other mechanisms. Many of these approaches aim to reduce the scale of the sound horizon at the baryon drag epoch, thereby allowing a higher inferred value of $H_0$. However, in models such as early dark energy, the accompanying shifts in cosmological parameters required to fit the CMB data tend to increase the predicted amplitude of matter clustering, potentially exacerbating the $S_8$ tension ($S_8 \equiv \sigma_8(\Omega_{\rm m}/0.3)^{0.5}$) \citep{CosmoVerseNetwork:2025alb, Poulin:2023lkg, Jedamzik:2020zmd}. On the other hand, late-time solutions are severely restricted by the wealth of data in the local Universe. Ultimately, a growing consensus appears to point to a degree of new physics at both extremes of the Universe.

The CosmoVerse network has created a community where these discussions can take place frequently between observational physicists, data analysts, and fundamental physicists. By bringing these researchers from these subcommunities together, the CosmoVerse network has contributed to the success of this Data Challenge and its aim of clarifying the processes by which local measurements of $H_0$ are made.

\subsection{Motivation for the CosmoVerse Data Challenge and School}

The discussions at the CosmoVerse workshop in Naples highlighted a practical opportunity to make
distance-ladder analyses more accessible to the wider cosmology community.
Community data challenges have an established history in weak lensing,
where initiatives such as GREAT3 \citep{Mandelbaum:2013esa} have enabled researchers to analyze
shared datasets, compare methods, and investigate sources of systematic
uncertainty. These exercises also provide an
entry point for researchers with different backgrounds, who can contribute
to an observational problem without first developing an entire analysis
pipeline. The CosmoVerse Data Challenge was conceived to bring this
collaborative approach to TRGB-based distance-ladder measurements,
combining a methodological investigation with an opportunity for
practical training.

A central motivation was the difficulty of interpreting differences
between the results of established TRGB research programs. As discussed
in Sec.~\ref{sec:Hubb_ten}, several groups use this distance indicator to
infer $H_0$, obtaining different central values and uncertainty estimates.
For cosmologists who do not work directly on these measurements, such
differences can create the impression that the analyses are contradictory,
or that one group must have made an error. Understanding these differences,
and where possible reconciling them, requires more than a comparison of the
final values: it requires an understanding of the data entering each
analysis, the choices made in extracting distances, and the statistical
treatment through which these distances lead to a constraint on $H_0$.

The challenge was motivated by the need to make these connections visible
and testable. Access to the intermediate steps allows researchers to
examine why a particular choice is made, how alternatives affect the
measurement, and how the resulting changes propagate through the distance
ladder. Equally important is understanding the uncertainties and
correlations entering the likelihood that connects distances to $H_0$ (in the Data Challenge, the generalized least-squares fit with the Pantheon+ covariance matrix described in Section~\ref{sec:sn_notebook}), since a shift in a central value
cannot be interpreted independently of its statistical context. Providing
a common framework in which these ingredients can be inspected helps
turn a discussion of apparently discrepant results into an investigation
of identifiable methodological differences. Through this process, the
initiative aims to strengthen confidence in distance-ladder measurements
by making their assumptions, strengths, and limitations easier to assess.

The scientific and educational objectives therefore informed the project
from its inception. The comparison required a clearly defined framework
within which participants could exercise their own judgment, while the
educational component required sufficient explanation for those choices
to be understood. The notebooks were consequently designed to expose
the reasoning behind the calculations, rather than simply return a final
value of $H_0$. Participants could follow the connection between stellar
photometry, a TRGB distance estimate, the calibration of SNe~Ia, and the
inferred $H_0$. The statistical framework on the SN~Ia side was
made explicit and held fixed for the comparison, allowing participants to
understand this stage while concentrating their independent analysis on
the TRGB measurements.

Developing these materials required a coordinated preparation program.
The work was divided into distinct tasks covering data curation, the
TRGB analysis notebook, and the SN~Ia notebook used to infer $H_0$.
Members of the community volunteered to join the preparation groups and
contribute to these defined tasks under the guidance of their respective
coordinators. Kayla Owens coordinated the collection and organization of
the photometric data, while Siyang Li coordinated the development of the
TRGB notebook and the associated algorithm contributions. Yukei Murakami
prepared the SN~Ia notebook. This structure brought together expertise
in the observations and in their analysis, producing a connected set of
materials with clearly assigned responsibilities. The individual
contributions and the resulting resources are described in Section~\ref{Sec3}.

An accompanying school provided the training needed to open the exercise
to students and researchers entering the field. CosmoVerseSchool@Sofia
2026 was held on 25--29 May 2026 at the Institute of Mechanics of the
Bulgarian Academy of Sciences in Sofia, Bulgaria. The school was organized
in the context of Working Group~1, Observational Cosmology and
Systematics, with a close connection to Working Group~2, Data Analysis
in Cosmology. Its program linked the physical basis of the distance
indicators to the practical steps involved in analyzing observations.
Lectures on Cepheids, the TRGB, stellar photometry, and SNe~Ia were
combined with an introduction to the challenge and practical work with
the notebooks.

This combination gave students the opportunity to begin using the code
while discussing the underlying measurements with researchers experienced
in the field. Working through an analysis makes questions concrete:
which features of the data support a measurement, how stable the result
is under changes in the analysis, and how the adopted uncertainties enter
the inference. The school provided a setting in which these questions
could be explored alongside the relevant astrophysics and statistical
methods. It also helped participants develop the judgment needed to
interpret diagnostic plots, recognize ambiguous measurements, and
explain the choices made in their own analyses. Appendix~\ref{school} describes the
school program and its connection to the challenge in more detail.

Following this preparation and training, the online challenge was opened
to the wider community through the call for submissions on 30 June 2026.
Participation extended beyond those who had attended the school. The
notebooks were written in a textbook style, assuming a general background
in astronomy and physics but no previous experience with distance-ladder
measurements. This approach allowed the same resources to support both
independent participation in the challenge and subsequent use for
teaching or individual study.

Transparency and accessibility extend to the materials themselves. The
photometric catalogs come from public sources, and the analysis code
and supporting documentation are available in a public
repository.\footnote{\url{https://github.com/cosmoversecost/Data-Challenge}}
Researchers can therefore inspect the implementation, repeat the
calculations, and explore how alternative choices affect their results
without requiring access to private software or proprietary data.
The challenge was not blinded: participants could run the SN~Ia notebook and see the value of $H_0$ implied by their distances, because one goal was to make every step of the measurement, including the final one, transparent.
Participants were also asked to explain their analysis choices, making
the reasoning behind a submission part of the scientific record.
Together, the challenge and the school establish a shared resource
through which a broader community can engage directly with
distance-ladder measurements and contribute to their continued scrutiny
and development.

\subsection{The Data Challenge}\label{sec:data_challenge}

To translate the motivations outlined above into a practical comparison,
CosmoVerse organized a community-wide Data Challenge in which participants
inferred $H_0$ using a common set of publicly available photometric
catalogs and a shared analysis framework. Within this framework,
participants were free to make independent choices at the TRGB stage
of the distance ladder. These included the selection of galaxies and
photometric catalogs, together with the smoothing bandwidth,
color-selection window, spatial clipping, and edge-detection weighting
used to identify the tip and estimate its uncertainty. The notebook locates
the tip by Sobel edge detection only, with a choice of three weighting
schemes; luminosity-function fitting and maximum-likelihood methods were
not implemented to constrain the scope of this challenge, and no submission
used them.

The SN~Ia data compilation, its associated covariance information,
and the fitting code that converts distances into $H_0$ were supplied
to all participants and held fixed. Each participant therefore used
the same procedure to propagate their TRGB distances into a constraint
on $H_0$, although the subset of calibrating SN~Ia hosts entering
the fit depended on which distances they submitted. This arrangement
provides a common basis for comparing the results while preserving
the freedom to exercise judgment in the selection and analysis of
the TRGB observations.

The commitment to transparency described in the preceding section
is implemented through public access to the analysis materials and
the release of the submissions in the same repository.
Together with the participants' descriptions of their methods, these
resources will allow readers to inspect both the inputs and the
decisions behind each reported value of $H_0$. Reproducibility extends
to the final inference: running the shared SN~Ia notebook on a
submitted set of distances and uncertainties recovers the corresponding
$H_0$ result.

The spread in the values reported in Section~\ref{Sec4} can therefore be
examined in terms of TRGB measurement methodology, photometric
catalog selection, and the composition of the calibrator sample,
within a common SN analysis framework. Documenting these
choices alongside the final measurements provides the information
needed to investigate the origin of differences between submissions
and assess how they propagate through the distance ladder.

Section~\ref{Sec3} presents the shared data and
the two-notebook pipeline, while Section~\ref{Sec4} introduces the submissions,
with individual methodological descriptions written by the participants collected in Appendix~B.
In Section~\ref{sec:analysis}, we examine the variation in the inferred $H_0$ values and its
relationship to the participants' choices, and our conclusions are presented in Section~\ref{sec:conclusions}. Appendix~\ref{school} describes the associated school and the lectures, as well as
the practical training that introduced participants to the relevant
concepts and analysis tools.

\section{Data Challenge Materials} \label{Sec3}

In keeping with the goal of transparency, all of the data challenge materials are publicly available at \url{https://github.com/cosmoversecost/Data-Challenge}, and the photometry comes entirely from public archives, so the full measurement can be repeated without access to proprietary data or private software. Some datasets from the literature, such as those from \cite{Freedman2019}, are not publicly available and thus are not included in the Data Challenge repository.

The Data Challenge materials consist of three parts which follow the rungs of the distance ladder. The first is the \emph{data}: HST photometry catalogs of resolved stars in the anchor galaxy NGC~4258 and in SN~Ia host galaxies, taken from two public sources, EDD \citep{Tully2009, Anand2022} and CATs \citep{Wu2022,Li2023,Scolnic2023}, together with a table that records for each catalog its source, HST program, instrument, filters, and notes on data quality. The second is the \emph{TRGB notebook}, in which participants select red giant stars, smooth the luminosity function, detect the tip with a Sobel filter \citep{Lee1993}, estimate its uncertainty by bootstrap resampling, and convert the tip magnitude into a distance modulus relative to NGC~4258. This is the part of the analysis that participants were free to change. The third is the \emph{SN~Ia notebook}, a simplified distance-ladder solver that combines the submitted TRGB distances with a fixed Pantheon+ SN sample to give $H_0$. Participants were asked not to modify the SN~Ia notebook, so that differences in $H_0$ can be traced to the TRGB measurements alone. Both notebooks are written in a textbook style, with thorough explanations of each step, and assume a general background in astronomy and physics but no prior knowledge of the distance ladder or of $H_0$ measurements.

The data and the TRGB notebook were prepared with contributions from volunteers in the community, with each part coordinated by one person. The data curation was coordinated by Kayla Owens, with contributions from Ayush Hazarika, Vasiliki Karanasou, Trisha Khan, Yoelsy Leyva, Masroor C. Pookkillath, Antonio Quintana-Estellés, and Jess Worsley. The TRGB notebook was coordinated by Siyang Li, with algorithm contributions from Luis A. Escamilla, Jéferson A. S. Fortunato, Payam Ghafari, Mahdi Najafi, Swayamtrupta Panda, Masoume Reyhani, Nandan Roy, Chandra Shekhar Saraf, and Rahul Shah. The SN~Ia notebook was prepared by Yukei Murakami. We describe each part in turn below.

\subsection{Data}
\label{sec:data}

\subsubsection*{Preparation and compilation}

The challenge uses HST photometry of resolved stars in 18 galaxies: the geometric anchor NGC~4258 and 17 other galaxies, 14 of which have SNe~Ia in the challenge's SN~Ia sample (Table~\ref{tab:catalogs}). Every galaxy is covered by two separately compiled sets of public catalogs, and participants could use either one:
EDD or CATs.
The EDD set contains 22 catalogs, one for each HST program provided for a galaxy, so M66 has three catalogs and M101 and NGC~4258 have two each. The CATs set contains 28 catalogs, one for each host galaxy and 11 separate fields for NGC~4258. Every catalog contains F814W, the band in which the tip is measured, and one bluer band used to select stars by color: F606W in 15 EDD and 19 CATs catalogs, and F555W in 7 EDD and 9 CATs catalogs.

\begin{table}[!htbp]
\centering
\footnotesize
\caption{Photometry catalogs provided for the Data Challenge.$^a$}
\label{tab:catalogs}
\begin{tabular}{lr lr lr}
\hline
 & & \multicolumn{2}{c}{EDD} & \multicolumn{2}{c}{CATs} \\
Galaxy & HST program & Blue filter & Sources & Blue filter & Sources \\
\hline
NGC~4258 & 9477 & F555W & 48,277 & F555W & 34,823 (5 fields) \\
 & 16198 & F606W & 23,410 & F606W & 45,414 (6 fields) \\
M66 & 10433 & F555W & 2,271 & -- & -- \\
 & 13691 & F606W & 16,322 & F606W & 16,322 \\
 & 15654 & F606W & 40,407 & -- & -- \\
M96 & 13691 & F606W & 8,948 & F606W & 8,916 \\
M101 & 9492 & F555W & 30,474 & -- & -- \\
 & 13691 & F606W & 8,653 & F606W & 8,113 \\
NGC~1309 & 10497 & F555W & 15,864 & F555W & 3,470 \\
NGC~1316 & 13691 & F606W & 8,812 & F606W & 8,758 \\
NGC~1365 & 13691 & F606W & 9,921 & F606W & 9,395 \\
NGC~1404 & 15642 & F606W & 5,724 & F606W & 1,077 \\
NGC~1448 & 13691 & F606W & 5,913 & F606W & 3,850 \\
NGC~3021 & 10497 & F555W & 6,732 & F555W & 1,998 \\
NGC~3370 & 9351 & F555W & 20,682 & F555W & 2,011 \\
NGC~4038 & 10580 & F606W & 40,038 & F606W & 21,612 \\
NGC~4424 & 13691 & F606W & 15,993 & F606W & 12,323 \\
NGC~4526 & 13691 & F606W & 22,854 & F606W & 20,063 \\
NGC~4536 & 13691 & F606W & 13,869 & F606W & 5,112 \\
NGC~5584 & 11570 & F555W & 13,097 & F555W & 1,837 \\
NGC~5643 & 15642 & F606W & 13,192 & F606W & 6,334 \\
NGC~7814 & 10889 & F606W & 25,236 & F606W & 12,893 \\
\hline
\end{tabular}
\par\smallskip
\begin{minipage}{\linewidth}\footnotesize
$^a$For each galaxy, every HST program with a catalog in the EDD or CATs set is listed with the catalog's blue filter (all catalogs also contain F814W) and its number of sources; a dash means that the set has no catalog from that program. The CATs photometry of NGC~4258 is split into 11 fields; for each program, the number of fields and the total number of sources over those fields are given. All fields were observed with ACS/WFC except those of NGC~5584, which were observed with WFC3/UVIS.
\end{minipage}
\end{table}

Quality cuts had already been applied to the photometry by each source. Because these cuts can differ slightly between EDD and CATs, participants were advised to use catalogs from only one source within a single $H_0$ measurement.

\subsubsection*{Analysis materials}

Each row of a catalog is one detected source, with its Vega magnitudes and uncertainties in the two bands and the photometric quality parameters of the source. The EDD catalogs provided also give each source's right ascension and declination, which the notebook uses to remove stars projected onto the disk of the galaxy. The CATs catalogs do not include sky coordinates, so the notebook does not apply this mask to them; instead, they already contain the spatial selection that CATs made when producing the catalogs \citep{Wu2022}.

A table distributed with the catalogs (\texttt{photometry/catalog\_info.csv}) records for each catalog its source, galaxy, HST program, field center, instrument, and filters, together with notes on data quality for 13 of the 50 catalogs. These notes flag sparse color-magnitude diagrams (M66 in program 10433 and NGC~3021), fields centered on the disk rather than the halo (M101 in program 13691, NGC~3370, and NGC~5584), contamination from the galaxy center (NGC~1404), streaks in the images (M66 in program 13691), and the four hosts (NGC~1309, NGC~3021, NGC~3370, and NGC~5584) that \citet{Scolnic2023} excluded from their $H_0$ analysis because no reliable tip was detected. Alongside the photometry, the notebook provides for each galaxy the Milky Way foreground reddening $E(B-V)$ from the maps of \citet{SchlaflyFinkbeiner2011}, retrieved through NED, which it converts into an extinction correction for each band, and the galaxy's isophotal radius, axis ratio, and position angle, which define the region of the disk that is masked for the EDD catalogs.

\subsubsection*{Validation and reproducibility}

Members of the data team converted the catalogs to the common format, checked them, and documented it in a shared spreadsheet, which records for each catalog the field center, instrument, filters, and exposure times, any problems with the field, and whether the sky coordinates are correct.

\subsubsection*{Final materials and purpose}

The data products give every participant the same inputs: the 22 EDD and 28 CATs catalogs (\texttt{photometry/EDD/} and \texttt{photometry/CATS/}), the table \texttt{photometry/catalog\_info.csv}, and the reddening and disk geometry of each galaxy. Because the dataset is fixed, differences between submitted TRGB distances that use the same photometry source arise from analysis choices in the TRGB notebook rather than from the input photometry. The notebook warns that mixing sources between the anchor and the targets introduces pipeline-dependent systematics that its error budget does not capture. The datasets are released in the public challenge repository.

\subsection{TRGB Notebook}

\subsubsection*{Preparation and compilation}

The notebook was built around the photometry assembled by the data team. It reads the EDD and CATs catalogs directly from \texttt{photometry/}, identifies the blue filter (F555W or F606W) of each catalog, and combines the separate CATs fields of NGC~4258. For every galaxy, it stores the Milky Way foreground reddening $E(B-V)$ from \citet{SchlaflyFinkbeiner2011}, retrieved through NED and converted into an extinction in each HST band, and the isophotal radius, axis ratio, and position angle from RC3 and HyperLEDA, which define the disk mask for the EDD catalogs. NGC~4258 is fixed as the anchor, with $\mu = 29.397 \pm 0.032$~mag \citep{Reid2019}. Volunteers from the community contributed to algorithm development.

\subsubsection*{Analysis materials}

The TRGB notebook (\texttt{1\_trgb/TRGB\_distance\_measurement.ipynb}) is a 71-cell walkthrough that takes public photometry for a chosen target galaxy, drawn from EDD \citep{Tully2009, Anand2022} or CATs \citep{Wu2022,Li2023,Scolnic2023}, and applies foreground extinction correction, spatial clipping, color/RGB selection, GLOESS smoothing of the luminosity function (bandwidth $\tau$), Sobel edge detection, and bootstrap resampling to estimate the uncertainty on the tip magnitude. At each stage, the notebook produces a diagnostic plot (the color-magnitude diagram, the smoothed luminosity function with its Sobel response, and the bootstrap distribution of recovered tip magnitudes) so that a participant can check, before exporting a result, whether their chosen parameters produce a single clean peak rather than an ambiguous or bimodal detection.

Participants are free to vary the target galaxy and photometry source, the color-selection window, the smoothing bandwidth $\tau$, the Sobel edge detector weighting, the spatial-clipping scale factor, and the number of bootstrap resamples. The notebook also provides a stability checklist: participants rerun the measurement with $\tau = 0.05$, $0.10$, $0.15$, and $0.20$~mag, with the color limits shifted by $\pm0.2$~mag, and with spatial-clipping scale factors of $0.8$, $1.0$, and $1.2$; a TRGB shift of more than $0.03$~mag under any of these variations indicates that the result should be examined more thoroughly by the participant and encourages them to consider the effects of these measurement choices on their final result. The notebook computes and prints quality checks, such as the contrast ratio from \citet{Wu2022} and the number of stars 1\,mag fainter than the TRGB from \citet{Madore2009}, for the participant to gauge the robustness of their measurement.

\subsubsection*{Validation and reproducibility}

The notebook lists the package versions with which it was tested, reports any missing photometry files, and stops if a galaxy is absent from its database or has an invalid reddening or axis ratio. The bootstrap uses a fixed random seed, so a rerun with the same parameters returns the same tip magnitude and uncertainty. Each row written by the single-galaxy export to \texttt{trgb\_results.csv} records the photometry source, the target's analysis parameters, and the anchor and target tip magnitudes, so that the distance modulus can be recomputed as $\mu = m_{\rm TRGB,target} - m_{\rm TRGB,anchor} + 29.397$~mag and the measurement reproduced by rerunning the notebook.

\subsubsection*{Final materials and purpose}

Once a participant is satisfied with their measurement, the notebook exports the resulting distance modulus and its uncertainty to a \texttt{trgb\_results.csv} file; the distance moduli and their uncertainties are then entered as the input table \texttt{data/TRGB\_distances.csv} of the SN~Ia notebook described in Section~2.3.


\subsection{SN~Ia Notebook}
\label{sec:sn_notebook}

\subsubsection*{Preparation and Compilation}

The SN~Ia notebook (\texttt{2\_sn\_to\_h0/}) converts the TRGB distances measured by participants into $H_0$. It is \texttt{mu\_to\_H0}\footnote{\url{https://github.com/SterlingYM/mu_to_H0}}, a simplified distance-ladder solver that Y.~S.~Murakami wrote for a Cepheid-based version of the SH0ES ladder \citep{Riess2022} and used in the coding session at the School, integrated unchanged into the challenge repository. For the TRGB challenge, we added an SN~Ia dataset built from the public Pantheon+ release \citep{Scolnic2022,Brout2022b}. The calibrators are the 37 SH0ES Cepheid hosts \citep{Riess2022} and three CATs hosts without Cepheid distances (NGC~1316, NGC~1404, and NGC~4526; \citealt{Scolnic2023}), with 82 light curves of 47 SNe~Ia matched to Pantheon+ by SN name and survey. Because TRGB calibrators include early-type hosts, the Hubble-flow sample follows the all-host-type selection of \citet{Riess2022} adopted by CATs \citep{Scolnic2023}: all noncalibrator SNe~Ia at $0.023<z<0.15$ with $|x_1|<3$ and $|c|<0.3$, giving 496 light curves of 434 SNe~Ia.

\subsubsection*{Analysis Materials}

The observables are the Pantheon+ magnitudes $m_B^{\rm corr}$, which include the Tripp standardization \citep{Tripp1998} and the simulation-based bias corrections, and the uncertainties are the corresponding $578\times578$ block of the Pantheon+ statistical plus systematic covariance matrix \citep{Brout2022b}. For a calibrator SN~Ia in host $h$, $m_{B,i}^{\rm corr} = \mu_h + M_B$; for a Hubble-flow SN~Ia, $m_{B,j}^{\rm corr} - 5\log_{10}[cz_j\{1+\tfrac{1}{2}(1-q_0)z_j - \tfrac{1}{6}(1-q_0-3q_0^2+j_0)z_j^2\}] - 25 = M_B - 5\log_{10}H_0$, with $q_0=-0.55$ and $j_0=1$ \citep{Riess2022,Scolnic2023}. The solver takes the participant's table of $\mu_h$ and $\sigma_{\mu,h}$, adds $\sigma_{\mu,h}^2$ as a fully correlated term to all SNe~Ia in the same host, and solves for $M_B$ and $5\log_{10}H_0$ by generalized least squares. Hosts without a submitted distance are removed from the fit. The notebook reports $H_0$, $M_B$, their uncertainties, and the $\chi^2$, draws a Hubble diagram that compares the SN-based and input distances for each calibrator host, and shows how $H_0$ changes under a uniform shift of all host distances ($\mathrm{d}H_0/\mathrm{d}\Delta\mu \approx -34$~km~s$^{-1}$~Mpc$^{-1}$~mag$^{-1}$). Participants supply only the distance table; the SN~Ia data, the covariance matrix, and the solver are fixed. The reported uncertainty does not include covariance between host distances, such as that from the common NGC~4258 anchor.

\subsubsection*{Validation and Reproducibility}

We validated the solver on the Cepheid ladder. Using the SN~Ia rows of the SH0ES data vector (75 calibrator light curves in 37 hosts and 277 Hubble-flow SNe~Ia) and the Cepheid host distances of \citet{Riess2022}, it gives $H_0=73.16\pm0.78$~km~s$^{-1}$~Mpc$^{-1}$, compared with $73.04\pm1.04$ for the full SH0ES fit. The smaller uncertainty is expected, because the Cepheid calibration and its covariance are not part of the fit. With the same Cepheid distances as placeholders for the 37 shared hosts, the TRGB dataset gives $H_0=73.71\pm1.15$~km~s$^{-1}$~Mpc$^{-1}$ with $\chi^2/\mathrm{d.o.f.}=490.8/569$. The numerical response of $H_0$ to a uniform distance shift matches the analytic $-H_0\ln(10)/5$. The notebook was tested on a fresh clone and runs from pre-built data arrays; the Pantheon+ covariance file is needed only to rebuild the dataset.

\subsubsection*{Final Materials and Purpose}

The SN~Ia notebook gives every participant the same mapping from a set of TRGB host distances to $H_0$: the same SNe~Ia, the same covariance matrix, and the same solver. Each submitted $H_0$ is therefore a function of the submitted distances only, and the differences between submissions measure the effect of TRGB analysis choices. Because the notebook writes out the equations of each rung and the code is short, participants can follow the step from distances to $H_0$ and see directly how a change in one host distance, or in all of them, propagates to $H_0$.

\section{Data Challenge Submissions} \label{Sec4}

With these materials in place, the Data Challenge was announced by email on 30 June 2026 and was open to anyone in the community. Results were due by 15 August 2026 and were submitted by email to the organizers. Every participant worked from the same photometry catalogs and passed their distances through the same, unmodified SN~Ia notebook, but was free to make their own choices in the TRGB measurement: which galaxies to measure, which photometry source to use, and the settings of the tip detection, such as the color selection, the smoothing bandwidth, the spatial clipping, and the edge-detection weighting.

Table~\ref{tab:master_submissions} lists the final value of $H_0$ when the TRGB distances from each submission are run through the SN~Ia notebook, together with the photometric catalog used in the TRGB analysis. Consistent with the transparency goal of the challenge, all submissions are made public in the Data Challenge repository, so that each value of $H_0$ in the table can be reproduced by running the submitted distances through the SN~Ia notebook. Each participant also wrote a short description of their analysis, explaining the reasoning behind these choices and any particular features they encountered. These descriptions are collected in Appendix~\ref{sec:individual_submissions}. Section~\ref{sec:analysis} then examines the spread of the submitted values of $H_0$ and where it comes from.


\begin{table*}[!htbp]
\centering
\small
\begin{tabular}{lcc}
\hline
\textbf{Participant} &
\textbf{Phot.} &
\boldmath{$H_0$}
\\
&
\textbf{source} &
\textbf{[km\,s$^{-1}$\,Mpc$^{-1}$]}
\\
\hline

Alejandro Aguirre         & EDD   & $99.32\pm4.56$ \\
Miguel Barroso Varela     & CATs  & $73.69\pm2.23$ \\
Kathan Dave               & EDD   & $67.91\pm9.22$ \\
Francesca Di Nella        & CATs  & $76.81\pm2.07$ \\
Elliot Dutnall            & EDD/CATs  & $80.59\pm1.73$ \\
Hussain Gohar (1)         & CATs  & $69.50\pm2.31$ \\
Hussain Gohar (2)         & CATs  & $76.72\pm1.85$ \\
Hussain Gohar (3)         & EDD   & $73.13\pm2.41$ \\
Hussain Gohar (4)         & EDD   & $82.00\pm2.17$ \\
Paridhi Jain              & EDD   & $69.93\pm6.57$ \\
Gracelynn Jost            & EDD   & $78.23\pm2.14$ \\
Shivani Kulkarni          & EDD   & $75.79\pm5.78$ \\
Dhiraj Kuniyal            & EDD   & $79.76\pm2.47$ \\
Maria Marinichenko        & EDD   & $80.37\pm1.70$ \\
Yash Mehta                & EDD   & $85.69\pm2.15$ \\
Vasudev Mittal            & CATs  & $73.90\pm1.88$ \\
Lily Myers                & EDD   & $93.45\pm2.00$ \\
Devaraja Mallesha Naik    & EDD   & $75.89\pm2.11$ \\
Jose Antonio Najera (1)   & EDD   & $84.98\pm1.88$ \\
Jose Antonio Najera (2)   & EDD   & $71.61\pm2.32$ \\
Antonio Quintana-Estellés & CATs  & $77.39\pm1.83$ \\
Pilar Ruiz-Lapuente       & EDD   & $74.08\pm2.25$ \\
Elaina Sadler             & EDD   & $85.31\pm1.98$ \\
Gurnoor Sandhu            & CATs  & $75.24\pm1.64$ \\
Sarang Shah               & EDD   & $78.67\pm2.34$ \\
Petar Suman               & EDD   & $88.87\pm2.19$ \\

\hline
\end{tabular}
\par\smallskip
\begin{minipage}{0.6\linewidth}\footnotesize
$^a$For each participant, we report the adopted photometric catalog (``Phot. source'') and the value of $H_0$, with its uncertainty, obtained by running the submitted distances through the SN~Ia notebook. Participants who sent more than one submission have one numbered row per submission.
\end{minipage}
\caption{Summary of the Data Challenge submissions.$^a$}
\label{tab:master_submissions}
\end{table*}

\section{Analysis}
\label{sec:analysis}

In this section, we examine how the choices made in TRGB measurements move $H_0$. We describe the submitted values, identify which choices are responsible for the differences between them, and compare the results with an earlier study of algorithmic variants of the TRGB measurement.

Throughout, a \emph{submission} is one set of TRGB distance moduli to SN~Ia host galaxies sent in by a participant. The 22 participants
sent 26 submissions: one participant sent four and another sent two. For each
submission, $H_0$ is the value returned by the Data Challenge SN~Ia notebook (\texttt{demo\_TRGB\_to\_H0.ipynb})
when run on that set of distances without modification, so all values share
the same supernova sample and the same fitting code. As the submissions will be made public in the Data Challenge repository, each of these values can be reproduced by running the corresponding submission through the notebook. A calibrator galaxy that
is missing from a submission is simply left out of its fit. Participants could
measure distances to 17 galaxies. Fourteen of them host the calibrating
SNe~Ia in the notebook's dataset; the other three (M66, M96, and NGC~7814)
do not have an SN~Ia in Pantheon+ (though they can exist in other SN compilations), so their distances do not affect $H_0$.

The aim of this section is to characterize the submissions: how their values
of $H_0$ are distributed and which measurement choices produce the differences
between them. Interpreting the values, and in particular judging whether any
one value of $H_0$ is more correct than another, is beyond the intended scope
of this paper.

\subsection{The Submitted Values of \texorpdfstring{$H_0$}{H0}}

\begin{figure}[!htbp]
\centering
\includegraphics[width=\linewidth]{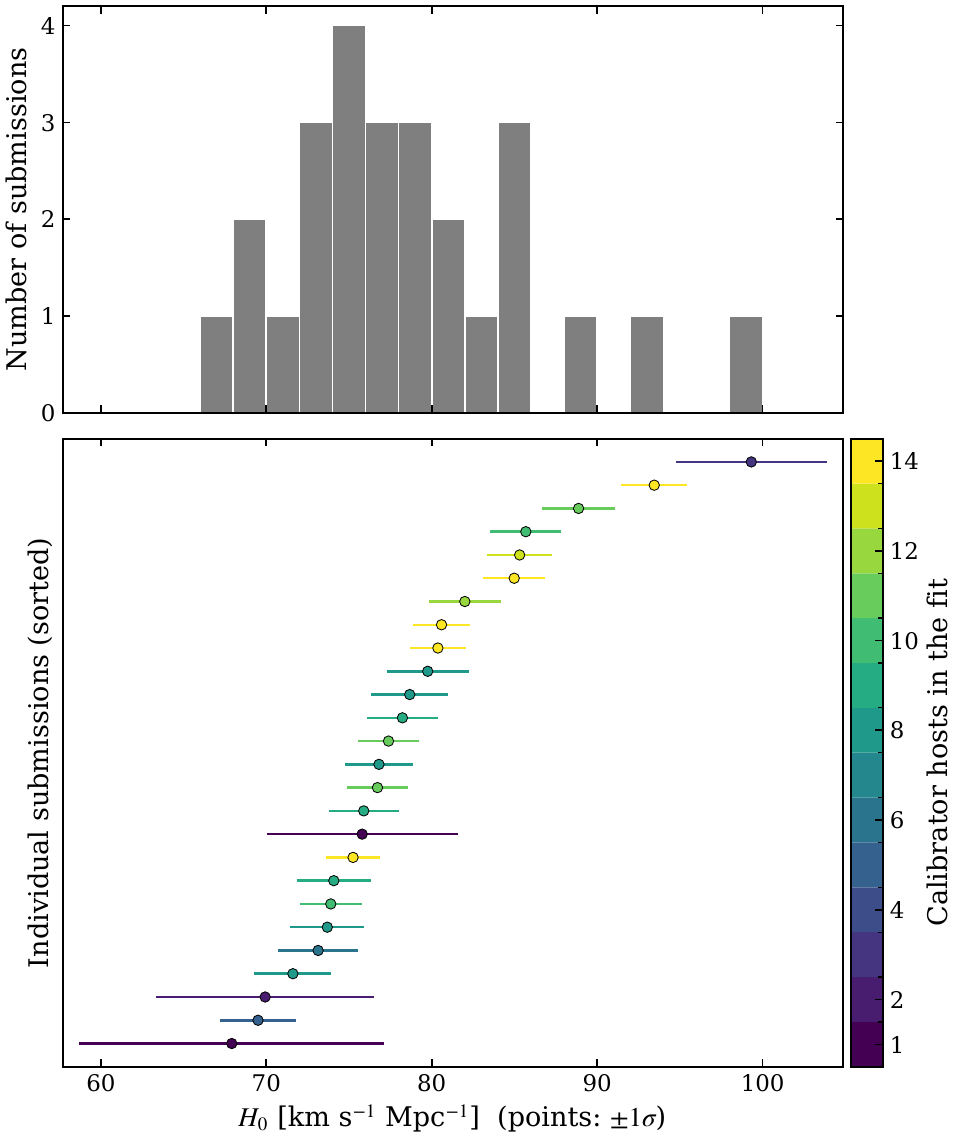}
\caption{$H_0$ from each of the 26 submissions (sets of TRGB distances) of the 22
participants, each obtained by running the Data Challenge SN~Ia notebook on the submitted distances.
\emph{Top:} histogram in bins of 2\,km\,s$^{-1}$\,Mpc$^{-1}$.
\emph{Bottom:} the same 26 values, sorted, with the formal $\pm1\sigma$
uncertainty returned by the notebook, colored by the number of calibrator
galaxies in each fit. Participants are not identified.}
\label{fig:h0hist}
\end{figure}

Each submission lists a set of TRGB distances to SN~Ia host galaxies. These distances are fed into the SN~Ia notebook, which outputs a value of $H_0$ and an uncertainty. This uncertainty is the formal error of the notebook's fit: it comes from the uncertainties of the submitted distances and from the statistical and systematic covariance of the supernova magnitudes, and it does not account for the different choices that
could have been made in the TRGB measurement. Figure~\ref{fig:h0hist} shows
the distribution of the 26 values of $H_0$ listed in
Table~\ref{tab:master_submissions}. They have a median of $77.10$\,km\,s$^{-1}$\,Mpc$^{-1}$ and
a sample standard deviation (normalized by $N-1$, with $N=26$) of
$7.38$\,km\,s$^{-1}$\,Mpc$^{-1}$, and range from $67.91$ to $99.32$\,km\,s$^{-1}$\,Mpc$^{-1}$. Because the standard deviation is sensitive to the highest values, we also
give the median absolute deviation scaled by 1.4826, which equals the
standard deviation for a normal distribution but is insensitive to outliers:
it is $5.11$\,km\,s$^{-1}$\,Mpc$^{-1}$, and removing the two highest values
alone reduces the standard deviation normalized by $N-1$ to $5.42$\,km\,s$^{-1}$\,Mpc$^{-1}$. The standard
deviation measures how much the submissions differ from one another. If they
differed only because of statistical errors, this spread would be comparable
to the uncertainty of a single submission, since all submissions share the same supernova data and fitting code. The typical uncertainty,
taken as the median of the 26 uncertainties in
Table~\ref{tab:master_submissions}, is $2.16$\,km\,s$^{-1}$\,Mpc$^{-1}$, so the observed
spread is
$\sim 3.4$ times (2.4 times for the scaled median absolute deviation) as large: the submissions disagree with
each other far more than their individual error bars would suggest, and most
of the differences come from the choices made in the measurements rather
than from statistical noise. Part of the spread is nevertheless statistical, because submissions that use
different calibrator galaxies draw on different supernovae. If every
submission adopted the median submitted distance to each galaxy, so that only
the lists of galaxies differed, the supernova covariance alone would give a
standard deviation of about $2.2$\,km\,s$^{-1}$\,Mpc$^{-1}$ among the 26
values (mean of 400 random realizations), compared with
$5.8$\,km\,s$^{-1}$\,Mpc$^{-1}$ for the same values computed from the real
supernova data. The quoted uncertainties also treat the distance errors of
different galaxies as independent, although all distances are measured
relative to the same anchor, so they underestimate the true uncertainty of
each submission. Finally, because all submissions analyze the same catalogs
and supernovae, their values are correlated, and the median summarizes the
submissions rather than combining independent measurements.

The distribution is also asymmetric, with a longer tail to high values. To
show this, we use the 16th and 84th percentiles, the values below which 16
and 84 percent of the submissions lie; for a normal (Gaussian) distribution,
they lie one standard deviation below and above the median and enclose the
central 68 percent of the values. Computed by linear interpolation between
the sorted values (the default of \texttt{numpy.percentile}), they are
$73.13$ and $85.31$\,km\,s$^{-1}$\,Mpc$^{-1}$, and they lie $4.0$\,km\,s$^{-1}$\,Mpc$^{-1}$ below and
$8.2$\,km\,s$^{-1}$\,Mpc$^{-1}$ above the median, respectively; in a symmetric distribution,
these two distances would be equal. Finally, two participants sent more than
one submission (the numbered rows in Table~\ref{tab:master_submissions}), so
the statistics above count them more than once. To check that these
participants do not drive the result, we count each participant once,
replacing their submissions by the unweighted mean of their values. This
gives 22 values with a median of $77.81$\,km\,s$^{-1}$\,Mpc$^{-1}$ and a sample standard
deviation (again normalized by $N-1$, with $N=22$) of $7.39$\,km\,s$^{-1}$\,Mpc$^{-1}$, close
to the values for all 26 submissions.

\subsection{Why the Values Differ}
\label{sec:spread_origin}

Two things change from one submission to the next: which calibrator galaxies
are included in the fit, and the distance measured to each of them. A
calibrator galaxy is one that hosts an SN~Ia and has a TRGB distance; the
calibrators fix the intrinsic brightness of SNe~Ia used to calibrate more distant
SNe~Ia in the Hubble flow to measure $H_0$. To find out how much of
the spread in $H_0$ comes from each of the two, we recompute $H_0$ for each
submission, keeping its own set of galaxies but replacing its
distances and their uncertainties with the medians over all submissions for
each galaxy. In these recomputed values, every submission uses the same
distance to a given galaxy, so they differ only in which galaxies they
include. Comparing their variance with
that of the submitted values shows how much of the spread comes from the
choice of galaxies alone. The variance of the recomputed values is about 60 percent of that of the
submitted $H_0$ (between 50 and 87 percent when any single submission is left
out), or about 50 percent after subtracting the variance expected from
supernova sampling alone. The variance of the differences between the
submitted and recomputed values is about 45 percent, and the remaining
covariance term is about $-7$ percent, so the two parts are not independent.
Because the recomputed values use the median distance to each galaxy, which
for four galaxies appears biased (see below), these fractions depend largely
on whether a submission includes those galaxies. We discuss the two effects in turn, and then the settings of the tip
detection. These comparisons describe how the submitted values depend on each
choice; they do not indicate which choices give the more correct value of
$H_0$.

\paragraph{Which Galaxies are Included.} Submissions used between 1 and 14
of the 14 calibrator galaxies, and those that used more galaxies tend to give
a higher $H_0$ (Spearman rank correlation $\rho=0.56$, $p=0.003$, bootstrap 68\% interval 0.37--0.72; this coefficient
measures how consistently one quantity rises with another, from $-1$ to $+1$,
with 0 meaning no such trend). We give for each coefficient its $p$-value and the range containing 68 percent of 5000 bootstrap
resamples of the submissions; because several correlations are tested, values with $p$ above
about 0.01 should be read as tentative. Much of this trend comes from four galaxies
(NGC~1309, NGC~3021, NGC~3370, and NGC~5584), in which the TRGB is hard to
detect. The notebook checks each detection with the tip contrast, the ratio
of the number of stars in the $0.5$\,mag half below the tip to the number in
the $0.5$\,mag above it. A clear tip, where the red giant branch begins
abruptly, gives a large ratio, and the notebook treats a ratio below 3 as a
weak detection, following \citet{Wu2022}, \citet{Li2023}, and \citet{Scolnic2023}. Of the 22 submitted measurements with a tip contrast below 3,
17 are of these four galaxies, and the CATs analysis \citep{Scolnic2023}
found no tip of sufficient contrast in any of them and excluded them. Their
distances also scatter more between submissions than those of most other
calibrators, by $0.22$--$0.36$\,mag in distance modulus (about 10--16 percent
in distance), compared with a typical $0.08$\,mag (about 4 percent) for the
other ten calibrators (Table~\ref{tab:host_scatter}); among those ten, only
NGC~1404 ($0.43$\,mag) and NGC~4038 ($0.29$\,mag) scatter as much. Only one of the 16 submitted measurements of NGC~1404 falls below either threshold of Table~\ref{tab:host_scatter}, although the catalog notes for both photometry sources warn that stars from its center contaminate the sample; the two thresholds therefore do not flag all of the measurements that disagree. These four
galaxies matter for $H_0$: using the median distance to every galaxy, the fit
with all 14 calibrators gives $83.5$\,km\,s$^{-1}$\,Mpc$^{-1}$, and the fit without the four
gives $76.0$\,km\,s$^{-1}$\,Mpc$^{-1}$. For the 14 submissions that include at least one of
them, removing them lowers $H_0$ by a median of $6.5$\,km\,s$^{-1}$\,Mpc$^{-1}$.
Compared with the Cepheid distances of \citet{Riess2022} ($\mu_{\rm Ceph}$ in their Table~6), the median submitted
distance moduli of NGC~1309, NGC~3021, NGC~5584 and NGC~3370 are smaller by
$0.98$, $1.13$, $0.40$ and $0.14$\,mag, whereas for the seven other
calibrators with Cepheid distances the median difference is $0.01$\,mag
(range $-0.16$ to $+0.09$\,mag). A galaxy placed too close makes its
supernova appear intrinsically fainter, which raises $H_0$. This pattern
would arise if the detected edge in these galaxies lies brighter than the
true tip, as can happen when the tip is close to the detection limit of the
data, but the challenge does not establish the cause. A dependence
on the sample of calibrators is also found in published TRGB calibrations:
\citet{Li2026} show that subsamples of TRGB-calibrated SNe~Ia can give
noticeably different values of $H_0$, both through the statistics of small
samples, since each SN~Ia scatters by about 7 percent in $H_0$, and through
differences in how the samples are selected. Both effects are present among
the submissions. Small samples are the most exposed to the first: the lowest
value, $67.91$\,km\,s$^{-1}$\,Mpc$^{-1}$, comes from a submission that used a single
calibrator galaxy and has an uncertainty of $9.2$\,km\,s$^{-1}$\,Mpc$^{-1}$. The four
galaxies above illustrate the second, shifting $H_0$ in the same direction
whenever they are included.

\begin{table}[!htbp]
\centering
\caption{Scatter of the submitted distance moduli for each galaxy.$^a$}
\label{tab:host_scatter}
\small
\vspace{6pt}
\resizebox{\textwidth}{!}{%
\begin{tabular}{lrcccccccc}
\hline
Galaxy & $N$ & median $\mu$ & $\sigma(\mu)$ & $\sigma_{\rm MAD}$ & $\sigma(m_{\rm TRGB})$ & $\sigma_{\rm MAD}(m_{\rm TRGB})$ & $\tilde\sigma_\mu$ & $R<3$ & $N_{\rm faint}<400$ \\
 & & [mag] & [mag] & [mag] & [mag] & [mag] & [mag] & rows & rows \\
\hline
NGC~1404 & 16 & 31.106 & 0.432 & 0.368 & 0.403 & 0.295 & 0.069 & 0 & 1 \\
NGC~3370$^b$ & 6 & 31.980 & 0.356 & 0.059 & 0.288 & 0.059 & 0.094 & 6 & 4 \\
NGC~5584$^b$ & 6 & 31.379 & 0.320 & 0.376 & 0.340 & 0.303 & 0.094 & 6 & 5 \\
NGC~4038 & 15 & 31.456 & 0.289 & 0.163 & 0.292 & 0.119 & 0.056 & 3 & 0 \\
NGC~3021$^b$ & 9 & 31.341 & 0.249 & 0.096 & 0.229 & 0.022 & 0.088 & 4 & 1 \\
NGC~1309$^b$ & 14 & 31.569 & 0.218 & 0.067 & 0.203 & 0.013 & 0.062 & 1 & 0 \\
NGC~1365 & 22 & 31.386 & 0.102 & 0.040 & 0.058 & 0.030 & 0.063 & 0 & 0 \\
NGC~4424 & 20 & 30.921 & 0.093 & 0.058 & 0.064 & 0.039 & 0.059 & 0 & 0 \\
NGC~4536 & 21 & 30.923 & 0.082 & 0.051 & 0.046 & 0.015 & 0.063 & 0 & 0 \\
NGC~4526 & 23 & 30.985 & 0.080 & 0.040 & 0.018 & 0.015 & 0.049 & 0 & 0 \\
NGC~1316 & 23 & 31.443 & 0.079 & 0.024 & 0.012 & 0.006 & 0.047 & 0 & 0 \\
NGC~1448 & 21 & 31.357 & 0.079 & 0.035 & 0.032 & 0.021 & 0.053 & 0 & 0 \\
M101 & 15 & 29.118 & 0.074 & 0.058 & 0.012 & 0.015 & 0.046 & 2 & 0 \\
NGC~5643 & 22 & 30.492 & 0.046 & 0.036 & 0.019 & 0.023 & 0.081 & 0 & 0 \\
M66$^c$ & 16 & 30.136 & 0.172 & 0.203 & 0.152 & 0.158 & 0.052 & 0 & 0 \\
M96$^c$ & 18 & 30.471 & 0.103 & 0.031 & 0.048 & 0.030 & 0.065 & 0 & 0 \\
NGC~7814$^c$ & 17 & 30.781 & 0.042 & 0.014 & 0.017 & 0.015 & 0.049 & 0 & 0 \\
\hline
\end{tabular}}
\par\smallskip
\begin{minipage}{\linewidth}\footnotesize
$^a$$N$ is the number of files containing the galaxy; $\sigma(\mu)$ is the standard deviation of the submitted $\mu$, and $\sigma_{\rm MAD}$ is the scaled median absolute deviation ($1.4826\times$MAD); $\sigma(m_{\rm TRGB})$ and $\sigma_{\rm MAD}(m_{\rm TRGB})$ are the same statistics for the submitted apparent tip magnitudes of the galaxy, which do not include the anchor tip; $\tilde\sigma_\mu$ is the median of the formal uncertainties in the submissions. The last two columns count submitted rows with a distance below the two detection diagnostics of the TRGB notebook (tip contrast $R<3$; fewer than 400 stars within 1\,mag below the tip).\\
$^b$No tip with $R>2$ in CATs (Scolnic et al.~2023, Appendix~B).\\
$^c$Not a calibrator host in the SN notebook; does not enter $H_0$.
\end{minipage}
\end{table}

\paragraph{The Distances to the Galaxies.} The remaining variation comes from
the distances themselves, and part of it is shared by all galaxies in a submission: some submissions place every galaxy slightly closer or farther
than the others do. This common shift is set by the anchor. The distance to
each galaxy is not measured on its own but relative to NGC~4258, whose
distance is known independently from water masers \citep{Reid2019}: the
apparent magnitude of the tip in each galaxy is compared with the tip in
NGC~4258, so a brighter or fainter anchor tip moves all of a submission's
distances together. The anchor tip magnitudes adopted in the submissions
differ by $0.079$\,mag (standard deviation), whereas a submission's typical offset in the other galaxies, the median over its galaxies of the difference between its tip and the median tip of all submissions for that galaxy, has a standard deviation of only $0.018$\,mag between submissions. For the galaxies whose distance moduli scatter by $0.07$ to $0.10$\,mag (Table~\ref{tab:host_scatter}), the submitted tip magnitudes scatter by only $0.01$ to $0.06$\,mag, so most of their distance scatter comes from the anchor. The notebook measures the anchor tip from all eleven CATs fields of NGC~4258 or from the two EDD catalogs of NGC~4258, in each case combining the fields observed with F555W and with F606W by inverse-variance weighting; only with EDD photometry can a participant restrict the anchor to one catalog, and the submitted files do not record whether anyone did. The adopted anchor tips partly follow the photometry source: the six submissions that used only CATs photometry have a median anchor tip of $25.285$\,mag and a standard deviation of $0.035$\,mag, compared with $25.364$\,mag and $0.071$\,mag for the 18 that used only EDD photometry. For comparison, \citet{Gibbon2026} find an intrinsic field-to-field scatter of $0.067^{+0.023}_{-0.017}$\,mag among eleven HST fields of NGC~4258. The
shift of each submission's distances follows its anchor tip almost one for
one ($\rho=-0.91$, $p=5\times10^{-10}$, with a slope of $-1.01$): a submission whose anchor tip is
$0.1$\,mag fainter places its galaxies, on average, $0.1$\,mag closer in
distance modulus. A common shift of all distances by 1\,mag changes $H_0$ by about $-35.5$\,km\,s$^{-1}$\,Mpc$^{-1}$ for these submissions (the SN~Ia notebook quotes the approximation $-H_0\ln(10)/5\approx-34$\,km\,s$^{-1}$\,Mpc$^{-1}$). Replacing each submission's anchor tip with the median anchor tip changes its $H_0$ by a standard deviation of $3.0$\,km\,s$^{-1}$\,Mpc$^{-1}$, so the anchor alone accounts for a scatter of about $3$\,km\,s$^{-1}$\,Mpc$^{-1}$ in $H_0$. Published analyses of the F814W tip of NGC~4258 quote uncertainties of $0.014$ to $0.044$\,mag \citep{Mager2008, JangLee2017, Jang2021, Anand2022, Li2023, Gibbon2026}, so the scatter of the anchor tips between submissions is about two to six times these uncertainties.

\begin{figure*}[!htbp]
\centering
\includegraphics[width=\textwidth]{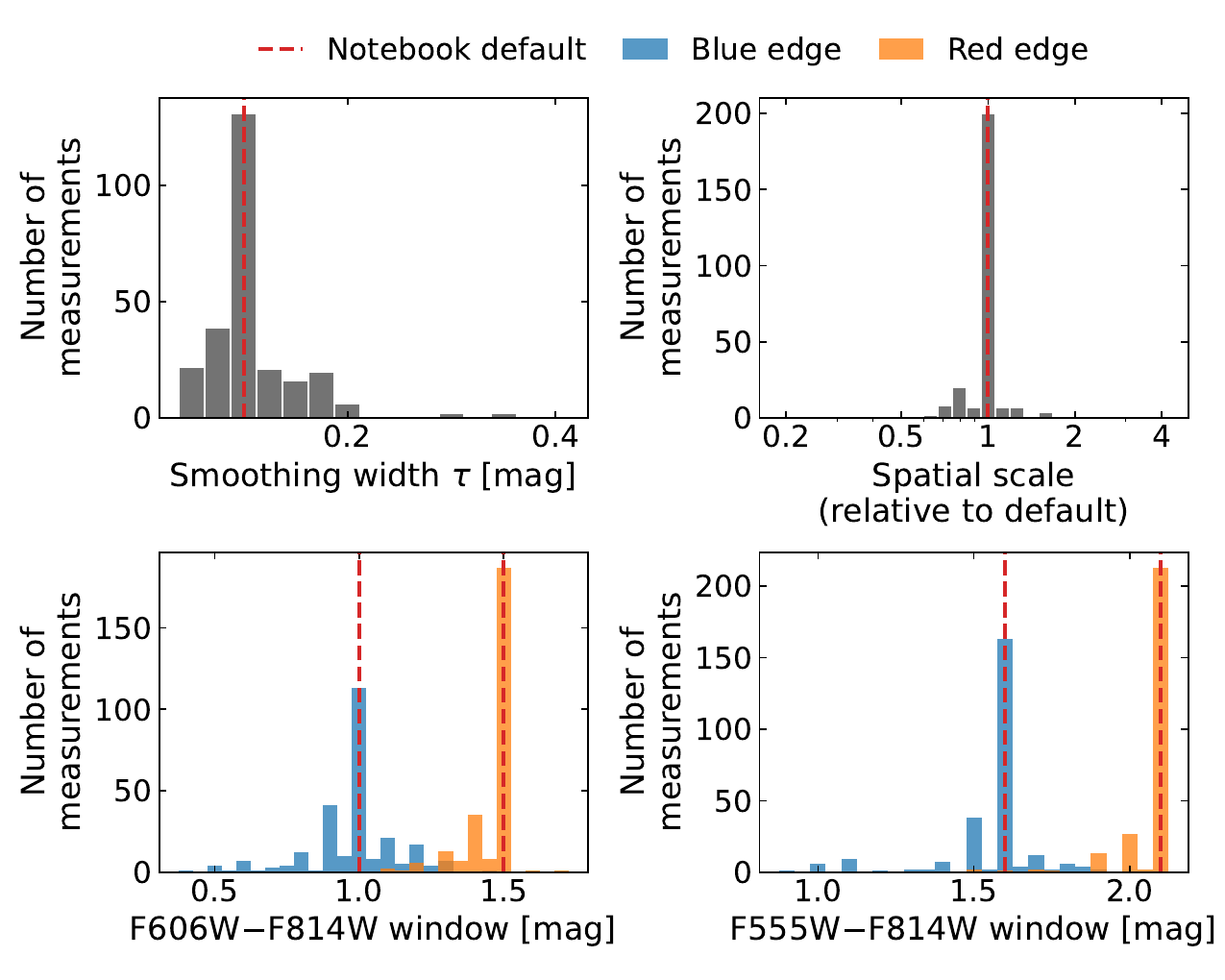}
\caption{Distribution of settings of the TRGB measurement over the 261
submitted galaxy measurements that recorded them (of 284 total): the
smoothing width $\tau$ of the luminosity function, the spatial scale relative
to the default, and the blue and red edges of the color window in
F606W$-$F814W and F555W$-$F814W. Dashed red lines mark the notebook defaults.
One measurement each used a spatial scale of 0.2 and of 4.0; their bars are too small to see.
Both color windows are recorded for every measurement, and each is applied
to the photometry taken in that filter. The remaining settings were rarely
changed: the edge-detection weighting, the luminosity-function bin width, and
the Poisson floor kept their default values in 256, 257, and 261 of the 261
measurements, respectively; magnitude limits were set in 12, and the number of bootstrap
iterations, which affects only the uncertainty, ranged from 500 to 10\,000.}
\label{fig:choices}
\end{figure*}

\paragraph{Settings of the Tip Detection.} Participants also changed the
settings of the TRGB measurement itself (Fig.~\ref{fig:choices}). The
smoothing width, which sets how strongly the luminosity function is smoothed
before the tip is located, ranged from $0.05$ to $0.40$\,mag; of the 261 measurements whose settings were recorded, $41$ percent used $0.10$\,mag, the value in the notebook's batch mode, and $8$ percent used $0.05$\,mag, the value that its step-by-step mode applies to every galaxy. The spatial scale sets the size of the ellipse around the center of each galaxy inside which stars are excluded; the notebook applies this cut only to EDD photometry. It ranged from $0.2$ to $4.0$ times the default, which $65$ percent of the 183 recorded EDD measurements kept, and the color window, the range of colors
used to select red giant stars, was often moved. These settings can change individual distances. We find no clear trend of
$H_0$ with the median smoothing width of a submission ($\rho=0.08$, $p=0.71$, interval $-0.14$ to $0.30$), but
submissions that changed the spatial scale for more of their galaxies tend to
give a higher $H_0$ ($\rho=0.50$, $p=0.01$, interval 0.31--0.67, over the 25 submissions with recorded
settings; $\rho=0.41$, $p=0.09$, interval 0.16--0.65, for EDD measurements alone, where the cut is applied).
This trend is related more to the part of $H_0$ due to the distances
($\rho=0.55$, $p=0.004$, interval 0.38--0.70) than to the part due to the choice of galaxies ($\rho=0.22$, $p=0.3$, interval $-0.02$ to $0.45$).
With 26 submissions, we cannot tell whether changing the spatial scale itself
moves the distances or whether it goes together with other choices that do.

\paragraph{Photometry source.} The seven submissions that used only CATs
photometry have a median of $75.24$\,km\,s$^{-1}$\,Mpc$^{-1}$ and a standard
deviation of $2.73$\,km\,s$^{-1}$\,Mpc$^{-1}$, compared with $79.22$ and
$8.26$\,km\,s$^{-1}$\,Mpc$^{-1}$ for the 18 that used only EDD photometry.
The four galaxies with hard-to-detect tips do not explain this difference:
they appear in four of the seven CATs submissions and in nine of the 18 EDD
submissions, and removing them from the EDD submissions still leaves a
standard deviation of $7.11$\,km\,s$^{-1}$\,Mpc$^{-1}$. The two groups do
differ in the spatial cut, which the notebook applies only to EDD photometry:
none of the 78 recorded CATs measurements changed the spatial scale, compared
with 64 of the 183 recorded EDD measurements. With 26 submissions, we cannot
tell whether the spatial cut, other differences between the catalogs, or
differences between participants account for the broader spread of the EDD
submissions.

\subsection{Comparison with the CATs Algorithm Variants}
\label{sec:cats}

\citet{Scolnic2023} explored the same question in a different way. Using a
single unsupervised algorithm, they varied four of its parameters (the
spatial clipping, the width of the color band, the smoothing of the
luminosity function, and the minimum edge response) and obtained $H_0$ for all
108 combinations, applying each combination to every galaxy in the same way.
Table~\ref{tab:cats_compare} compares their 108
values, taken from Table~6 of \citet{Scolnic2023}, with the 26 Data Challenge submissions.

\begin{table}[!htbp]
\centering
\caption{Summary statistics of the Data Challenge $H_0$ values and of the 108 algorithm variants of CATs.$^a$}
\label{tab:cats_compare}
\small
\begin{tabular}{lcc}
\hline
 & Data Challenge & CATs variants \\
\hline
Number of values & 26 files & 108 variants \\
Median & 77.10 & 72.94 \\
Mean & 78.80 & 73.71 \\
Standard deviation & 7.38 & 3.15 \\
16th--84th percentile & 73.13--85.31 & 71.67--75.81 \\
Median $-$ 16th percentile & 3.97 & 1.27 \\
84th percentile $-$ median & 8.21 & 2.87 \\
Minimum & 67.91 & 67.91 \\
Maximum & 99.32 & 86.22 \\
Sample skewness (bias-adjusted) & 1.06 & 1.80 \\
Values below 70 & 3 & 1 \\
Values above 80 & 9 & 10 \\
\hline
\end{tabular}
\par\smallskip
\begin{minipage}{\linewidth}\footnotesize
$^a$CATs values are computed from Table~6 of Scolnic et al.~(2023). Units are km\,s$^{-1}$\,Mpc$^{-1}$. The two sets differ in what was varied, in the supernova analysis, and in the calibrator galaxies (see text), so the comparison is descriptive.
\end{minipage}
\end{table}

The two distributions have a similar shape: most values form a core, with a
longer tail toward high $H_0$ and few values below 70\,km\,s$^{-1}$\,Mpc$^{-1}$ (1 of the 108 CATs
variants and 3 of the 26 submissions). They differ in location and width.
The Data Challenge values are centered about $4$\,km\,s$^{-1}$\,Mpc$^{-1}$ higher (median $77.1$ against $72.9$\,km\,s$^{-1}$\,Mpc$^{-1}$); their standard deviations are $7.4$ and $3.2$\,km\,s$^{-1}$\,Mpc$^{-1}$, but the CATs value is set by the ranges and spacing chosen for its grid of variants, so the two are not directly comparable.

The origin of the high tail also differs. CATs attributes its high tail to variants that
produce spurious tips, and its variants that keep fewer SN host galaxies give
higher $H_0$; keeping only the variants that retain all hosts and anchor
fields largely removes the tail. In the Data Challenge, keeping only the six submissions that use at least 13 of the 14 calibrators does not remove the tail: they still range from $75.24$ to $93.45$\,km\,s$^{-1}$\,Mpc$^{-1}$. All six include the four galaxies with hard-to-detect tips, which CATs excludes; without those four, their own distances give $72.7$ to $84.0$\,km\,s$^{-1}$\,Mpc$^{-1}$. The five submissions above $85$\,km\,s$^{-1}$\,Mpc$^{-1}$ would give between $83.1$ and $84.7$\,km\,s$^{-1}$\,Mpc$^{-1}$ with the median distance to each of their own galaxies, and their own distances add between $0.6$ and $15.8$\,km\,s$^{-1}$\,Mpc$^{-1}$ to that. The highest value, $99.32$\,km\,s$^{-1}$\,Mpc$^{-1}$, uses none of the four galaxies with hard-to-detect tips. The
broader spread is expected given the greater freedom of the challenge:
participants could also choose which galaxies to include, how to measure the
anchor, which photometry to use, and different settings for each galaxy. The
two analyses also differ in method. The notebook takes the single strongest
edge as the tip and does not standardize tips by their contrast, whereas CATs
weights multiple candidate edges by contrast, and the two treat the
Hubble-flow supernovae differently. With 26 submissions, we cannot separate
how much each of these differences contributes.

\section{Conclusions}
\label{sec:conclusions}

The CosmoVerse Data Challenge asked participants to measure TRGB distances to SN~Ia host galaxies from the same photometry catalogs to convert them to $H_0$ with the same supernova framework, leaving them free to choose which galaxies to measure, which photometry source to use, and the settings of the tip detection, using public data and open code so that the analysis may be inspected by anyone. The 26 submissions from 22 participants give values of $H_0$ with a
median of $77.10$\,km\,s$^{-1}$\,Mpc$^{-1}$ and a standard deviation of
$7.38$\,km\,s$^{-1}$\,Mpc$^{-1}$ ($5.11$\,km\,s$^{-1}$\,Mpc$^{-1}$ for the scaled median absolute deviation, which is less sensitive to the highest values), about 3.4 (2.4) times the
uncertainty of a typical single submission. Because all submissions analyze the same catalogs and supernovae, these values are correlated, and the median summarizes the submissions rather than providing a combined measurement of $H_0$. The findings here characterize
the submissions; interpreting them, including judging whether any one value
of $H_0$ is more accurate than another, is beyond the intended scope of this
paper. Our main findings are as follows.

\begin{enumerate}
\item \textbf{TRGB analysis and sample-selection choices produce a
wide spread in $H_0$ among the submissions.}
This illustrates the sensitivity of the inferred
$H_0$ to the selection and analysis of the TRGB observations when these
choices are left to individual participants.
\item \textbf{Which galaxies are included matters most.} The choice of
  calibrator galaxies accounts for a large part of the variance of $H_0$, but the
  exact share depends on a few submissions and on four galaxies whose median
  distances appear biased. Their distances also differ more between submissions than those of most other calibrators. This sensitivity to which calibrators are included matches
  that found for TRGB-calibrated SNe~Ia by \citet{Li2026}, who trace
  differences between subsamples to both small-sample statistics and sample
  selection.
\item \textbf{One setting of the tip detection is related to $H_0$.}
  Participants used smoothing widths from $0.05$ to $0.40$\,mag and changed
  the spatial and color selections often. The smoothing width shows no clear
  relation to $H_0$, but submissions that changed the spatial scale, which
  sets the central region of each galaxy whose stars are excluded, for more
  of their galaxies tend to give a higher $H_0$; whether the spatial scale
  itself causes this is not established.
\item \textbf{The result resembles the CATs algorithm
  variants.} Like the 108 variants of \citet{Scolnic2023}, the submissions form
  a core with a tail to high $H_0$. The challenge values are centered about
  $4$\,km\,s$^{-1}$\,Mpc$^{-1}$ higher. In CATs, keeping only the variants that retain all host galaxies largely removes the high tail. The analogous cut here, the submissions that use at least 13 of the 14 calibrators, does not: these submissions span 75.24–93.45 km s$^{-1}$ Mpc$^{-1}$, and all of them include the four galaxies that are not in the CATs sample. 
\end{enumerate}

More broadly, the challenge shows that a community exercise with shared data
and code can measure how the choices made within one TRGB analysis notebook
propagate into $H_0$. Methods that model the luminosity function directly,
such as those of \citet{Makarov2006}, \citet{Wu2014}, \citet{Anand2022}, \citet{Li2022Gaia, Li2023Gaia}, and
\citet{Gibbon2026}, and references therein,  were not part of the challenge. Because the data and code are public, and the submissions are public as well, the comparison can be checked, extended with new measurements, or repeated with different choices by anyone in the community. This openness was a goal of the challenge in its own right, not only a means of running it. This approach can also be extended to other standard candles and approaches to measuring $H_0$ in the future. Future rounds could also be blinded,
for example by hiding an offset in the supernova absolute magnitude or in
the anchor distance until submission.

The Data Challenge's broader impact within the CosmoVerse network will be to provide a framework within which to identify different routes by which observations lead to measurements of the expansion rate and other cosmological parameters. TRGB observations provide a key probe for constraining the Hubble constant, but other important aspects of the distance ladder may also benefit from this approach. The CosmoVerse network offers a platform on which to consider these future challenges and opportunities.

\section*{Acknowledgments}
Siyang Li is supported by the Brinson Foundation as a Brinson Prize Postdoctoral Fellow.
Eleonora Di Valentino is supported by a Royal Society Dorothy Hodgkin Research Fellowship. 
Jackson Levi Said would like to acknowledge funding from ``Xjenza Malta'' as part of the ``Technology Development Programme'' DTP-2024-014 (CosmicLearning) Project. 
Ayush Hazarika acknowledges support from the National Natural Science Foundation of China (NSFC) under grants 12347103 and 12547104, as well as the Fundamental Research Funds for the Central Universities.
Swayamtrupta Panda is supported by the International Gemini Observatory, a program of NSF NOIRLab, which is managed by the Association of Universities for Research in Astronomy (AURA) under a cooperative agreement with the U.S. National Science Foundation, on behalf of the Gemini partnership of Argentina, Brazil, Canada, Chile, the Republic of Korea, and the United States of America.
Masroor C. Pookkillath acknowledges support from JSPS KAKENHI Grant No. 26KF0150.
Antonio Quintana-Estellés and Pilar Ruiz-Lapuente are very grateful for the support of the COST Action CA21136 -- ``Addressing observational tensions in cosmology with systematics and fundamental physics (CosmoVerse).''
Chandra Shekhar Saraf is supported by the KASI project ``Research on the principles of the accelerating expansion of the Universe'' (project 2026183201).
The work of Miguel Barroso Varela is supported by FCT (Fundação para a Ciência e Tecnologia, Portugal) through grant 2024.00457.BD with DOI identifier \href{https://doi.org/10.54499/2024.00457.BD}{10.54499/2024.00457.BD}. Centro de Fisica do Porto is partially funded by Fundação para a Ciência e Tecnologia (FCT) under the grant UID04650-FCUP.
Vasudev Mittal acknowledges financial support from the Physics Foundation Scholarship and the University of Sydney's Postgraduate Research Support Scheme.
José Antonio Nájera is supported by Royal Society University Research Fellowship grant 211046.
Alexei V. Filippenko is grateful for financial assistance from many donors.
Richard Stiskalek is supported by a Career Development Research Fellowship at St John's College, Oxford, and a Hintze Prize Fellowship in Cosmology at the Oxford Centre for Astrophysical Surveys, funded through generous support from the Hintze Family Charitable Foundation.
Du\v{s}ko Borka acknowledges funding from the Ministry of Science, Technological Development and Innovations of the Republic of Serbia, Project contract  451-03-33/2026-03/200017.
Vesna Borka Jovanovi\'{c} acknowledges funding from the Ministry of Science, Technological Development and Innovations of the Republic of Serbia, Project contract 451-03-33/2026-03/200017.
Umut Demirbozan was supported by the ErUM Data Project 05D25PE2 (PhysicsLLM), funded by the German Federal Ministry for Research, Technology and Space (BMFTR), and by the state of North Rhine-Westphalia as part of the Lamarr Institute for Machine Learning and Artificial Intelligence.
Predrag Jovanovi\'{c} acknowledges support from the Ministry of Science, Technological Development and Innovations of the Republic of Serbia through the Project contract  451-03-33/2026-03/200002.
Nils A. Nilsson was supported the Institute for Basic Science under the project code IBS-49 R018-D3 and acknowledges support from PSL/Observatoire de Paris.
Marika Asgari acknowledges the UK Science and Technology Facilities Council (STFC) under grant number ST/Y002652/1 and the Royal Society under grant number ICAR1231094.
Nihan Kat{\i}rc{\i} is supported by the COST Action CA21136 ``CosmoVerse'' (COST, European Cooperation in Science and Technology) and by The Scientific and Technological Research Council of T\"{u}rkiye (T\"{U}B\.{I}TAK) under the 1001 program, project no. 126F213.
Claude (Anthropic) was used to assist in preparing this manuscript.
This article is based upon work from the COST Action CA21136 -- ``Addressing observational tensions in cosmology with systematics and fundamental physics (CosmoVerse),'' supported by COST -- ``European Cooperation in Science and Technology''.

\newpage
\appendix
\section{The CosmoVerse School} \label{school}

Measuring a TRGB distance requires both an understanding of the
underlying stellar physics and practical judgment in analyzing
observational data. The school associated with the CosmoVerse Data
Challenge was designed to connect these aspects, providing participants
with a common grounding in the distance ladder and opportunities to
apply what they learned using the challenge notebooks. This preparation
was intended to help participants make informed methodological choices
and interpret their consequences for the inferred value of $H_0$.

CosmoVerseSchool@Sofia 2026, the second school of the CosmoVerse COST
Action (CA21136), was held from 25 to 29 May 2026 at the Institute of
Mechanics of the Bulgarian Academy of Sciences (IMech-BAS) in Sofia,
Bulgaria, with 70 registered participants. Organized in the context of
Working Group~1 (Observational Cosmology and Systematics), in close
connection with Working Group~2 (Data Analysis in Cosmology), the school
focused on the interface between observational measurements and their
statistical analysis. Its program combined lectures, practical
exercises, coding demonstrations, and dedicated time to work on the
Data Challenge.

The scientific organizing committee comprised S.~Li, J.~Levi Said,
E.~Di Valentino, N.~Frusciante, A.~Pollo, and M.~Asgari. The local
organizing committee comprised D.~Staicova, L.~Anguelova, M.~Stoilov,
P.~Danev, and H.~Tonchev, from the Institute for Nuclear Research and
Nuclear Energy of the Bulgarian Academy of Sciences. There was no
registration fee, and the CosmoVerse COST Action provided travel and
subsistence support to a number of student participants, helping to
make the training accessible to researchers with different levels of
institutional support.

A distinctive feature of the school was that all five lecturers were
young researchers actively working on the observations and methods
being taught. Their proximity to the students in career stage helped
create an approachable atmosphere in which participants felt
comfortable asking questions and engaging in discussion. This was
particularly valuable during the practical sessions, where students
could raise questions about individual analysis steps, discuss
difficulties with the code, and connect the lecture material to their
own measurements. The interaction between lecturers and participants
was therefore an important part of the learning experience.

The scientific content was organized around three complementary
themes. Theme~A introduced the distance ladder and Cepheid variables.
L.~Breuval and T.~Sicignano covered the Hubble tension, the Cepheid
period--luminosity relation, its metallicity dependence, and the
measurement of stellar abundances from spectroscopy. These lectures
established the observational and physical foundations of stellar
distance measurements and provided context for the alternative
calibration based on the TRGB.

Theme~B focused on the TRGB and the photometric measurements required
to use it as a distance indicator. S.~Li and K.~Owens discussed the
physics of the tip, stellar photometry, the available datasets, and
the practical measurement procedure, including the construction and
smoothing of luminosity functions, edge detection, and uncertainty
estimation. The lectures connected these measurements to the
calibration of distances to SN~Ia host galaxies through the geometric
distance to NGC~4258. They also introduced the principal sources of
systematic uncertainty and the notebooks prepared for the Data
Challenge. Theme~C, delivered by Y.~Murakami, completed the connection
to $H_0$ by treating SNe~Ia as standardizable candles, explaining their
role in the distance ladder, and discussing systematic uncertainties
and prospects for LSST, Roman, and Euclid.

The timetable followed a progression from background concepts to
practical application. Monday was devoted to the distance ladder,
Cepheids, and stellar abundances, concluding with exercises and
notebook work. Tuesday focused on TRGB physics, photometry, the
challenge datasets, and distances to SN~Ia hosts. Wednesday connected
the TRGB tutorial to SN~Ia standardization and a distance-ladder coding
demonstration. Thursday addressed SN~Ia systematics and future surveys,
followed by a synthesis session and dedicated work on the challenge.
This progression allowed participants to encounter the relevant
concepts before applying them within the analysis pipeline.

The practical component gave students an opportunity to move beyond
following a calculation and begin assessing the measurements
themselves. By working with the notebooks, participants could inspect
the stellar populations selected for analysis, examine the features
used to identify the TRGB, and explore the response to changes in
their choices. They could then follow how the resulting distances
entered the SN~Ia calibration and the determination of $H_0$.
Discussion with the lecturers helped connect these numerical steps
to their physical interpretation and to the treatment of uncertainty.

The program also included lightning talks, giving participants an
opportunity to present their own research and learn about the
interests of others attending the school. The best lightning-talk
award went to Umut Emek Demirbozan for his presentation,
\emph{An AI-Assisted Personalized Literature Explorer: Contextual
Language Model Assisted arXiv Retrieval (CLArXivR)}. The presentation
introduced a tool for identifying and prioritizing relevant research
papers using a researcher's publication history and a selected
literature collection. These contributions broadened the scientific
exchange beyond the lecture topics and gave students an active role
in the program.

The school thus provided both preparation for the Data Challenge and
a setting in which participants could develop practical familiarity
with distance-ladder analyses through direct interaction with the
researchers performing them. In the following subsections, we
summarize the principal lecture content, moving from Cepheid-based
distance measurements to the TRGB, the challenge tutorial, and the
SN~Ia analysis, before concluding with the synthesis lecture.

\subsection{Lecture 1 -- The Hubble Tension \& the distance ladder(s)}
\textbf{Lecturer: Louise Breuval}

The opening lecture of the school introduced the measurement of the Hubble constant ($H_0$) and the tension that motivates the rest of the program. We began with history. A century ago, it was unclear whether the Milky Way constituted the entire Universe. Edwin Hubble's distance measurements showed that the spiral nebulae are separate galaxies. By relating their redshift velocities to their distances, he also showed that the Universe is expanding, with $H_0$ as the slope of that relation. We emphasize that $H_0$ matters beyond its own value, because its inverse sets the age of the Universe. For a flat universe with $\Omega_{\rm m}=0.3$ and $\Omega_\Lambda=0.7$, a value around 73~km~s$^{-1}$~Mpc$^{-1}$ implies an age of roughly 12.9 billion years, while a value around 67 implies about 14.1 billion years. The 1998 discovery of the accelerated expansion from Type Ia supernovae (SNe~Ia), attributed to dark energy, was presented as a major result from a very active field.

Two independent routes to $H_0$ were described during this lecture. The early-Universe route combines measurements of the cosmic microwave background (CMB), principally from Planck, with the $\Lambda$CDM model to predict the current expansion rate, giving 67.4 $\pm$ 0.5 km/s/Mpc. Ground-based experiments such as the South Pole Telescope and the Atacama Cosmology Telescope corroborate this value. The late-Universe route is the distance ladder, which combines distances and redshifts empirically without assuming a cosmological model. Its most precise result, from the SH0ES team, is about 73 km/s/Mpc with an uncertainty near 1 km/s/Mpc. The two values now differ at almost six sigma. Because the Universe should have only one expansion rate, the discrepancy implies a possible error in the CMB measurement, in the distance ladder, or in $\Lambda$CDM itself. We highlight that the CMB result is well cross-checked and that the ladder has been extensively tested as well. That leaves $\Lambda$CDM, although proposed modifications have so far tended to spoil agreement elsewhere.

The lecture then builds up the ladder from familiar ideas. Parallax, measured by the Gaia satellite, provides geometric distances. Standard candles, objects of known luminosity, yield distances from their observed brightness. Objects of known size, such as eclipsing binaries, yield distances from orbital modeling. The two principal standard candles in the SH0ES ladder are Cepheid variables, which obey a tight period--luminosity relation, and SNe~Ia, which are much brighter; the TRGB, discussed below, is an alternative to Cepheids. The ladder has three rungs. First, Cepheids are calibrated in four anchor galaxies: the Milky Way with Gaia parallaxes, the Large and Small Magellanic Clouds with eclipsing binaries, and NGC 4258 with its maser. Second, Cepheids in galaxies that also host SNe~Ia set the absolute brightness of the SNe~Ia. That sample has grown from about 19 SNe~Ia a decade ago to 42. Third, SNe~Ia in the Hubble flow, at redshifts roughly from 0.02 to 0.2, give $H_0$. Observing all these Cepheids with the Hubble Space Telescope (HST) keeps the photometry homogeneous, and near-infrared observations limit the effect of dust.

The lecture then addressed possible systematics. All Gaia parallaxes require a zero-point correction, which depends on magnitude and color; because Cepheids are comparatively bright, this offset is difficult to determine for them from external information, so we described an alternative approach based on Cepheids in open clusters. By averaging the parallaxes of the fainter member stars, seventeen cluster Cepheids matched the precision of seventy-five individual ones. This brought the $H_0$ uncertainty below 1 km/s/Mpc for the first time. Crowding by neighboring stars had long been a suspected bias, and JWST's finer resolution allowed a direct test. It reduced the scatter in the period--luminosity relation by a factor of 2.5, and its distances agreed with HST's for about half the host galaxies. A host with a tidal companion free of old stars gave the same conclusion. Crowding therefore does not explain the tension, although the smaller JWST sample constrains $H_0$ less precisely than the HST sample.

Finally, we described independent efforts that do not involve Cepheids and SNe~Ia. These include a ladder based on the tip of the red giant branch and surface-brightness fluctuations, and a combined network of methods that accounts for their covariances. We closed by explaining that Edwin Hubble's original $H_0$ value was about a factor of seven too high because he confused classical and Type II Cepheids.

\subsection{Lecture 2 -- Measuring astronomical distances with Cepheid variables}
\textbf{Lecturer: Teresa Sicignano}

\subsubsection*{Scientific Motivation}

Pulsating variable stars are primary distance indicators, linking stellar astrophysics to precision cosmology. Classical Cepheids provide a key rung of the local distance ladder, connecting geometric anchors to galaxies hosting Type~Ia supernovae (SNe~Ia). Other pulsators, including Type~II and anomalous Cepheids, follow their own period--luminosity (PL) relations and offer complementary routes to the distance scale. As distance measurements approach percent-level precision, understanding the physical origin of these relations and controlling their intrinsic dispersion and systematic uncertainties become increasingly important. Accurate calibration of pulsating stars is therefore essential for robust distance measurements and, ultimately, for the determination of $H_0$.

\subsubsection*{Main Concepts and Methodology}

Pulsating stars become useful distance indicators because their variability is linked to fundamental stellar properties. In radial pulsators, the pulsation period is related to the mean stellar density and therefore to mass, luminosity, and effective temperature. Stars pulsate while crossing the instability strip, but different stellar populations and evolutionary phases produce different classes of variables, including classical, Type~II, and anomalous Cepheids, each following distinct PL relations.

Observationally, the main steps are to identify variable stars, measure their pulsation periods from well-sampled light curves, determine intensity-averaged magnitudes, and construct PL or Period--Wesenheit (PW) relations. The PL relation is a projection of the more general Period--Luminosity--Color relation and therefore has an intrinsic width. Wesenheit magnitudes combine photometric bands to reduce the effects of extinction, assuming an adopted extinction law.

For Galactic Cepheids, the absolute zero-point can be calibrated using Gaia parallaxes and multiband photometry. Once calibrated, PL/PW relations provide absolute magnitudes and hence distance moduli. Independent geometric distances, such as the eclipsing-binary distance to the LMC, do more than validate the resulting scale: they contribute measurably to the absolute calibration, and, because the LMC is more metal-poor than the Milky Way, they also constrain the metallicity dependence of that calibration.

\subsubsection*{Analysis Choices and Sources of Uncertainty}

Several methodological choices affect distances inferred from pulsating stars. The nonzero width of the instability strip introduces intrinsic scatter in PL relations, while light-curve sampling and the determination of intensity-averaged magnitudes contribute measurement uncertainties. Interstellar extinction changes both magnitudes and colors; Wesenheit relations reduce this effect, but depend on the adopted extinction law and coefficient. Chemical composition can modify both the slope and zero-point of pulsation relations and therefore represents an additional astrophysical systematic. Further choices include the selection of pulsator class and pulsation mode, sample selection, outlier rejection, the adopted fitting relation, and the treatment of Gaia parallaxes. Different assumptions in these steps can lead analysts using the same underlying observations to infer different calibrations and distances.

\subsubsection*{Connection to the Data Challenge}

The lecture provided the physical and observational framework needed for the pulsating-star stage of the Data Challenge. Participants worked with periods, multiband photometry, and Gaia parallaxes to calibrate Galactic Cepheid PL/PW relations, explored the effect of including metallicity, and then applied the calibrated relation to infer an external-galaxy distance. A further exercise used LMC pulsators to recover its distance and compare it with an independent geometric calibration. While the input data were common, participants could make different choices in fitting, sample treatment, and adopted relations. The resulting calibrated distance information was then used as input for subsequent stages of the distance ladder.

\subsubsection*{Key Takeaway}

Pulsating stars provide distance measurements because their periods encode information about their intrinsic luminosities, but each class follows relations shaped by its stellar properties and evolutionary state. Reliable distances therefore require both a physical understanding of pulsation and careful control of extinction, calibration choices, sample definition, and other sources of systematic uncertainty.

\subsubsection*{Sources}

The lecture material was based on and adapted from \citet{Leavitt1912}, \citet{Hubble1925}, \citet{Sandage1958}, \citet{MadoreFreedman1991}, \citet{Udalski2015}, \citet{Riess2018}, \citet{Cusano2021}, \citet{Riess2022}, \citet{Bhardwaj2022}, \citet{Ripepi2022}, \citet{GaiaDR3}, \citet{DeSomma2024}, \citet{Sicignano2024}, \citet{Trentin2024}, \citet{Breuval2025}, \citet{Sicignano2025}, \citet{Lengen2026}, and \citet{CruzReyes2026}.

\subsection{Lecture 3 -- The Cepheid Metallicity Dependence}
\textbf{Lecturer: Louise Breuval}
\medskip

The third lecture continued the discussion of Cepheid systematics with the effect that used to dominate the Cepheid error budget: the dependence of Cepheid luminosity on metallicity. The central concern is consistency between the first and second rungs of the distance ladder. Cepheids in the Milky Way, in NGC 4258, and in the SN~Ia host galaxies are metal-rich, whereas those in the Large and Small Magellanic Clouds are metal-poor. The period--luminosity relation is therefore extended with a term $\gamma$*[Fe/H] (or [O/H]), where a negative $\gamma$ (which is the observed direction) means that metal-rich Cepheids are more luminous.

Two empirical approaches were described. The first uses the metallicity gradient within a single galaxy, comparing Cepheids near the center with those in the outskirts. The second compares period--luminosity intercepts between galaxies of different metallicities. Both give $\gamma$ between about $-0.2$ and $-0.3$ mag/dex, and recent theoretical models, including rotating Geneva models, also predict a negative value. We showed that even an implausibly extreme value of $\gamma$ would leave $H_0$ above $\sim 72.6$ km/s/Mpc, so metallicity alone cannot resolve the tension. Including the Milky Way and NGC 4258 as anchors balances the weight of the metal-poor Magellanic Clouds and limits the sensitivity of $H_0$ to $\gamma$ to below 0.8 km/s/Mpc. The correction still matters: it makes the four anchors mutually consistent, and it affects distances to metal-poor Local Group galaxies. For example, measuring the distance to M33 while calibrating Cepheids in the Milky Way rather than in the LMC changes the distance by about 30 kpc, or roughly 3.6\% of its 840 kpc distance \citep{Breuval2023}.

The lecture also reviewed published measurements that deviate from this range and explained why. A zero value from the Magellanic Clouds was traced to the SMC's elongated geometry along the line of sight. Restricting the sample to the SMC core or applying a geometric correction restores a negative $\gamma$. LMC-only studies span a limited range of metallicity, which weakens their leverage on $\gamma$. Baade-Wesselink distances depend on the projection factor, whose calibration remains uncertain. Some Milky Way samples reach low metallicity only through distant stars with a low parallax signal-to-noise ratio (S/N). Comparisons with TRGB distances assume that the TRGB's own, weaker metallicity dependence is negligible or correctly calibrated.

The recommendations that followed were to use a metallicity range wider than 0.5 dex by combining the Milky Way, LMC, and SMC, and to use full light curves when possible, which give more accurate intensity-averaged magnitudes than sparsely sampled or phase-corrected data. They also included applying the official Gaia parallax corrections, avoiding a mix of photometric systems, and adopting consistent metallicity scales. Gaia DR4, expected in December 2026, should improve the Milky Way calibration further. Metallicities come from direct spectroscopy for nearby Cepheids and from H II regions in the supernova hosts, and the two agree in the Milky Way. The metallicity term, once a leading contribution to the $H_0$ uncertainty, is now well controlled, and there is a broad consensus on its sign and value.

\subsection{Lecture 4 -- The Cepheid metallicity dependence \& how to measure stellar abundances}
\textbf{Lecturers: Louise Breuval \& Teresa Sicignano}

\subsubsection*{Scientific Motivation}

Metallicity is an important systematic in the use of Cepheids as distance indicators because stellar chemical composition affects their luminosity at a fixed period. This is particularly relevant when comparing Cepheids across the distance ladder, since the Milky Way, Magellanic Clouds, NGC~4258, and SN~Ia hosts span different metallicity regimes. A metallicity-dependent correction must therefore be included to obtain a consistent calibration. The lecture addressed both sides of this problem: how the metallicity dependence of Cepheid luminosities is constrained and, in its second part, how stellar metallicities and chemical abundances are actually measured from high-resolution spectra.

\subsubsection*{Main Concepts and Methodology}

The Cepheid period--luminosity relation can be extended with a metallicity term, commonly written as $\gamma[\mathrm{Fe/H}]$ or $\gamma[\mathrm{O/H}]$, where observations generally indicate a negative $\gamma$, corresponding to metal-rich Cepheids being brighter at a fixed period. Two main empirical approaches were discussed. One exploits metallicity gradients within individual galaxies, comparing Cepheids at different galactocentric radii, while the other compares PL/PW intercepts among galaxies spanning different mean metallicities. Current empirical determinations typically yield $\gamma \approx -0.2$ to $-0.3$~mag,dex$^{-1}$, broadly consistent with recent theoretical predictions.

The second part of the lecture focused on how the required stellar abundances are measured spectroscopically. After continuum normalization and radial-velocity correction, stellar parameters and abundances can be derived using either equivalent widths through the curve-of-growth method or direct comparison with synthetic spectra. Effective temperature, surface gravity, and microturbulence are constrained through excitation and ionization equilibria and the absence of abundance trends with line strength. Spectral synthesis instead models complete line profiles and is particularly useful for blended regions. For Cepheids, additional care is required because atmospheric parameters vary throughout the pulsation cycle and static atmosphere models are applied to intrinsically dynamic atmospheres.

\subsubsection*{Analysis Choices and Sources of Uncertainty}

Metallicity corrections depend on the adopted abundance scale, metallicity baseline, geometry, and distance calibration of the Cepheid samples. Narrow metallicity ranges, the line-of-sight structure of the SMC, uncertain Baade-Wesselink projection factors, and low-S/N parallaxes can bias empirical estimates of $\gamma$. Spectroscopic abundances introduce additional methodological choices, including continuum placement, line selection, atomic data, atmosphere models, local thermodynamic equilibrium (LTE) and non-LTE assumptions, stellar-parameter determination, and the adopted solar reference composition. These choices can produce systematic differences between analyses. Cepheids introduce further complexity because temperature, gravity, and velocity fields vary with pulsation phase, making the treatment of their dynamic atmospheres particularly important for homogeneous metallicity measurements.

\subsubsection*{Connection to the Data Challenge}

This lecture provided the physical and observational basis for treating metallicity in the Cepheid context of the Data Challenge. Participants could compare calibrations obtained with and without a metallicity term and examine how this choice changes the scatter and inferred distances. The discussion of spectroscopy explained how the metallicities entering such relations are obtained and why measurements from different analyses may not be directly interchangeable. Particular emphasis was placed on using homogeneous metallicity scales and sufficiently broad abundance baselines. The resulting metallicity-corrected Cepheid calibration can then be propagated to subsequent stages of the distance ladder.

\subsubsection*{Key Takeaway}

Cepheid metallicity is not simply an additional fitting parameter: its effect must be measured over a sufficiently broad abundance range using reliable and homogeneous chemical abundances. Spectroscopic choices can themselves introduce systematic offsets, so controlling both the luminosity--metallicity relation and the abundance analysis is essential for a consistent distance-scale calibration.

\subsubsection*{Sources}

The lecture material was based on and adapted from \citet{Ripepi2021}, \citet{Breuval2021}, \citet{Breuval2022}, \citet{Romaniello2022}, \citet{Trentin2024}, and \citet{Breuval2025b}.

\subsection{Session: Exercises \& Notebooks}
\textbf{Session leaders: Louise Breuval and Teresa Sicignano}

At the end of the first day, Louise Breuval and Teresa Sicignano led a session of exercises and notebooks. It is not summarized separately here.

\subsection{Lecture 5 -- The Tip of the Red Giant Branch: Theoretical Foundations \& Observational Practice}
\textbf{Lecturer: Siyang Li}

\subsubsection*{Scientific motivation}

The tip of the red giant branch (TRGB) offers an alternative standard candle to Cepheid variables for anchoring the extragalactic distance ladder. Because it relies on old, low-mass stars, it can be measured in the stellar halos of galaxies, where crowding and dust extinction are also low. Establishing why the TRGB has a nearly fixed luminosity, and how that luminosity is measured in practice, is a prerequisite for using it as an independent rung of the distance ladder feeding into $H_0$.

\subsubsection*{Main concepts and methodology}

Low-mass ($M\lesssim2\,M_\odot$), old ($>1$--$2$~Gyr) stars ascending the red giant branch (RGB) support their hydrogen-shell burning on an electron-degenerate helium core. Because degeneracy pressure depends on density but not temperature, the core can heat without expanding, so once the core mass reaches $M_{c,\mathrm{flash}}\approx0.50\,M_\odot$, the triple-alpha reaction ignites explosively in a helium flash ($T_\mathrm{core}\approx10^8$~K, $L_\mathrm{core}\approx10^{11}\,L_\odot$ for hours, unobservable from outside) \citep{Iben_1983A}. The onset of the flash is set by fixed nuclear and quantum physics, essentially independent of stellar age and mass and only weakly dependent on metallicity. The luminosity observed at the photosphere, however, depends on the envelope opacity and hence on metallicity; in the $I$ band, this dependence is small for metal-poor populations, giving a nearly constant tip luminosity, $M_I(\mathrm{TRGB})\approx-4$~mag. Observationally, this appears as a sharp discontinuity (edge) in the RGB luminosity function (LF): a steep power-law rise toward fainter magnitudes truncated abruptly at the tip. Because raw star counts near the tip are Poisson-noisy, the LF can first be smoothed with a Gaussian-windowed local regression (GLOESS; \citealt{Persson2004}); then a Sobel edge-detection filter \citep{Lee1993}, most commonly the three-point kernel $S(m)=\phi(m+1)-\phi(m-1)$ (wider variants also exist, e.g., \citealt{Sakai1996}), is applied to the smoothed LF; the tip magnitude is the peak of the Sobel response, $m_\mathrm{tip}=\mathrm{argmax}[S(m)]$. The statistical uncertainty is estimated by bootstrap resampling of the full color-magnitude diagram (CMD). The measurement is carried out separately for each galaxy. Target fields are chosen in the stellar halo of each galaxy, away from bright foreground stars (to avoid diffraction spikes) and with little dust in the target galaxy itself. A measured tip is then checked with two robustness metrics defined relative to it: a tip-contrast ratio above 3 \citep{Wu2022,Li2023,Scolnic2023}, and at least about 400 RGB stars within 1\,mag fainter than the tip \citep{Madore2009}.

\subsubsection*{Analysis Choices and Sources of Uncertainty}

Several explicit choices enter a TRGB measurement and can shift the result: the GLOESS smoothing bandwidth $\tau$ (too small leaves the LF noisy and produces false edge detections; too large blurs the edge and biases the result, so the tip should be checked for stability over several values of $\tau$); the RGB color-selection window (the color dependence of the tip requires a calibration such as that of \citealp{Rizzi2007, JangLee2017, Newman2024a, Newman2024b}); the spatial clipping used to isolate halo fields from crowded or younger populations; and the Sobel weighting scheme (simple, Poisson, or Hatt-weighted), which can itself introduce a bias. Real data add further complications: finite star counts, photometric scatter across the edge, incompleteness near the detection limit, and asymptotic giant branch (AGB) stars that appear above the tip. In theoretical models, the $I$-band tip magnitude follows $M_I^{\rm tip}=-3.732+0.588\,[\mathrm{M/H}]+0.193\,[\mathrm{M/H}]^2$ for $-2.35\le[\mathrm{M/H}]\le-0.57$, while age changes the bolometric tip by $\le0.05$~mag between 10 and 20~Gyr \citep{SalarisCassisi1997}.

\subsubsection*{Connection to the Data Challenge}

This lecture supplies the physical and algorithmic background for the nine-step pipeline that turns raw Hubble Space Telescope (HST) imaging into a TRGB distance modulus: data download, point-spread-function (PSF) photometry, quality cuts, CMD/RGB selection, extinction correction, LF construction, GLOESS smoothing plus Sobel filtering, and final distance-modulus and uncertainty calculation. The CosmoVerse Data Challenge notebook begins at the quality-cut stage using an already-reduced photometric catalog, so participants directly exercise the smoothing, color-selection, and Sobel-filter choices introduced here.

\subsubsection*{Key takeaway}

The TRGB is a standard candle whose nearly universal luminosity is set by well-understood helium-flash physics, but recovering that luminosity from real, noisy star counts requires a specific measurement pipeline (smoothing, edge detection, bootstrap uncertainty) whose choices, such as the smoothing bandwidth, can bias the result and should be tested for stability.

\subsection{Lecture 6 -- How to Measure Stars: Methods and Challenges of Photometry}
\textbf{Lecturer: Kayla Owens}

\subsubsection*{Scientific Motivation}

Photometry, the measurement of the brightness of a source in an image, underlies every step of the distance ladder. This lecture explained how the stellar catalogs used in distance-ladder work are produced from telescope images.

\subsubsection*{Main Concepts and Methodology}

A telescope projects light onto a detector, where photons excite electrons that are collected in pixels and counted at readout, so an image records detail only down to the pixel scale. In aperture photometry, the counts inside a circular aperture are summed, the sky background measured in a surrounding annulus is subtracted for the same area, and the result is calibrated by dividing by a reference flux. This works well for isolated stars. The extragalactic distance scale instead relies on point-spread-function (PSF) photometry, in which counts are weighted by the shape of a point source in the image. The PSF depends on the telescope (the shape of the mirror and its obstructions), on the instrument (the light path through the optics and the detector response), and on wavelength, since the angular size scales as $\lambda/D$. The basic procedure is to prepare the images, measure or adopt a PSF, detect sources, fit the brightness and position of each source with the PSF, subtract the fitted sources and search the residual image again, and finally apply aperture corrections and the photometric zero-point. Preparing the images involves aligning exposures taken over several visits to better than 0.1~pixel and measuring stars either on each exposure, which is often necessary for variable stars, or on a drizzled stack, which improves the signal-to-noise ratio and resolution, removes cosmic rays, and applies geometric and pixel-area corrections.

\subsubsection*{Analysis Choices and Sources of Uncertainty}

The PSF can be a model provided by the telescope team (simple but sometimes inaccurate), a model adjusted to stars in the image, or an empirical PSF built entirely from the image (ideal but not always possible). Because a PSF fit recovers only about 70--80\% of the flux of a star, an aperture correction is derived from the growth curves of bright, isolated stars as the difference between their PSF and aperture magnitudes. Artificial stars, injected into the images and measured with exactly the same procedure as the real ones, give the recovery rate as a function of magnitude, realistic uncertainties (photometry software tends to underestimate them), and the size of crowding and blending biases, which are important for Cepheids but usually small for the TRGB. They also reveal errors in the sky measurement and blends of stars too close to be separated. Independent reductions of the same images can agree closely, as in a comparison of ALLFRAME and DOLPHOT \citep{Dolphin2000} magnitudes shown in the lecture, but in an extreme example, two reductions differed in zero-point by $-0.120$~mag in F555W and $-0.072$~mag in F814W.

\subsubsection*{Connection to the Data Challenge}

Participants did not perform photometry themselves: the challenge started from catalogs that had already been reduced and cleaned with quality cuts (Section~\ref{sec:data}). The lecture described the steps that produce such catalogs and the uncertainties they carry, including incompleteness at faint magnitudes, crowding, and the photometric zero-point.

\subsubsection*{Key Takeaway}

A magnitude in a stellar catalog is the outcome of a sequence of choices, from image alignment and the PSF model to aperture corrections and the zero-point. Artificial-star tests are needed to know the true uncertainties and completeness, and different reductions of the same images do not always agree.

\subsection{Lecture 7 -- An Overview of TRGB Data Sets for the Data Challenge}
\textbf{Lecturer: Kayla Owens}

\subsubsection*{Scientific Motivation}

This lecture introduced the datasets used in the Data Challenge. It followed the description of how stellar photometry is measured in Lecture~6.

\subsubsection*{Main Concepts and Methodology}

The challenge provides HST photometry of resolved stars in 18 galaxies: the geometric anchor NGC~4258 and 17 other galaxies, 14 of which have SNe~Ia in the challenge's SN~Ia sample. Every galaxy is covered by two separately compiled sets of public catalogs, the Extragalactic Distance Database (EDD; \citealt{Tully2009}) and the Comparative Analysis of TRGBs (CATs; \citealt{Wu2022,Li2023,Scolnic2023}). The EDD set has one catalog for each HST program provided for a galaxy (22 catalogs), and the CATs set has one catalog for each host galaxy plus 11 separate fields for NGC~4258 (28 catalogs). All catalogs come from ACS/WFC images, except the two for NGC~5584, which come from WFC3/UVIS. Every catalog contains F814W, the band in which the tip is measured, and one bluer band, F606W or F555W, used to select red giants by color. Each row of a catalog is one detected source, with its Vega magnitudes, uncertainties, and photometric quality parameters. The EDD catalogs also give sky coordinates, which the notebook uses to mask stars projected onto the disk of the galaxy, whereas the CATs catalogs already contain the spatial selection made by CATs. Alongside the photometry, each galaxy comes with its Milky Way foreground reddening from the maps of \citet{SchlaflyFinkbeiner2011} and the isophotal radius, axis ratio, and position angle that define the masked disk region.

\subsubsection*{Analysis Choices and Sources of Uncertainty}

Quality cuts were applied to the photometry by each source and can differ slightly between EDD and CATs, so participants were advised to use catalogs from only one source within a single $H_0$ measurement. A table later added to the challenge repository records notes on data quality for 13 of the 50 catalogs: sparse color-magnitude diagrams (M66 in program 10433 and NGC~3021), fields centered on the disk rather than the halo (M101 in program 13691, NGC~3370, and NGC~5584), contamination from the galaxy center (NGC~1404), streaks in the images (M66 in program 13691), and the four hosts (NGC~1309, NGC~3021, NGC~3370, and NGC~5584) that \citet{Scolnic2023} excluded from their $H_0$ analysis because no reliable tip was detected. Participants chose the photometry source and the galaxies to include. Within EDD, they could also choose among the catalogs available for M66, M101, and NGC~4258, whereas all CATs fields of a galaxy are always combined.

\subsubsection*{Connection to the Data Challenge}

These catalogs are the starting point of every submission: participants did not perform photometry themselves, but analyzed the provided catalogs (Section~\ref{sec:data} and Table~\ref{tab:catalogs}).

\subsubsection*{Key Takeaway}

The same galaxy can be represented by different catalogs with different fields, filters, and quality cuts.

\subsection{Lecture 8 -- Tip of the Red Giant Branch in SN Ia Host Galaxies}
\textbf{Lecturer: Siyang (Sean) Li}

\subsubsection*{Scientific Motivation}

Having established the TRGB as a standard candle, the next step is to use it to measure distances to the host galaxies of Type~Ia supernovae (SNe~Ia), which sets the absolute-magnitude zero-point of the SN~Ia Hubble diagram and hence $H_0$. Because the TRGB traces old halo populations rather than young Cepheids, it offers a route to the SN~Ia calibration that is largely independent of Cepheid-specific systematics, while still sharing the same geometric anchors -- making it a valuable cross-check in the distance-ladder approach to the Hubble tension.

\subsubsection*{Main Concepts and Methodology}

The distance ladder proceeds from geometric anchors -- the megamaser disk in NGC~4258, whose distance needs no stellar physics, standard candles, or calibration chain, mapped by very long baseline interferometry to $d=7.576\pm0.082\,(\mathrm{stat})\pm0.076\,(\mathrm{sys})$~Mpc \citep{Reid2019}, and detached eclipsing binaries in the Large Magellanic Cloud (LMC), $d=49.59\pm0.09\,(\mathrm{stat})\pm0.54\,(\mathrm{sys})$~kpc \citep{Pietrzynski2019} -- to the TRGB absolute magnitude, and finally to SN~Ia host galaxies. Concretely, one solves $m-M_\mathrm{anchor}=\mu_\mathrm{anchor}$ for the absolute tip magnitude $M_I(\mathrm{TRGB})$, then applies the same apparent-magnitude measurement in an SN~Ia host to solve for that host's distance modulus. The current HST/JWST sample has roughly 35 observed SNe~Ia in $\sim 30$ host galaxies with TRGB measurements, with target fields typically designed to observe the halos of galaxies. Because TRGB magnitudes measured in different fields of the same galaxy can disagree (the lecture showed a field-to-field dispersion of about 0.3~mag in NGC~4258), the Comparative Analysis of TRGBs (CATs) program standardizes measurements using the tip-contrast ratio $R$ (the ratio of the number of stars within 0.5~mag fainter than the tip to the number within 0.5~mag brighter), fitting a tip--contrast relation to correct each measurement to a common fiducial contrast before it enters the distance ladder \citep{Wu2022,Li2023,Scolnic2023}.

\subsubsection*{Analysis Choices and Sources of Uncertainty}

Population effects, with the relative numbers of RGB and AGB stars among the possible causes, can produce this field-to-field scatter, which motivated the contrast-ratio standardization above. Blending and crowding bias measured magnitudes brighter, which makes distances too small; this is mitigated by restricting measurements to uncrowded outer-halo fields, artificial-star tests, and (for JWST) a smaller PSF. Foreground dust extinction is corrected using $A_I=R_I\,E(B-V)$ with dust maps \citep{SchlegelFinkbeinerDavis1998,SchlaflyFinkbeiner2011}; halo fields typically have $A_I<0.05$~mag, inner-disk fields can exceed $A_I\approx0.1$~mag, and on high-extinction sightlines ($E(B-V)>0.1$ mag), the dust maps carry an uncertainty of $\sim 20$\%, so field selection itself is an analysis choice with a direct effect on the extinction systematic. Finally, the same methodological choices from Lecture~5 ($\tau$, color cuts, spatial clipping, Sobel weighting) propagate into the final $H_0$ and are recommended to be reported as a separate $\sigma_\mathrm{method}$ term rather than absorbed silently into the statistical error.

\subsubsection*{Connection to the Data Challenge}

The EDD \citep{Anand2022} and CATs measurements both use the public DOLPHOT photometry of the Extragalactic Distance Database (EDD; \citealt{Tully2009}), whereas the CCHP \citep{Freedman2019} used its own photometry with DAOPHOT. The EDD photometry is one of the two photometry sets that participants could use in the Data Challenge notebooks, the other being the CATs set. Participants measure their own TRGB magnitude and tip contrast for one or more SN~Ia host galaxies, convert to a distance modulus, and feed that result into the companion SN~Ia notebook, directly reproducing the anchor-to-host TRGB step described in this lecture using their own choice of smoothing, color cuts, and spatial clipping.

\subsubsection*{Key Takeaway}

Published differences among independent TRGB-based $H_0$ measurements are traceable to specific, identifiable analysis choices, such as SN~Ia and host-galaxy selection, peculiar-velocity corrections, and contrast-ratio standardization.

\subsection{Lecture 9 -- CosmoVerse Data Challenge TRGB Tutorial}
\textbf{Lecturer: Siyang (Sean) Li}

\subsubsection*{Scientific Motivation}

If two groups analyze the same data with the same underlying physics but reach different values of $H_0$, the discrepancy must be traced to analysis choices rather than to the data or the physics. The Data Challenge is designed to isolate exactly that: by fixing the supernova side of the calibration (dataset, covariance matrix, and solving code) identically for every participant and leaving only the TRGB-measurement choices free, any spread in the resulting $H_0$ values can be attributed specifically to TRGB methodology, answering how much of the published scatter in TRGB-based $H_0$ is a real methodological effect rather than a difference in underlying data.

\subsubsection*{Main Concepts and Methodology}

The challenge is built as a two-notebook pipeline. Notebook~1 (\texttt{1\_trgb/}) is a 71-cell walkthrough that takes public photometry (from the EDD or CATs catalogs) for a chosen target galaxy through spatial clipping, color/RGB selection, GLOESS smoothing (bandwidth $\tau$), Sobel edge detection, and bootstrap uncertainty estimation, then exports the resulting distance modulus to a \texttt{trgb\_results.csv} file (Section~9 of the notebook). Notebook~2 (\texttt{2\_sn\_to\_h0/}) reads the host distances from its input table \texttt{data/TRGB\_distances.csv}, which lists 40 calibrator hosts (the 37 SH0ES hosts of \citealt{Riess2022} plus 3 CATs-only hosts). In the distributed table, 37 rows carry Cepheid-based placeholder distances from \citet{Riess2022} and 3 are blank; participants replace these values with their own TRGB measurements, and hosts with a blank distance are dropped from the fit. The notebook combines these distances with a fixed SN~Ia dataset, covariance matrix, and solver script (\texttt{mu\_to\_H0.py}), and returns $H_0\pm\sigma$, the SN~Ia absolute magnitude $M_B\pm\sigma$, and a Hubble diagram. Diagnostic plots at each stage (the CMD, the luminosity function and Sobel response, and the bootstrap distribution) let a participant judge, before exporting a result, whether their chosen parameters have produced a clean single-peaked detection rather than an ambiguous or bimodal one.

\subsubsection*{Analysis Choices and Sources of Uncertainty}

The ground rules make the free parameters explicit: participants may vary the smoothing bandwidth $\tau$, the color-selection window, the spatial-clipping radius, and the Sobel weighting scheme, and they choose the target galaxies and the photometry source. The notebook also lets them change the number of bootstrap resamples, which affects the estimated statistical uncertainty but not the central value. Participants are asked to run a stability check that re-measures the tip with $\tau=0.05$, 0.10, 0.15, and 0.20, with the color limits shifted by $\pm0.2$~mag, and with spatial-clipping scale factors of 0.8, 1.0, and 1.2, and to flag any TRGB shift larger than 0.03~mag. By contrast, the supernova dataset, its covariance matrix, and the fitting code are fixed for everyone and must not be altered, so that all reported variation in $H_0$ can be unambiguously attributed to the participant's TRGB-side choices rather than to changes in the SN~Ia analysis.

\subsubsection*{Connection to the Data Challenge}

The notebook's submission checklist asks for the target galaxy and a justification of the parameter choices, the distance modulus and its uncertainty, the tip-contrast values for both the anchor and the target galaxy, a CMD figure, and a systematics paragraph. Participants could compare their results with the published EDD \citep{Anand2022}, CATs \citep{Scolnic2023}, and CCHP \citep{Freedman2019} distances, and all results and pipelines are to be collected in a public repository.

\subsubsection*{Key Takeaway}

A community data challenge that fixes the supernova side of the calibration and exposes only the TRGB-measurement choices as free parameters turns ``why do different groups get different $H_0$ values?'' from a qualitative debate into a directly testable, reproducible question.

\subsection{Lecture 10 -- Type Ia Supernovae as Standardizable Candles}
\textbf{Lecturer: Yukei S. Murakami}

\subsubsection*{Scientific Motivation}

The last rung of the distance ladder requires a distance indicator observed far enough away that peculiar velocities, typically a few hundred km~s$^{-1}$ \citep{Carr2022}, are small compared with the recession velocity (about 700~km~s$^{-1}$ at 10~Mpc), yet close enough that the dependence on the expansion history is weak \citep{Riess2022}. A Type~Ia supernova (SN~Ia) reaches $M_B\approx-19.3$~mag at peak \citep{Riess2022}, a luminosity that can rival that of its host galaxy, so SNe~Ia can be observed both in Cepheid or TRGB hosts and in the Hubble flow. We described what SNe~Ia are and how the standardized SN~Ia magnitude used in the Data Challenge is obtained.

\subsubsection*{Main Concepts and Methodology}

An SN~Ia is the thermonuclear explosion of a carbon-oxygen white dwarf, most likely in a binary system, although the progenitor systems have not been identified definitively \citep{Maoz2014}. SN~Ia spectra show no hydrogen and strong Si\,\textsc{ii} and S\,\textsc{ii} absorption, and the light curve is powered by the radioactive decay $^{56}$Ni$\,\rightarrow\,^{56}$Co$\,\rightarrow\,^{56}$Fe. Before standardization, SN~Ia peak luminosities have a dispersion of about 0.4~mag \citep{Hamuy1996}. Most of this dispersion correlates with two quantities measured from the light curve without knowledge of the distance: the light-curve width (stretch), with slower-declining SNe~Ia being brighter \citep{Phillips1993,Riess1996,Kasen2007}, and the color, with redder SNe~Ia being fainter because of both intrinsic color variation and host-galaxy dust \citep{Riess1996,Brout2021}. The modified Tripp relation \citep{Tripp1998,Scolnic2022},
\begin{equation}
\mu = m_B + \alpha x_1 - \beta c - M_B ,
\end{equation}
reduces the dispersion of the Hubble residuals to about 0.10--0.16~mag. The parameters $m_B$, $x_1$, and $c$ are obtained by fitting a model of the spectral energy distribution (SED) to multiband photometry. In SALT2 and SALT3 \citep{Guy2007,Kenworthy2021}, this model is a mean spectral time series plus a component scaled by $x_1$, multiplied by a color law; other models (e.g., MLCS2k2, SNooPy, BayeSN; \citealt{Jha2007,Burns2011,Mandel2022}) use analogous shape and color or extinction parameters. Normal and 91T-like SNe~Ia are routinely included in cosmological samples, whereas subluminous 91bg-like SNe~Ia deviate from the width--luminosity relation and are excluded \citep{Taubenberger2017}.

\subsubsection*{Analysis Choices and Sources of Uncertainty}

Each step from detection to $m_B$ involves choices that differ between surveys. The host-galaxy light is removed either by subtracting a template image or by modeling the host and the SN together. Each survey calibrates to its own standard stars, and these systems are cross-calibrated using Pan-STARRS1 photometry and HST CALSPEC standards \citep{Scolnic2015,Brout2022a}. SNe~Ia are classified from spectra or, in DES-SN5YR \citep{DES2024}, from their light curves. The host is identified visually or with the directional light radius \citep{Sullivan2006,Gupta2016}, and the redshift is preferably measured from the host spectrum, since SN spectra give $\sigma_z\approx 0.005$ \citep{Carr2022}. Compilations combine surveys that differ in these respects, such as targeted or untargeted searches and spectroscopic or photometric classification. Pantheon+ \citep{Scolnic2022} and Union3 \citep{Rubin2025} draw on largely overlapping samples but differ in calibration, light-curve model, selection, and standardization.

\subsubsection*{Connection to the Data Challenge}

The SN~Ia notebook uses the Pantheon+ magnitudes $m_B^{\rm corr}$, which include the light-curve fit, the Tripp standardization, and the bias corrections discussed in the following lecture. These inputs were identical for all participants, so that differences in $H_0$ reflect the TRGB measurements. The lecture showed what this fixed input contains and that the SN~Ia side involves analysis choices of its own, held constant by design. A coding session on the same day used the Pantheon+ data release with \texttt{mu\_to\_H0}, the distance-ladder solver later integrated into the SN~Ia notebook (Section~\ref{sec:sn_notebook}).

\subsubsection*{Key Takeaway}

An SN~Ia magnitude in a cosmological sample is the result of a chain of measurements and choices: photometry, calibration, classification, host association, redshift, and standardization. Compilations built largely from the same SNe~Ia differ in these choices, so a published SN~Ia dataset should be used with an understanding of how it was constructed.

\subsection{Session: Type Ia Supernovae in the Distance Ladder + Coding Demo}
\textbf{Lecturer: Yukei S. Murakami}

\medskip

This session combined a lecture on SNe~Ia in the distance ladder with a coding demonstration. As noted in Lecture~10, the coding part used the Pantheon+ data release with \texttt{mu\_to\_H0}, the distance-ladder solver later integrated into the SN~Ia notebook (Section~\ref{sec:sn_notebook}). The session is not summarized separately here.

\subsection{Lecture 11 -- Systematic Uncertainties in SN~Ia Cosmology}
\textbf{Lecturer: Yukei S. Murakami}

\subsubsection*{Scientific Motivation}

SN~Ia searches are limited in apparent magnitude, not in redshift, and the limiting magnitude changes with observing conditions. Selection effects therefore skew the observed sample toward brighter, bluer, and higher-stretch SNe~Ia \citep{Scolnic2016}; without correction, they bias the inferred distances and cosmological parameters (Malmquist bias). The lecture focused on the full redshift range used to constrain dark energy, where these corrections are needed for a cosmological fit. We described how current analyses correct for selection effects and model the host-galaxy dependence of SN~Ia luminosities, how the uncertainties of these corrections are propagated, and how upcoming surveys will change the field.

\subsubsection*{Main Concepts and Methodology}

The BEAMS with Bias Corrections (BBC) method \citep{Kessler2017}, used by Pantheon+ and DES-SN5YR \citep{Brout2022b,DES2024}, corrects each SN~Ia by the mean bias in $m_B$, $x_1$, and $c$ that simulations predict for SNe~Ia with similar redshift, stretch, and color. The simulations \citep{Kessler2009} reproduce each survey's cadence, observing conditions, photometric noise, detection and spectroscopic follow-up efficiency, and light-curve fitting. UNITY \citep{Rubin2015} and its update used for Union3 \citep{Rubin2025} instead model the selection within a hierarchical Bayesian likelihood. Bias corrections reach a few tenths of a magnitude for the bluest and reddest SNe~Ia \citep{Kessler2017}, and they also account for intrinsic scatter and noise in the fitted stretch and color, which would otherwise make the Tripp relation overcorrect or undercorrect individual SNe~Ia \citep{Scolnic2016}. Standardized magnitudes also correlate with host-galaxy properties, which was first corrected for with an empirical step of 0.05--0.1~mag between low-mass and high-mass hosts \citep{Kelly2010,Sullivan2010}. \citet{Brout2021} attributed this step to differences in dust properties, mainly $R_V$, between these hosts. Their model is part of the simulated population in Pantheon+, after which the fitted residual mass step is consistent with zero \citep{Brout2022b}. Upcoming samples from ZTF and ATLAS at low redshift, LSST at intermediate redshift, and Roman at high redshift will be 1--2 orders of magnitude larger \citep{Rigault2025,Ivezic2019,Rose2025}. We discussed how each addresses a different systematic: the intrinsic population at low redshift, a single photometric system across redshift, and SNe~Ia in the matter-dominated era.

\subsubsection*{Analysis Choices and Sources of Uncertainty}

The survey properties are measured, but the input population of the simulations (the intrinsic stretch and color distributions, their dependence on redshift and host properties, and the dust model) is chosen to reproduce the observed sample. This choice is a systematic uncertainty, propagated by repeating the simulations with alternative population and scatter models \citep{Brout2022b,Vincenzi2024}; in the DES-SN5YR reanalysis, the resulting uncertainty in $w$ is about twice that from calibration \citep{Popovic2026a}. As an example of assessing a new claim within this framework, we discussed the proposed dependence of standardized magnitudes on progenitor age \citep{Son2025}. Bias corrections depend on observables rather than on redshift alone, DES-SN5YR already includes a host-age-dependent model as a systematic \citep{Wiseman2022,Vincenzi2024}, and low-redshift data disfavor the proposed dependence once the host-mass correction is applied \citep{Murakami2026a,Wiseman2026}. For $H_0$, SH0ES restricts its baseline Hubble-flow sample to late-type hosts like the calibrators \citep{Riess2022}, and the expected difference in mean progenitor age between the two rungs is $\sim$0.15~Gyr \citep{Murakami2026a}.

\subsubsection*{Connection to the Data Challenge}

The Pantheon+ magnitudes in the SN~Ia notebook include BBC bias corrections computed with the dust-based population model, which also absorb the host-mass dependence, and the covariance matrix used by the solver includes the systematic uncertainties described above \citep{Brout2022b}. Participants did not change these inputs. The lecture explained what this fixed SN~Ia side contains and which choices a future challenge on the SN~Ia analysis could vary.

\subsubsection*{Key Takeaway}

SN~Ia magnitudes used in cosmology are corrected for selection effects and host-galaxy dependence with survey simulations whose input population is itself a modeling choice, and this modeling is a leading systematic uncertainty in recent analyses such as DES-SN5YR. For $H_0$, the SN~Ia systematic uncertainty in Pantheon+ is 0.29~km~s$^{-1}$~Mpc$^{-1}$, which cannot explain the Hubble tension \citep{Brout2022b}.

\subsection{Lecture 12 -- Wrapping Up (May 27) and Synthesis \& Wrap-Up: What Have We Learned? (May 28)}
\textbf{Lecturer: Siyang (Sean) Li}

\subsubsection*{Scientific Motivation}

These two closing sessions, given on May 27 and May 28, followed the lectures on Cepheids, stellar photometry, the TRGB, and SN~Ia standardization, preceding the day of work on the Data Challenge (May 29), and their purpose is to reassemble those individually taught pieces into the single distance-ladder pipeline participants are about to run themselves. They revisit each rung's role and its dominant systematics side by side, so that a choice made in the TRGB notebook can be understood in the context of the full chain from geometric anchors to $H_0$.

\subsubsection*{Main Concepts and Methodology}

The lecture recaps the distance ladder rung by rung: Rung~1 geometric anchors (Gaia parallaxes, the NGC~4258 megamaser, and LMC/SMC detached eclipsing binaries; \citealt{Pietrzynski2019,Graczyk2020}), Rung~2 standard candles (Cepheid period--luminosity relations and the TRGB, calibrated on those same anchors; the TRGB reaches about 5--30~Mpc, and Cepheids reach SN~Ia hosts out to $\sim 50$~Mpc), and Rung~3 SNe~Ia, standardized via light-curve shape and color (the Phillips relation \citep{Phillips1993} and the SALT2 parameterization \citep{Guy2007}) to a scatter of $\sim 0.15$~mag and extended to the Hubble flow ($z\lesssim0.15$) to measure $H_0$. It also reviews the basics of stellar photometry (aperture photometry, which is simple but struggles with overlapping stars, versus PSF fitting, which handles crowding but requires an accurate PSF model for the detector) and the physical picture of the TRGB (degenerate helium core, helium flash, the resulting sharp edge in the luminosity function, GLOESS smoothing, and Sobel edge detection) taught earlier in the week, framing the TRGB as the direct input that sets the SN~Ia zero-point used in the Data Challenge. Each rung is explicitly tied back to its calibrating rung below, reinforcing that no single measurement in the chain stands on its own.

\subsubsection*{Analysis Choices and Sources of Uncertainty}

The systematics recap groups the week's material into four categories: population effects (field-to-field TRGB scatter, addressed via contrast-ratio standardization), blending and crowding (a one-sided bright bias, mitigated by halo-field selection and artificial-star tests), dust extinction (corrected with reddening maps, worse on high-extinction sightlines), and methodological variations (smoothing bandwidth, color cuts, spatial clipping, and edge-weighting scheme, each a legitimate but consequential analyst choice). SN~Ia standardization contributes its own residual intrinsic scatter after light-curve correction. The lecture also presents the Local Distance Network \citep{Casertano2025}, which combines distance indicators in one network and models the correlations created by shared calibrators (NGC~4258, the LMC, and Milky Way parallaxes), since treating these as independent would underestimate the errors.

\subsubsection*{Connection to the Data Challenge}

The lecture states the Data Challenge's ground rules explicitly: the supernova dataset, the supernova fitting code, and any file in \texttt{2\_hubble\_constant/data/} are fixed and identical for every participant, while the TRGB-side choices (smoothing bandwidth, color cuts, spatial clipping radius, and Sobel weighting) are free to vary, so that any spread in a participant's final $H_0$ can be traced only to TRGB methodology. It walks through the two-notebook pipeline -- a TRGB notebook that exports a distance modulus to a shared results file, and an $H_0$ notebook that reads that file and returns $H_0$ -- and the stability check, in which participants vary their parameters and flag any TRGB shift of more than $0.03$~mag under small parameter changes.

\subsubsection*{Key Takeaway}

The distance ladder is only as strong as its weakest rung: a small fractional error at any one step propagates through every rung built on top of it, which is why the Data Challenge asks participants to document every measurement choice as carefully as the measurement itself.

\section{Individual Data Challenge Submissions}
\label{sec:individual_submissions}

This appendix collects the descriptions written by the Data Challenge participants. Each subsection summarizes one participant's analysis and the reasoning behind their measurement choices; the resulting values of $H_0$ and the photometric catalogs used are listed in Table~\ref{tab:master_submissions}. Descriptions were submitted by 21 of the 22 participants. The descriptions are reproduced in the participants' own words, with light editing for clarity and consistency of notation.


\subsection{Alejandro Aguirre}

\subsubsection*{Analysis Approach}

For smoothing of bootstrap samples, I used median levels to eliminate bias and noise. I mostly used the CMD to determine where a clear line appeared between high/low density of stars for the higher end of the RGB, while on the lower end I identified a vertical line between high/low density of stars along with a sufficiently large sampling size. With luminosity function edge detection, I looked for one high, isolated peak and minimal other peaks, readjusting my smoothing and RGB color band parameters if this was not observed. For the bootstrap sample, I made sure to have distinct, large peaks, and minimal error.

\subsubsection*{Analysis Choices}

I chose EDD photometry, allowing greater control over which stars to eliminate per galaxy, accounting for the galactic bulge, higher precision, and reliability. I decided to treat individual galaxies differently. For catalogs with F606W, I set the lower and upper limits on the F606W$-$F814W color to 0.7--0.9 and 1.4--1.5 mag, respectively, depending on the galaxy, the sample size, and how clearly the RGB edges stood out in the CMD. For catalogs with F555W, I used lower and upper limits on F555W$-$F814W of 1.0--1.5 and 2.0--2.1 mag, keeping a color range of similar width. For smoothing, I chose Poisson edge detection, and values of 0.1--0.13, eliminating both noise and bias with a middle ground. For the spatial-clipping scale factor (the size of the excluded inner ellipse in units of $R_{25}$), I used values of 0.8--1; I did not go above 1 so as to keep enough stars, and I used the lower values for galaxies with few stars or small bulges. All else was left default.

\subsubsection*{Robustness and Particular Cases}

I tested variability by varying smoothing, color range, and scale factor per galaxy, treating shifts beyond 0.03 mag as unhealthy. Most measurements were stable, with Poisson and Hatt edge detection agreeing. NGC 1365 and M96 were smoothing-dependent, shifting under low or high smoothing. NGC 1404 was sensitive to spatial clipping, showing noise with low smoothing and large error with high scale factor, a consequence of undersampling. Notably, lowering the upper limit on the F606W$-$F814W color collapsed the tip contrast in NGC~1316, which otherwise had a healthy uncertainty.

\subsubsection*{Main Takeaway}

Spatial clipping mattered most, with scale factor driving undersampling errors. Per-galaxy systematics dominated the remaining uncertainty, especially in NGC 1404, whose passage through the Fornax intracluster medium via halo-star contamination and possible extinction likely biased its TRGB.


\subsection{Miguel Barroso Varela}

\subsubsection*{Analysis Approach}

I inspected the extinction-corrected color-magnitude diagrams alongside the luminosity functions, Sobel responses, and bootstrap diagnostics to assess whether the detected edge corresponded to the upper RGB. Tip contrast provided an additional quality diagnostic, interpreted alongside sensitivity to the analysis parameters. I used a common measurement procedure for the CATs photometry, calibrated against NGC~4258, while adapting the color selection and occasionally the smoothing and magnitude limits to individual galaxies. These adjustments aimed to isolate the RGB while retaining sufficient stars near the tip.

\subsubsection*{Analysis Choices}

For $(F606W-F814W)_0$, I generally selected $1.2$--$1.5$ mag, narrowing this to $1.3$--$1.5$ mag for NGC~4038 and NGC~4536. NGC~5643 used $1.0$--$1.5$ mag to retain the RGB under color variations; M101 used $1.1$--$1.4$ mag because its sequence was bluer. The anchor selections were $1.3$--$1.5$ mag in $(F606W-F814W)_0$ and $1.6$--$2.1$ mag in $(F555W-F814W)_0$. These windows respected the notebook's adopted flat-TRGB regime.

I used $\tau=0.10$ mag, except for NGC~4038, where I adopted $0.08$ mag. Smoothing balanced suppression of count fluctuations against broadening of the edge. I retained the supplied binning and Hatt weighting. To avoid selecting fainter features, I imposed faint limits of $27.3$ mag for NGC~4424 and $27.0$ mag for NGC~4038. Otherwise, magnitude bounds were determined automatically from the photometry. Spatial selection was inherited from CATs.

\subsubsection*{Robustness and Particular Cases}

I assessed smoothing and color sensitivity separately from bootstrap uncertainty. NGC~5643 remained stable, although extinction dominated its quoted uncertainty. NGC~4424 and NGC~4038 showed the strongest parameter dependence. NGC~4536 was sensitive mainly to smoothing, while NGC~1365 showed color sensitivity and a larger bootstrap contribution. NGC~3370 and NGC~1309 were excluded because their detections were insufficiently reliable. Additional spatial clipping could not be tested with these catalogs.

\subsubsection*{Main Takeaway}

The interaction between RGB color selection and smoothing was the main methodological concern. High tip contrast did not guarantee a stable measurement, making parameter sensitivity essential to interpreting the precision of the final distances.

\subsection{Kathan Dave}

\subsubsection*{Analysis Approach}
 
I worked on NGC 1365, using the EDD photometry given for the challenge. Instead of trying many galaxies, I considered one target and spent most of my time checking how much my result depended on the parameter choices. I checked and looked at the color-magnitude diagram, the smoothed luminosity function, and the Sobel edge response together, and generated a clear single peak with a roughly unimodal bootstrap distribution as the sign of a good detection. Since I only worked with one galaxy, there was no need for me to determine whether to maintain consistent settings across different targets.

\subsubsection*{Analysis Choices}

I used a relatively narrow color window to isolate the red giant branch while minimizing contamination from nearby AGB stars and the red clump, both of which can affect the edge response. The smoothing bandwidth was based on smoothing out the noise in the luminosity function without smearing the location and form of the TRBG feature or without blurring the tip. The luminosity function was calculated with Hatt weighting as it was more stable than flat weighting to compensate for increasing photometric uncertainty and decreasing numbers of stars at the fainter end. The HST field is located in the outer halo of NGC 1365, so crowding should be relatively limited and the default spatial selection was thus adequate for the analysis.

\subsubsection*{Robustness and Particular Cases}

I explored the sensitivity of the measurement by varying the smoothing bandwidth, color limits, spatial selection, and bin width over a broader range than the standard notebook checks. The TRGB location remained stable across these choices, although the measured magnitude showed shifts that were larger than would be expected from the statistical uncertainty alone.
 
\subsubsection*{Main Takeaway}
 
The main result for me was that reasonable analysis choices, each physically and statistically defensible on their own, could shift the inferred TRGB magnitude by an amount comparable to the formal uncertainty. This suggests that the bootstrap error captures only part of the uncertainty in the measurement, and that the sensitivity to analysis choices is important to keep in mind when interpreting the final distance.
    
\subsubsection*{Master Table Values (if required) }

\begin{table}[H]
\centering
\small
\resizebox{\textwidth}{!}{
\begin{tabular}{lcccccccccc}
\hline
\textbf{Participant} &
\boldmath{$H_0$} &
\textbf{Phot.} &
\textbf{Color cut} &
\boldmath{$\tau$} &
\boldmath{$\Delta m$} &
\textbf{Edge} &
\textbf{P. floor} &
\textbf{Spatial} &
\textbf{Mag. range} &
\boldmath{$N_{\rm boot}$}
\\
&
\textbf{[km\,s$^{-1}$\,Mpc$^{-1}$]} &
\textbf{source} &
&
\textbf{[mag]} &
\textbf{[mag]} &
\textbf{weight} &
&
\textbf{scale} &
&
\\
\hline

Kathan Dave & 67.91$\pm$9.22 & EDD & 1.0--1.5 & 0.05 & 0.01 & Hatt & 1.0 & 1.0 & Auto & 500 \\

\hline
\end{tabular}
}
\caption{Measurement settings and resulting $H_0$ for the submitted galaxy, NGC~1365.}
\end{table}

\subsection{Francesca Di Nella}

\subsubsection*{Analysis Approach}

To obtain a robust TRGB measurement, my strategy was to ensure stability across varying smoothing scales and color selections. By inspecting the color-magnitude diagrams and luminosity functions, I aimed to identify a dominant peak in the edge response alongside a unimodal bootstrap distribution. Meeting these two requirements provided evidence for a clean and unambiguous detection of the TRGB horizontal line in the color-magnitude diagram. Stability checks and the resulting selections were evaluated individually for each galaxy.

\subsubsection*{Analysis Choices}

To determine the most suitable setup for TRGB detection, I mainly focused on tuning two key parameters: the GLOESS smoothing bandwidth and the color-selection range. For the GLOESS smoothing bandwidth, a value of $\tau=0.1$ proved to be the best compromise, as it successfully smoothed out the noise without artificially merging competing peaks. The RGB color ranges were initially identified ``by eye'' to fully enclose the horizontal TRGB feature and maximize sample size, then validated against the luminosity functions to verify clean edge responses. Despite performing custom stability checks for every galaxy, the resulting parameter configurations were highly consistent among the target galaxies: the only variable that changed slightly in some cases was the lower limit of the color band.

\subsubsection*{Robustness and Particular Cases}

I evaluated robustness by testing $\tau\in [0.05,0.2]$ and shifting color bounds by $\pm 0.2$ mag, requiring spreads $<0.05$ mag. Targets not mentioned below were stable, alongside NGC~1309, whose smoothing spread reached $0.07$ mag due to noise at $\tau=0.05$ while color shifts remained stable. Excluded galaxies failed for different reasons: M96 was extremely color-sensitive; NGC~1404, NGC~3021, NGC~3370, and NGC~5584 lacked contrast or faint stars; NGC~4424 exhibited strong bimodality; and NGC~4038 lacked a clear TRGB line, causing large spreads.

\subsubsection*{Main Takeaway}

Rigorous stability checks under parameter variations were the most decisive factor in TRGB detections. However, no choice of smoothing or color selection can compensate for poor edge contrast and insufficient faint-star counts in the underlying data.
\subsection{Elliot Dutnall}

\subsubsection*{Analysis Approach}

For this analysis, my approach to identifying the TRGB started with an
inspection of the diagnostic panel, which included a color-magnitude
diagram, luminosity function, and the TRGB results from 500 simulations.
I mainly used the color-magnitude diagram to find the rough location of
the TRGB using the point of a prominent stellar density dropoff, and
made sure it was a significant detection using the variance of the
simulations' results and the significance of the tip detection. Owing to
this, I made different choices of parameters for each galaxy, as using
the same parameters for each would return unstable measurements.

\subsubsection*{Analysis Choices}

The principal parameters I varied for this analysis were the range of
color magnitudes and the smoothing of the luminosity function, also
known as $\tau$. My methodology for choosing these parameters for each
galaxy was to attempt to have them be as close to the default values as
possible. However, in many cases this produced an unstable result that
had high uncertainty. In these cases, I generally solved this problem by
increasing $\tau$ or inspecting the color-magnitude diagram to determine
the true location of the TRGB for the specific galaxy. I then confirmed
for any set of my choices that it met minimum requirements of having
enough stars, having a significant tip detection, and all simulations
returning a TRGB magnitude around the same value.

\subsubsection*{Robustness and Particular Cases.}

The most important aspect that I considered while selecting the TRGB
detection parameters was the stability of the final detection to
variations in the parameters. If the detection was stable to these, then
my parameter choices were effectively arbitrary. For most galaxies, the
TRGB detection was largely stable and usable. However, for 
NGC~3021, NGC~3370, and NGC~5584, the measurement was unstable and did not
meet minimum requirements for any choice of parameters; thus, they
were rejected from my analysis.

\subsubsection*{Main Takeaway}

Overall, the most important parameter in affecting my final result was
the magnitude range I selected for the TRGB, as I selected it by hand
and different selection criteria or process would most likely change
my results.
\subsection{Hussain Gohar}

\subsubsection*{Analysis Approach}

I analyzed EDD and CATs separately, treating each galaxy individually: for every host I varied
color window, GLOESS bandwidth, spatial clipping, and magnitude range while inspecting the
color-magnitude diagram, luminosity function, edge response, and bootstrap distribution. A
detection was adopted only if one edge marked the top of the red giant branch, the tip contrast exceeded 3 in host and anchor, at least 400 stars lay within 1~mag below the tip, the
bootstrap error was below 0.10~mag, and the tip stayed within 0.03~mag when bandwidth, color
edges, and clipping were varied. Five CATs and six EDD hosts pass, contributing 12 and 14
calibrating supernovae.

\subsubsection*{Analysis Choices}

Color windows and bandwidths were chosen per galaxy from values meeting these criteria, with
red edges kept inside the flat-tip limits. The anchor windows mattered most: removing the
bluest 0.2~mag from the CATs windows narrows the gap between its two NGC~4258 fields from 0.05
to 0.02~mag and lowers $H_0$ by 0.8 km/s/Mpc. The same cut destabilizes the EDD anchor, so EDD keeps the
challenge's windows. Hatt weighting and 0.01~mag bins were used throughout; unweighted edges
move tips up to 0.87~mag fainter. EDD clipping was scaled per galaxy to avoid crowded centers. The search range is automatic, except in EDD NGC~4424, where I set the bright end to
26.5~mag; at the adopted bandwidth this moves the tip by only 0.004~mag.

\subsubsection*{Robustness and Particular Cases}

Accepted tips move by $\le0.02$~mag under the stability tests and agree with their supernovae
within $1.6\sigma$. EDD NGC~4424 is the exception: at larger bandwidths the smoothing fails near
its bright limit, so the tip jumps; without that limit its variation is 0.023~mag. Admitting hosts that fail only stability gives $H_0=76.7$ km/s/Mpc (CATs) and 82.0 (EDD);
most of that shift comes from NGC~1309, whose tip is among the most stable yet lies 0.9~mag
closer than its supernovae imply -- a disk rather than halo field. Excluding every host more than $3\sigma$ from its supernovae gives 72.0 and 73.8 km/s/Mpc. EDD NGC~1404 passes every criterion
in one setting, but that edge vanishes once its contaminated center is excluded.

\subsubsection*{Main Takeaway}

Two systematics dominate. The NGC~4258 anchor tip differs by 0.08~mag between catalogs and
shifts by 0.13~mag with bandwidth in EDD alone, setting the 2.1--3.8 EDD--CATs spread in $H_0$.
Passing every rule does not guarantee accuracy: NGC~1309 is among the most stable tips yet lies
0.9~mag from its supernovae, and one EDD NGC~1404 configuration passes every criterion despite
documented contamination.

Table~\ref{tab:master_submissions_hgohar} lists the settings of the two submissions.

\begin{table*}[t]
\centering
\small
\resizebox{\textwidth}{!}{
\begin{tabular}{lcccccccccc}
\hline
\textbf{Participant} & \boldmath{$H_0$} & \textbf{Phot.} & \textbf{Color cut} & \boldmath{$\tau$} &
\boldmath{$\Delta m$} & \textbf{Edge} & \textbf{P. floor} & \textbf{Spatial} & \textbf{Mag. range} &
\boldmath{$N_{\rm boot}$} \\
& \textbf{[km\,s$^{-1}$\,Mpc$^{-1}$]} & \textbf{source} & & \textbf{[mag]} & \textbf{[mag]} &
\textbf{weight} & & \textbf{scale} & & \\
\hline
Hussain Gohar (EDD)  & 73.13 $\pm$ 2.41  & EDD  & var & var. & 0.01 & Hatt & 1.0 & var. & var.: auto$^\dagger$ & 1000 \\
Hussain Gohar (CATs) & 69.50 $\pm$ 2.31 & CATs & var. & var. & 0.01 & Hatt & 1.0 & n/a & auto & 1000 \\
\hline
\end{tabular}
}
\caption{Excerpt of the summary of Data Challenge submissions (organizers' master table).  $^\dagger$EDD NGC~4424 uses a bright
limit of 26.5~mag.}
\label{tab:master_submissions_hgohar}
\end{table*}

\subsection{Paridhi Jain}

\subsubsection*{Analysis Approach}
I analyzed NGC 1316, NGC 4424, and NGC 4526 using the provided EDD photometry, with NGC 4258 as geometric anchor. The selection of these galaxies was based on their physical properties, minimum foreground extinction for more stable TRGB measurements, and on the cleaner RGB stars in the color-magnitude diagrams with minimum AGB star contamination. After systematically running a range of values of the provided parameters on all the EDD galaxies, I investigated these three targets more closely. For selecting parameter values I focused on tip contrast value, nearly unimodal bootstrap distributions, small dispersions based on MAD, and a sufficient number of stars below the tip.

\subsubsection*{Analysis Choices}
I retained the baseline RGB color limits because the selected interval covered the RGB population seen in the diagrams and contained enough stars for the measurement. The inferred distances also exhibited relatively small changes in the color tests; thus, my main adjustments were focused on smoothing and spatial selection. The smoothing value at 0.18 mag reduced the filter F555W/F606W discrepancy from 0.101 to 0.039 mag and changes broad or multimodal bootstrap distributions into approximately unimodal ones. The same settings were applied to both NGC 1316 and NGC 4526; also, the anchor used a smaller spatial exclusion ellipse. All this ended up improving the formal precision, too. 

\subsubsection*{Robustness and Particular Cases}
I assessed the robustness of the measurements by varying the smoothing bandwidth, color limits, and scale factor, and by examining the bootstrap distributions. NGC 1316 and NGC 4526 showed a clear TRGB measurement and the inferred distance was most sensitive to the smoothing bandwidth. For NGC 4424, the result varied significantly with the spatial scale factor, and it generated a secondary peak in the bootstrap distribution.

\subsubsection*{Main Takeaway}
I found that the default pipeline parameters provided were reliable for TRGB detections for 
most of the galaxies. Smoothing bandwidth and spatial scale factor affected on the TRGB measurements the most and can be considered for refinements.


\subsection{Gracelynn Jost}

\subsubsection*{Analysis Approach}

The EDD database was used consistently across all galaxies. To cancel systematic errors across the distance ladder, color cuts, smoothing, and spatial cuts were kept consistent when feasible. Notebook defaults were kept when checks passed. Each color-magnitude diagram (CMD) was inspected to visually verify the horizontal drop at the TRGB against the bootstrap median and edge-detection response. To avoid contaminants from other stellar populations into the TRGB magnitude calculation, the CMD was also inspected to ensure that the color cut did not extend bluer than the “slope” of the RGB.

\subsubsection*{Analysis Choices}

Consistency of variables across all galaxies was prioritized, as long as quality criteria (tip contrast, unimodal bootstrap, 1$\sigma$ MAD, $N$ faint) were met. Parameters were adjusted in moderation until these qualifications were met. The main departure from the notebook defaults came from the anchor, NGC 4258: the smoothing value gave a bimodal F606W bootstrap, so its value was increased and used for all galaxies to cancel choices across the ladder. In M101 (F606W), spatial clipping was reduced to have a reasonable number of stars in the calculation. The anchor's spatial clip for M101 was left unchanged because matching it would give a bimodal bootstrap with a high 1$\sigma$ MAD. In cases where no parameter adjustments yielded results that met the qualifications, the host galaxy was left out of the final $H_0$ calculation.

\subsubsection*{Robustness and Particular Cases}

For all adopted galaxies, different values of tau, spatial cuts, and color cuts were tested. The standard deviation across all tests was calculated and added in quadrature with other error measurements. Tests that resulted in a bimodal bootstrap were still included in the systematic uncertainty calculation to include the possibility of the bootstrap converging on the wrong peak. Galaxies with large spreads were kept to avoid small-number statistics. Specific cases include NGC 1309 (smoothing dependent) and NGC 4536 (smoothing, spatial clipping, and color-selection dependent).

\subsubsection*{Main Takeaway}

In the determination of the final result, in order to avoid human bias, the bootstrap median value was final say; no adjustments were made to favor one Sobel peak over the other.

\subsection{Shivani Kulkarni}

\subsubsection*{Analysis Approach}

In order to achieve a robust TRGB measurement, with EDD as my photometry source, my main approach was to stay consistent in my choices for analysis parameters such as inspection of the color-magnitude diagrams, smoothening scales, and the tip-contrast diagnostic. Upon inspection of the color-magnitude diagram, I adjusted the color limits such that it would favor enclosing the RGB tip cleanly. Overall, I made my analysis choices based on the desired features for a good TRGB measurement.

\subsubsection*{Analysis Choices}

The parameters that I found most important for making a reliable TRGB measurement include, but are not limited to, the RGB color selection, GLOESS smoothening ($\tau$), Sobel peak and edge detection, and spatial clipping. After experimenting with different color limits, I ended up tightening the windows in order to exclude younger, bluer stars, as well as stars that are not on the RGB. A smoothening bandwidth of $\tau = 0.2$ was chosen in order to smooth the luminosity function more because the Sobel peak was jagged, which corresponds to an unstable bootstrap. As for the spatial clipping, since the bootstrap distribution was bimodal initially, decreasing the scale factor down to 0.3 helped fix this issue. 

\subsubsection*{Robustness and Particular Cases}

The robustness of the final measurement was evaluated using the tip contrast value and the shape of the bootstrap distribution. For NGC 4536, the bootstrap distribution was approximately Gaussian and unimodal, indicating a good measurement, however, the tip contrast was 3.05, which is just barely above the threshold, making it the least secure of my detections. The Sobel response was a clear dominant peak, which overall, in combination with the rest of the factors, conveyed that the measurement was robust.

\subsubsection*{Main Takeaway}

Overall, I believe all parameter choices made a significant impact in affecting the robustness of the measurement, but I found that spatial clipping and smoothing played the largest role in determining my final result.

\subsection{Dhiraj Kuniyal}

\subsubsection*{Analysis Approach}

All 16 SN~Ia host galaxies were analyzed in batch mode using EDD photometry with a GLOESS-smoothed luminosity function (LF) and Sobel edge detection. For each host, plausible combinations of smoothing bandwidth, RGB color limits, and spatial clipping were examined. Each measurement was screened using its bootstrap uncertainty, tip contrast, and number of RGB stars within 1\,mag below the tip. Variation in the recovered TRGB across these configurations provided a separate measure of parameter sensitivity. The adopted measurement for each galaxy was selected by combining these checks with inspection of the color-magnitude diagram (CMD), LF, edge response, bootstrap distribution, and stellar-population characteristics.

\subsubsection*{Analysis Choices}

The configurations used the central, filter-appropriate RGB color windows: $1.0<(F606W-F814W)<1.5$ mag and $1.6<(F555W-F814W)<2.1$ mag. A GLOESS bandwidth of $\tau=0.10$ was adopted for NGC~1316, NGC~1448, and NGC~7814, whose edge detections remained clear with narrower smoothing. The other galaxies -- NGC~1309, NGC~1365, NGC~4424, NGC~4526, NGC~4536, NGC~5643, M66, and M96 -- used $\tau=0.15$, which better suppressed small-scale fluctuations while preserving the edge. All galaxies ultimately used the central spatial selection, with a scale factor of $1.0$. Tests with tighter and wider regions did not produce a consistent improvement in the CMD or edge-response diagnostics. LF bins of $0.01$ mag, Hatt edge weighting, and a unit Poisson floor were therefore kept fixed. The final choices were not based on optimizing a single quantity, such as tip contrast or bootstrap uncertainty. Instead, the selected configuration also had to show a plausible RGB termination, one clearly preferred Sobel peak, and a reasonably compact bootstrap distribution. For every target, NGC~4258 was reanalyzed with the matching configuration so that the target distance modulus was calibrated consistently relative to the anchor.

\subsubsection*{Robustness and Particular Cases}

Robustness was judged from two related questions: whether an individual measurement passed the basic quality checks, and how sensitive the TRGB is when the analysis settings were changed. Eleven galaxies were retained. NGC~1404, NGC~3021, NGC~3370, NGC~4038, and NGC~5584 were excluded because their detections repeatedly failed the diagnostics and/or depended strongly on the selected configuration. Occasional failures at an extreme parameter value were recorded as evidence of systematic sensitivity, but were not treated by themselves as grounds for rejection.

\subsubsection*{Main Takeaway}

The most important requirement was that a credible RGB termination remained visible across reasonable parameter choices. A strong edge peak alone was insufficient if the CMD, bootstrap distribution, or available stellar statistics did not support the same interpretation.
\subsection{Mariia Marinichenko}

\subsubsection*{Analysis Approach}

I analyzed the EDD sample, identifying the TRGB through joint inspection of the color-magnitude diagrams, luminosity functions, Sobel filter peaks, and bootstrap distributions. Preferred detections were selected by high tip contrast and a single, narrow bootstrap peak that remained stable under changes in the analysis parameters. I used the same quality criteria across the sample while allowing smoothing and stellar selections to vary between galaxies. My selected result are based on 14 galaxies, excluding NGC~3021, NGC~3370, and NGC~1309 owing to their unstable TRGB identifications.

\subsubsection*{Analysis Choices}

To identify stable edge peaks, I varied the GLOESS bandwidth  and required a tip contrast above 3. Detections with lower contrast were accepted only when the bootstrap distribution remained unimodal with a low median absolute deviation. Color cuts were placed near the upper limits from Jang \& Lee (2017). We reduced populations in inner regions by spatial cuts when they broadened or split the bootstrap distribution, while keeping sufficient stars to satisfy the \citet{Madore2009} criterion. Otherwise, the default spatial selection was preserved. We also applied cuts on bright magnitude end to reduce suspected AGB contamination. For M66 and M101, the two filter-based measurements were combined using inverse-variance weighting.

\subsubsection*{Robustness and Particular Cases}

Sensitivity to smoothing, spatial selection, color cuts, and Sobel weighting were assessed as the largest shifts in bootstrap means, and then combined in quadrature as a methodological uncertainty estimate. NGC~7814 showed a stable detection, but NGC~5584 had competing peaks whose relative prominence depended strongly on smoothing and spatial selection. M66 also showed noticeable sensitivity to smoothing and color cuts. Excluding the three unstable galaxies shifted $H_0$ by more than any quoted statistical uncertainty.

\subsubsection*{Main Takeaway}

The reliability of individual TRGB identifications, particularly when competing edge peaks were present, were controlling the final result. Bootstrap precision alone did not capture the sensitivity to analysis choices or the influence of problematic galaxies.

\subsection{Yash Mehta}

\subsubsection*{Analysis Approach}

I identified the TRGB by examining the color-magnitude diagram (CMD), luminosity function, Sobel edge response, and bootstrap distribution for each galaxy. I looked for a clear RGB sequence in the CMD and a dominant edge-response peak corresponding to a stable TRGB measurement. I also used the tip contrast, number of faint RGB stars, and bootstrap distribution as diagnostics of reliability. I began with the same default analysis choices for all galaxies and only adjusted individual parameters when the default measurement appeared ambiguous or unstable.

\subsubsection*{Analysis Choices}

I retained the default parameters whenever they produced a clear Sobel peak, a stable bootstrap distribution, and sufficient RGB statistics. The RGB color selection was the parameter I adjusted most often, primarily to reduce contamination while preserving enough stars for a reliable measurement. For several galaxies, shifting or tightening the color range produced a cleaner edge response and more stable bootstrap distribution. I adjusted the GLOESS smoothing bandwidth when changing the degree of smoothing improved the stability of the detected edge. Spatial selection was modified in a small number of cases when changing the region improved the measurement. I otherwise kept the luminosity-function binning and edge-weighting method consistent across the sample.

\subsubsection*{Robustness and Particular Cases}

I considered detections reliable when the tip contrast exceeded 3, the sample contained sufficient stars within 1\,mag below the tip, and the bootstrap distribution and Sobel response supported the same TRGB location. Most accepted galaxies remained stable under the parameter variations I explored. NGC~4038 and NGC~5584 did not produce sufficiently robust TRGB detections. M66 showed inconsistent TRGB measurements between its available filter datasets, while M101 had an unreliable F555W measurement, so these galaxies were excluded.

\subsubsection*{Main Takeaway}

The RGB color selection and GLOESS smoothing were the most important adjustable analysis choices, while the bootstrap distributions and tip contrast provided useful checks that the adopted TRGB measurements were stable and reliable.

\subsection{Vasudev Mittal}

\subsubsection*{Analysis Approach}
I analyzed each galaxy separately, and for each target used the same color cuts for both the anchor and target.
I started by identifying a broad color range in which the anchor had a stable TRGB.
Then, I studied the target to narrow down the range in which both the target and anchor had stable TRGBs.
For both the target and anchors, I maintained high tip contrast, ensured there was a sizable faint-star population, and verified that the bootstrap distribution was unimodal.
I performed these steps over a broad range of smoothing parameter $\tau$, to demonstrate that the TRGB was not an artifact of the smoothing function.

\subsubsection*{Analysis Choices}
My color-selection strategy involved keeping the lower limit below 1 and the upper limit above 1.2 because, in this range, the anchor always yielded a unimodal bootstrap distribution across a broad range of smoothing scales.
I used a $\tau$ range of $[0.5,0.15]$ to assess the robustness of the TRGB to the smoothing function.
This range was broad enough to rule out the sensitivity of the TRGB to the smoothing scale.
Finally, I kept my tip contrast above $3.8$, and the faint star count to roughly $10^3$.
These strict bounds ensured that the detected TRGB was not plagued by any edge effect.

\subsubsection*{Robustness and Particular Cases}
My results were mostly robust against the parameter variations; however, the following stood apart.
\begin{enumerate}
    \item NGC~1365 required an upper color cut of $\sim1.1$, and was unstable against variations in $\tau$. The results were stable if I reduced the upper color cut, but then the anchor TRGB became unstable.
    \item The bootstrap distribution of M66 had a subdominant second peak at $26.2$ mag, which made its TRGB somewhat unstable to variations in $\tau$.
    \item The TRGB of NGC~1404 was somewhat unstable to variations in $\tau$, which might be due to the lower color limit being at the very boundary of the acceptable range.
\end{enumerate}

\subsubsection*{Main Takeaway}
I found the smoothing scale to be the predominant factor in TRGB determination.
Note that it is better to first measure the TRGB at $\tau=0.8$, and then study the stability of TRGB against $\tau$, because the results determined for this value have been stable for my complete $\tau$ range.


\subsection{Lily Myers}

\subsubsection*{Analysis Approach}

I identified the TRGB using extinction-corrected color-magnitude diagrams (CMDs), smoothed luminosity functions, and Sobel edge-detection responses. For each galaxy, I systematically inspected the CMD to isolate the red giant branch from asymptotic giant branch (AGB) stars. Detections were evaluated using Sobel peak contrast, bootstrap median stability, and dispersion ($\sigma_{\text{MAD}}$). While baseline parameters were calibrated on the anchor NGC~4258, individual galaxies were adjusted as needed to resolve competing edge peaks, eliminate bootstrap bimodality, satisfy star-count thresholds ($N_{\text{faint}} \ge 400$), and ensure stable measurements. 

\subsubsection*{Analysis Choices}

Calibration with NGC~4258 established baseline smoothing ($\tau = 0.18$) and spatial clipping ($0.8\times D_{25}$) to suppress secondary peaks and achieve unimodal bootstrap distributions. For target hosts, I tightened color windows to isolate RGB populations and minimize AGB contamination. In sparsely populated or noisy targets (NGC~3021, NGC~3370, NGC~4038), I widened the GLOESS smoothing scale ($\tau = 0.20$) to smooth the luminosity function while maintaining $N_{\text{faint}} \ge 400$. Spatial clipping was adjusted between $0.7$ and $1.0\times D_{25}$ to excise crowded inner disks or preserve outer-halo stars. For NGC~1448 and NGC~3021, adopting Poisson weighting and bright-magnitude limits mitigated edge instability from sparse RGB sampling. 

\subsubsection*{Robustness and Particular Cases}

Robustness was evaluated by measuring tip stability across smoothing bandwidths ($\tau = 0.05\text{--}0.20$) and color shifts ($\pm 0.20\text{ mag}$). Nearby hosts like M101 and NGC~4526 showed stable, unimodal distributions with variations $<0.03\text{ mag}$, whereas distant targets (NGC~1309, NGC~3021, NGC~3370) displayed competing peaks and low contrast. In particular, NGC~5584 proved unresolvable and was excluded from distance-modulus synthesis owing to severe instability, low contrast, and bootstrap bimodality.  

\subsubsection*{Main Takeaway}

Photometric incompleteness and AGB contamination at $\mu \gtrsim 32\text{ mag}$ caused the Sobel filter to lock onto bright AGB features instead of the true tips. I believe that this artificially capped distance moduli to 31.5 mag, compressing the distance scale and ultimately leading to a very high $H_0$ measurement. 

\subsection{Devaraja Mallesha Naik}

\subsubsection*{Analysis Approach}
My analysis focused on selecting clean RGB populations and minimizing systematic effects from crowding, internal extinction, and spatially varying stellar populations. Using the provided EDD photometry, I inspected the color-magnitude diagrams, GLOESS-smoothed luminosity functions, Sobel edge responses, and bootstrap distributions for each target. Rather than imposing identical settings on all galaxies, I treated the targets individually, adjusting the spatial selection and smoothing where required by the morphology, photometric sampling, and stellar density of each host. The adopted TRGB was selected from the most distinct and stable edge feature supported by these diagnostics.

\subsubsection*{Analysis Choices}

To ensure a consistent statistical baseline, I fixed the bin width, edge weighting (``hatt''), and number of bootstrap iterations across all targets. However, color cuts, GLOESS smoothing ($\tau$), and spatial scales were tailored to each host's data quality. Most targets shared a common color window, while NGC 4038 and NGC 1404 required bluer-shifted cuts to better isolate their RGB sequences. Smoothing bandwidths were adjusted according to the quality and sampling of the luminosity function; sparser samples such as NGC 1404 required stronger smoothing to suppress noise and competing edge features, whereas well-populated fields allowed sharper edge localization. Spatial scale factors were similarly optimized per galaxy, balancing reduction of disk and central-region contamination against retention of sufficient halo stars.

\subsubsection*{Robustness and Particular Cases}
I assessed robustness using tip-contrast diagnostics, bootstrap distributions, and a leave-one-out jackknife analysis. Five targets were excluded: NGC 5584, NGC 3370, and M101 failed basic contrast checks, while NGC 1309 and NGC 3021 were removed after jackknife testing identified them as extreme statistical outliers. Among the twelve retained galaxies, NGC 1404 showed the strongest sensitivity and required careful spatial selection. Retained measurements remained generally stable under explored parameter variations.

\subsubsection*{Main Takeaway}
The analysis showed that spatial selection and the treatment of the RGB population were particularly important for controlling contamination and stabilizing the TRGB edge. Galaxy-dependent choices were therefore necessary to obtain consistent distance measurements across the heterogeneous host sample.
\subsection{Jos\'e Antonio N\'ajera}

\subsubsection*{Analysis Approach}

We consider the measurement as a reproducibility exercise, with a single parameter set, taken from the notebook defaults, applied to the anchor and all 17 targets. We check the color-magnitude diagram, smoothed luminosity function, and edge response of each galaxy, but we do not tune any parameter based on them. Instead, the diagnostics decide which detections enter the $H_0$ fit. We only retain a galaxy if its tip contrast exceeds 3 and at least 400 stars lie within 1 mag below the tip. Furthermore, its EDD distance must agree with the independent CATs distance to within 0.3 mag.

\subsubsection*{Analysis Choices}

The decisive choice is the color window, together with the filter to which it is applied. We restrict the red giant branch selection to the flat-TRGB regime of Jang \& Lee (2017), 
so that no color correction is needed. Moreover, the window follows the blue filter of each field, as in the notebook's Section~7. This matters because the notebook's batch mode applies the F606W window to every single-filter galaxy, which shifts the tip of the F555W-based NGC 5584 by 0.63 mag. We adopt the recommended Hatt weighting and the default smoothing. The only per-galaxy deviation is M101, since its clipping ellipse covers the whole EDD field at the default scale. Thus, a fixed fallback rule lowers its scale factor to 0.6.

\subsubsection*{Robustness and Particular Cases}

We run the full Section 11a stability grid for each galaxy and quote the scatter in $\mu$ as a separate $\sigma_{\rm method}$, which exceeds the bootstrap error in every galaxy except NGC 1404. It is largest for NGC 3370, NGC 5584, and NGC 3021, all flagged for low tip contrast. In contrast, NGC 1309 passes the contrast and star-count cuts and is stable under the scan, yet lies 0.59\,mag from its CATs distance. In fact, including or excluding this galaxy alone changes $H_0$ by $8\,\mathrm{km}\,\mathrm{s}^{-1}\,\mathrm{Mpc}^{-1}$.

\subsubsection*{Main Takeaway}

We highlight that the bootstrap error is a poor guide to accuracy. This is because, as a median-absolute-deviation width, it tracks only the dominant bootstrap peak and is nearly insensitive to draws on a competing edge. As a result, the least trustworthy distances can carry the most weight in the fit.


\subsection{Antonio Quintana-Estellés}

\subsubsection*{Analysis Approach}

The TRGB analysis was performed using the CATs photometry. Color-magnitude diagrams 
and luminosity functions were inspected individually to assess the RGB population 
and TRGB feature. The Sobel edge response was also evaluated, retaining the Hatt weighting 
since it provided a stable response across the sample. Many galaxies showed clear TRGB 
detections for similar analysis parameters, supporting the use of a common configuration 
approach rather than independent parameter selection for each target, thereby minimizing 
potential bias from subjective choices. This common analysis choice was applied whenever 
possible, yielding 14 target measurements out of the 17 initial targets.

\subsubsection*{Analysis Choices}

The adopted RGB color-selection ranges were 0.9--1.4 mag in F606W$-$F814W and 1.5--2.0 mag in F555W$-$F814W, with 
a GLOESS smoothing parameter of $\tau=0.12$. NGC~1404 was the only target treated 
differently: the common color selection gave $N_{\rm faint}=364<400$ despite a strong 
tip contrast. Since it was close to the \citet{Madore2009} threshold, the color 
selection was widened slightly to 0.9--1.5 mag in F606W$-$F814W and 1.5--2.1 mag in F555W$-$F814W, 
increasing $N_{\rm faint}$ to 429. This minor modification allowed us to retain NGC~1404 
as an additional measurement in the sample. Although NGC~1309 showed a marginal detection, 
with a tip contrast only slightly above threshold (3.15) and moderate statistical support 
($N_{\rm faint}=641\geq400$), no fine-tuning adjustments were made for this galaxy.

\subsubsection*{Robustness and Particular Cases}

M66 showed two competing TRGB peaks consistent with its large uncertainty. NGC~1404 and 
NGC~1309 were the most marginal cases: modest changes in smoothing or color 
selection may lower their statistics below threshold. The full 14-galaxy sample gives 
$H_0=77.38\pm1.83$ km s$^{-1}$ Mpc$^{-1}$. Removing both NGC~1404 and NGC~1309 lowers the 
inferred value to $H_0=70.23\pm2.00$ km s$^{-1}$ Mpc$^{-1}$, indicating that the
final value is noticeably sensitive to these two measurements. This suggests that
marginal TRGB identifications can propagate into the inferred Hubble constant.

\subsubsection*{Main Takeaway}

Fixing a common parameter set minimizes subjective choices, at the cost of retaining some 
ambiguous TRGB detections that can affect $H_0$. Overall, the results are robust to reasonable 
parameter variations for most targets, while marginal detections warrant careful assessment.


\subsection{Pilar Ruiz-Lapuente}

\subsubsection*{Analysis Approach}

We used the EDD photometry catalog and explored the color-magnitude
diagrams and luminosity functions for each galaxy. The main 
analysis parameters for individual targets have been varied
to assess which combinations 
produced a well-defined luminosity-function break and a stable TRGB 
detection. We then looked for a common range of parameters that worked 
reliably for most galaxies, rather than adopting the individually 
optimized configuration for every target. The final analysis therefore 
prioritizes a homogeneous treatment of the sample over fine-tuning 
the parameters separately for each galaxy.

\subsubsection*{Analysis Choices}

We fixed the RGB color selection to $0.9$--$1.5$ mag in F606W$-$F814W and
$1.5$--$2.1$ mag in F555W$-$F814W, and used a GLOESS smoothing parameter of
$\tau=0.11$, as these values gave satisfactory detections for most of
the sample. Optimizing these parameters separately for each galaxy 
can improve the local TRGB measurement, but a common configuration 
is preferable for consistency. Only  NGC~4038 required 
a different color selection ($0.7$--$1.2$ mag in F606W$-$F814W and $1.3$--$1.9$ mag in F555W$-$F814W), 
and NGC 1404 a slightly different smoothing parameter 
($\tau=0.13$). These modifications were necessary to obtain robust detections 
satisfying the adopted tip-contrast and \citet{Madore2009} criteria. 
We excluded NGC~3021, NGC~3370, NGC~5584, and NGC~1309 because their tip contrasts
were insufficient or marginal, and M101 because its low scale factor
could increase disk/bulge contamination.

\subsubsection*{Robustness and Particular Cases}

We assessed the robustness of the final TRGB measurements by testing their 
sensitivity to the color selection, the presence of competing edge-detection peaks, 
and the structure of the bootstrap distributions.
We noticed that using the color-magnitude diagram with the filter F555W
together with the one using filter F606W lowered the value of the
distance to M66 (as the code gives a weighted mean).  The mean value
of the distance went to $\mu$ $=$ 30.052 $\pm$ 0.056 mag. However the
color-magnitude diagram with 
F555W does not provide a good determination of the tip for M66 because 
the number of available stars for the edge detection is 233
(below 400 stars).
Thus, it is adequate to use only
the color-magnitude diagram with F606W.
Then the distance for this galaxy from a very well-defined tip (with a
contrast of 7.95 and 4388 stars)
turns out to be $\mu$ $=$ 30.217 $\pm$ 0.070 mag.

\subsubsection*{Main Takeaway}

We have seen how important it is to avoid
poorly sampled color-magnitude diagrams
(as in the case of M66). If combined with other well-sampled color-magnitude
diagrams to obtain distances, they put at risk the robustness of the distance measurement. Overall, we have good detections of the TRGBs for 12 galaxies and 
we obtain $H_0 = 74.07 \pm 2.2$~km~s$^{-1}$~Mpc$^{-1}$.

\subsection{Elaina Sadler}

\subsubsection*{Analysis Approach}

My approach focused primarily on adjusting magnitude selection cuts and spatial clipping regions individually for each galaxy. I identified the main body of each galaxy and altered the spatial region to lie along the edge of the main body. Starting with the default parameters, I tuned the magnitude cuts to isolate a plausible TRGB magnitude with the smallest uncertainty. I heavily relied on the Sobel filter response; if a specific magnitude selection failed to produce a singular, well-defined maximum peak, I inspected the smoothing filter to see if adjustments resolved the issue, or returned to color-cut adjustments to obtain a clear peak.

\subsubsection*{Analysis Choices}

I chose to adjust parameters uniquely from galaxy to galaxy. While I know it does not provide consistency for future measurement determinations, tailored parameters seemed the most efficient for minimizing TRGB errors due to each galaxy’s distinct stellar populations and properties. For most iterations, the optimal color selections, spatial cuts, and magnitude ranges were evident, as parameter changes affected the error and TRGB values greatly. For a few galaxies, independent adjustments resulted in notebook flags or failed checks, so for those I was left to slowly tune the spatial and magnitude selections until the diagnostics were satisfied.

\subsubsection*{Robustness and Particular Cases}

I deemed final measurements robust if distance moduli fell within the expected 30--40 mag range with relatively small errors. Substantial TRGB measurements were achieved for most of the galaxies, except M101, where no set of parameters provided a tip contrast above 3. Additionally, two other galaxies, NGC~3021 and NGC~1404, proved difficult in achieving a robust TRGB value. Both exhibited competing, dual Sobel peaks that largely shifted the TRGB measurement when magnitude cuts were adjusted, resulting in larger uncertainties.

\subsubsection*{Main Takeaway}

I found the spatial and magnitude selections crucial to the determination of final distances. Both variables greatly affected TRGB values, and changing them simultaneously shifts a galaxy’s measurement by multiple magnitudes.

\subsection{Sarang Shah}

\subsubsection*{Analysis approach}

I analyzed EDD files and treated each galaxy individually. For each galaxy I varied the color window, GLOESS bandwidth, spatial clipping, magnitude range, and luminosity function's bin width while inspecting the color--magnitude diagram, luminosity function, edge response and bootstrap distribution. A detection was adopted only if the luminosity function and Sobel filter response marked the tip of the red giant branch. While doing so, I also made sure the tip contrast exceeded 3 in host and anchor, there are at least 800 stars fainter than the tip, the bootstrap error was as low as possible.

\subsubsection*{Analysis choices}

Color windows and bandwidths were chosen per galaxy from values meeting these criteria, with (F606W - F814W) within 1.5 mag limits. Poisson floor,  Hatt edge-weighting and 0.01-0.05~mag bins were used for all sources. EDD clipping was scaled per galaxy to select halo regions while maintaining sufficient star counts and therefore I changed the scale factor parameter for each galaxy. The search range is automatic, except in NGC~1448, where I set the bright end to 26.7~mag and also change the number of bootstraps to 800.

\subsubsection*{Robustness and particular cases}

I assessed the robustness of the measurements by varying the smoothing band-width, color limits and scale factor and by examining the bootstrap distributions. NGC 1316 and NGC 4526 showed a clear TRGB measurement and the inferred distance was most sensitive to the smoothing bandwidth. For NGC 4424 and NGC 4536, the result varied significantly with the spatial scale factor and it generated a secondary peak in the bootstrap distribution.

\subsubsection*{Main take-away}

I found that the default pipeline parameters provided were reliable for TRGB detections for some of the galaxies with only mild tuning required. Smoothing bandwidth and spatial scale factor had most affect on the TRGB measurements and can be considered for refinements. But I noticed that too much tuning can lead to accidental detections. Even if one gets tip contrast >3 and number of faint stars above the threshold count of 400, bootstrapping plots also need to be inspected. A combination of additional filters could prove useful for increasing the robustness of the detection as halo was not clear in some cases.

\subsection{Petar Suman}

\subsubsection*{Analysis Approach}
EDD was used as a photometry source throughout the analysis, both for the anchor (NGC~4258) and for target galaxies. Out of the 20 target candidates within the EDD source, only 12 have been shown to be suitable for TRGB analysis. The choice was made to use only one photometry source in order to pin down any possible systematic effects within the single data source. 
\\
For each of the target galaxies, the visual inspection was first performed on the spatial selection. This would give possible hints about the RGB population size. Then, 
a preliminary run on a target galaxy with either starting default parameters or with the parameter settings of the previous galaxy was performed. The guiding diagnostic was the tip contrast and the number of stars below the tip. If either was too low (tip contrast $< 3.0$ or $N_{\rm faint} < 400$), parameters were changed before restarting. This analysis was accompanied by visually investigating the color-magnitude diagram (can a clear tip be spotted?) and the bootstrap distribution (is the distribution unimodal?). The varied parameters were smoothing, color selections, and scale factor, as appropriate, to generate a valid sample.

\subsubsection*{Analysis Choices}
The anchor galaxy was kept at the smoothing $\tau = 0.10$, and Hatt weighting with a bin width of 0.01 was used for all target galaxies. The upper limit on the F606W$-$F814W color was kept constant at 1.5 mag, and the upper limit on F555W$-$F814W at 2.1 mag (the exceptions being NGC~4424 and NGC~4526, whose upper limit was set to 1.5 mag to produce valid star populations). The lower limit on the F606W$-$F814W color varied between 1.0 and 1.3 mag, and that on F555W$-$F814W between 1.0 and 1.7 mag. Smoothing was selected for one of the values of $\tau\in \{0.05, 0.1, 0.2\}$ for all but one galaxy.
The general approach was to easily accept any measurements with tip contrast $\gtrsim5.0$ and $N_{\rm faint} \approx 10^3$. On the other hand, galaxies with low tip contrast were treated with more fine tuning, trying to find suitable parameters to satisfy the selection criteria. Ultimately, the motivation was to generate a larger number of galaxy samples which was expected to yield a better estimate of $H_0$, even at the expense of including spurious galaxies with tip contrast as low as $\sim 3$ or with fewer stars.

\subsubsection*{Robustness and Particular Cases}
As noted above, any significant detection was assumed to be robust, and minor quantitative robustness checks were performed to validate this (rerunning for arbitrary small changes in the parameters). No significant problems in robustness checks were observed in such cases.\\

Galaxies with lower tip contrast or low $N_{\rm faint}$ gave significantly less robust results. In particular, NGC~3021  only yielded a contrast tip of $3.07$ when the smoothing was increased to $\tau=0.35$. Another finely tuned galaxy was NGC~1309, for which a relatively large scale factor of 1.2 was needed to generate enough stars, with $N_{\rm faint} = 642$. Such galaxies could not be characterized as robust, and a larger effort was needed to find satisfying parameters. These (and possibly a few other) galaxies were included nonetheless with the aim of generating a larger data sample.

\subsubsection*{Main Takeaway}
The main conclusion is that robustness checks are more important than originally expected; one should be wary of possibly accidental detections and severe fine tuning of selection parameters. 
Some less robust measurements were included to expand the galaxy sample. 
Although the aim was to rely on the central limit theorem to produce an accurate measurement of $H_0$, the resulting value has been noticeably overestimated.
This suggests that a more stable choice of galaxies and fewer finely tuned parameters should be applied in future measurements.

\newpage
\bibliographystyle{apalike}
\bibliography{biblio,Murakami_refs}

\newpage

\section*{Author affiliations and ORCID iDs}
\addcontentsline{toc}{section}{Author affiliations and ORCID IDs}
\begingroup\small\setlength{\parindent}{0pt}\setlength{\parskip}{3pt}
\textbf{Siyang Li} (ORCID: \href{https://orcid.org/0000-0002-8623-1082}{0000-0002-8623-1082}). Department of Astronomy, University of California, Berkeley, CA 94720-3411, USA.

\textbf{Eleonora Di Valentino} (ORCID: \href{https://orcid.org/0000-0001-8408-6961}{0000-0001-8408-6961}). School of Mathematical and Physical Sciences, University of Sheffield, Hounsfield Road, Sheffield S3 7RH, United Kingdom.

\textbf{Jackson Levi Said} (ORCID: \href{https://orcid.org/0000-0002-7835-4365}{0000-0002-7835-4365}). Institute of Space Sciences and Astronomy, University of Malta, Malta, MSD 2080; Department of Physics, University of Malta, Malta.

\textbf{Kayla Owens} (ORCID: \href{https://orcid.org/0000-0003-3339-8820}{0000-0003-3339-8820}). Department of Astronomy and Astrophysics, University of Chicago, 5640 South Ellis Avenue, Chicago, IL 60637, USA.

\textbf{Yukei Murakami} (ORCID: \href{https://orcid.org/0000-0002-8342-3804}{0000-0002-8342-3804}). Department of Physics and Astronomy, Johns Hopkins University, Baltimore, Maryland 21218, USA; MIT Kavli Institute for Astrophysics and Space Research, Cambridge, MA, USA.

\textbf{Louise Breuval} (ORCID: \href{https://orcid.org/0000-0003-3889-7709}{0000-0003-3889-7709}). European Space Agency (ESA), ESA Office, Space Telescope Science Institute, 3700 San Martin Drive, Baltimore, MD 21218, USA.

\textbf{Teresa Sicignano} (ORCID: \href{https://orcid.org/0009-0004-7523-0799}{0009-0004-7523-0799}). European Southern Observatory, Karl-Schwarzschild-Strasse 2, 85748 Garching bei München, Germany; Scuola Superiore Meridionale, Largo San Marcellino10 I-80138 Napoli, Italy; INAF-Osservatorio Astronomico di Capodimonte, Salita Moiariello 16, 80131, Napoli, Italy.

\textbf{Luis A. Escamilla} (ORCID: \href{https://orcid.org/0000-0003-4334-5140}{0000-0003-4334-5140}). Department of Physics, Istanbul Technical University, Maslak, Istanbul, 34469, Turkey.

\textbf{Jéferson A. S. Fortunato} (ORCID: \href{https://orcid.org/0000-0001-7983-1891}{0000-0001-7983-1891}). High Energy Physics, Cosmology \& Astrophysics Theory Group, Department of Mathematics and Applied Mathematics, University of Cape Town, Cape Town 7700, South Africa; African Institute for Mathematical Sciences, 6 Melrose Road, Muizenberg, Cape Town 7945, South Africa.

\textbf{Payam Ghafari} (ORCID: \href{https://orcid.org/0009-0008-4650-6134}{0009-0008-4650-6134}). Shahid Beheshti University, Tehran, Iran.

\textbf{Ayush Hazarika} (ORCID: \href{https://orcid.org/0009-0004-5255-0730}{0009-0004-5255-0730}). International Center for Theoretical Physics Asia-Pacific (ICTP-AP), University of Chinese Academy of Sciences, Beijing 100190, China.

\textbf{Vasiliki Karanasou} (ORCID: \href{https://orcid.org/0000-0001-5407-875X}{0000-0001-5407-875X}). Laboratory of Theoretical Physics, Institute of Physics, University of Tartu, W. Ostwaldi 1, 50411, Tartu, Estonia.

\textbf{Trisha Khan} (ORCID: \href{https://orcid.org/0009-0004-6152-0664}{0009-0004-6152-0664}). Department of Physics and Electronics, CHRIST (Deemed to be University), Bengaluru 560029, India.

\textbf{Yoelsy Leyva} (ORCID: \href{https://orcid.org/0000-0002-4333-839X}{0000-0002-4333-839X}). Departamento de Física, Facultad de Ciencias, Universidad de Tarapacá, Casilla 7-D, Arica, Chile.

\textbf{Mahdi Najafi} (ORCID: \href{https://orcid.org/0009-0007-1230-9880}{0009-0007-1230-9880}). Physics Department, Sapienza University of Rome, P.le A. Moro 5, 00185 Roma, Italy.

\textbf{Swayamtrupta Panda} (ORCID: \href{https://orcid.org/0000-0002-5854-7426}{0000-0002-5854-7426}). International Gemini Observatory/NSF NOIRLab, Casilla 603, La Serena, Chile.

\textbf{Masroor C. Pookkillath} (ORCID: \href{https://orcid.org/0000-0002-7199-8037}{0000-0002-7199-8037}). Department of Physics, Waseda University, 3-4-1 Okubo, Shinjuku, Tokyo 169-8555, Japan.

\textbf{Antonio Quintana-Estellés} (ORCID: \href{https://orcid.org/0009-0003-7568-470X}{0009-0003-7568-470X}). Instituto de Física Fundamental, Consejo Superior de Investigaciones Científicas, c/. Serrano 121, E-28006, Madrid, Spain; Institut de Ciències del Cosmos (UB-IEEC), c/. Martí i Franqués 1, E-08028, Barcelona, Spain.

\textbf{Masoume Reyhani} (ORCID: \href{https://orcid.org/0009-0003-6718-3163}{0009-0003-6718-3163}). Department of Astronomy and High Energy Physics, Faculty of Physics, Kharazmi University, Tehran, Iran; PDAT Laboratory, Department of Physics, K.N. Toosi University of Technology, P.O. Box 15875-4416, Tehran, Iran.

\textbf{Nandan Roy} (ORCID: \href{https://orcid.org/0000-0001-7197-453X}{0000-0001-7197-453X}). Centre for Theoretical Physics and Natural Philosophy, Nakhonsawan Studiorum for Advanced Studies, Mahidol University, Nakhonsawan, 60130, Thailand.

\textbf{Chandra Shekhar Saraf} (ORCID: \href{https://orcid.org/0000-0002-5149-4042}{0000-0002-5149-4042}). Korea Astronomy and Space Science Institute, 776 Daedeok-daero, Yuseong-gu, Daejeon 34055, Republic of Korea.

\textbf{Rahul Shah} (ORCID: \href{https://orcid.org/0000-0001-7682-9219}{0000-0001-7682-9219}). Physics and Applied Mathematics Unit, Indian Statistical Institute, 203 B.T. Road, Kolkata 700 108, India.

\textbf{Jess Worsley} (ORCID: \href{https://orcid.org/0000-0003-2572-8965}{0000-0003-2572-8965}). Department of Mathematics and Applied Mathematics, University of Cape Town, 7 University Ave N, 7700, South Africa.

\textbf{Du\v{s}ko Borka} (ORCID: \href{https://orcid.org/0000-0001-9196-4515}{0000-0001-9196-4515}). Department of Theoretical Physics and Condensed Matter Physics (020), Vin\v{c}a Institute of Nuclear Sciences - National Institute of the Republic of Serbia, University of Belgrade, P.O. Box 522, 11001 Belgrade, Serbia.

\textbf{Vesna Borka Jovanovi\'{c}} (ORCID: \href{https://orcid.org/0000-0001-6764-1927}{0000-0001-6764-1927}). Department of Theoretical Physics and Condensed Matter Physics (020), Vin\v{c}a Institute of Nuclear Sciences - National Institute of the Republic of Serbia, University of Belgrade, P.O. Box 522, 11001 Belgrade, Serbia.

\textbf{Youri Carloni} (ORCID: \href{https://orcid.org/0009-0008-5234-0900}{0009-0008-5234-0900}). Università di Camerino, Via Madonna delle Carceri, 62032 Camerino, Italy; INAF – Osservatorio Astronomico di Brera, Milano, Italy; INFN – Sezione di Perugia, 06123 Perugia, Italy.

\textbf{Umut Demirbozan} (ORCID: \href{https://orcid.org/0000-0001-8053-1896}{0000-0001-8053-1896}). Bonn-Aachen International Center for Information Technology (b-it), University of Bonn, Friedrich-Hirzebruch-Allee 5, 53115 Bonn, Germany; Lamarr Institute for Machine Learning and Artificial Intelligence, Bonn, Germany.

\textbf{Predrag Jovanovi\'{c}} (ORCID: \href{https://orcid.org/0000-0003-4259-0101}{0000-0003-4259-0101}). Astronomical Observatory, Volgina 7, P.O. Box 74, 11060 Belgrade, Serbia.

\textbf{Mohammad Malekjani} (ORCID: \href{https://orcid.org/0000-0001-8846-7593}{0000-0001-8846-7593}). Department of Physics, Institute for Advanced Studies in Basic Sciences (IASBS), P.O. Box 45137-66731, Zanjan, Iran.

\textbf{Nils A. Nilsson} (ORCID: \href{https://orcid.org/0000-0001-6949-3956}{0000-0001-6949-3956}). Cosmology, Gravity and Astroparticle Physics Group, Center for Theoretical Physics of the Universe, Institute for Basic Science, Daejeon 34126, Korea; LTE, Observatoire de Paris, Universite, PSL, CNRS, LNE, Sorbonne Universite, 61 avenue de l’Observatoire, 75 014 Paris, France.

\textbf{Saeed Pourojaghi} (ORCID: \href{https://orcid.org/0000-0002-4845-7590}{0000-0002-4845-7590}). School of Physics, Institute for Research in Fundamental Sciences (IPM), P.O.Box 19395-5531, Tehran, Iran.

\textbf{Jorge Venzor} (ORCID: \href{https://orcid.org/0000-0003-4510-2089}{0000-0003-4510-2089}). Departamento de F\'isica, Centro de Investigaci\'on y de Estudios Avanzados del Instituto Politécnico Nacional, Apartado Postal 14-740, 07000, Ciudad de M\'exico, M\'exico.

\textbf{Alejandro Aguirre} (ORCID: \href{https://orcid.org/0009-0006-4481-6935}{0009-0006-4481-6935}). Department of Astronomy, University of California, Berkeley, CA 94720-3411, USA.

\textbf{Miguel Barroso Varela} (ORCID: \href{https://orcid.org/0009-0006-9844-7661}{0009-0006-9844-7661}). Centro de Física das Universidades do Minho e do Porto, Rua do Campo Alegre s/n, 4169-007 Porto, Portugal.

\textbf{Anupam Bhardwaj} (ORCID: \href{https://orcid.org/0000-0001-6147-3360}{0000-0001-6147-3360}). Inter-University Centre for Astronomy and Astrophysics (IUCAA), Post Bag 4, Ganeshkhind, Pune 411007, India.

\textbf{Elliot Dutnall} (ORCID: \href{https://orcid.org/0009-0000-4368-3460}{0009-0000-4368-3460}). Department of Astronomy, University of California, Berkeley, CA 94720-3411, USA.

\textbf{Francesca Di Nella} (ORCID: \href{https://orcid.org/0009-0009-5550-6539}{0009-0009-5550-6539}). Physics Department, Sapienza University of Rome, P.le A. Moro 5, 00185 Roma, Italy; Istituto Nazionale di Fisica Nucleare (INFN), Sezione di Roma, P.le A. Moro 5, 00185 Roma, Italy; Physics Department, Tor Vergata University of Rome, Via della Ricerca Scientifica 1, 00133 Roma, Italy.

\textbf{Hussain Gohar} (ORCID: \href{https://orcid.org/0000-0002-6862-0869}{0000-0002-6862-0869}). Institute of Physics, University of Szczecin, Wielkopolska 15, 70-451 Szczecin, Poland.

\textbf{Paridhi Jain} (ORCID: \href{https://orcid.org/0009-0003-6551-8767}{0009-0003-6551-8767}). Virtual Institute of Astroparticle Physics, 75018 Paris, France.

\textbf{Gracelynn Jost} (ORCID: \href{https://orcid.org/0009-0004-9450-2489}{0009-0004-9450-2489}). Department of Astronomy, University of California, Berkeley, CA 94720-3411, USA.

\textbf{Shivani Kulkarni} (ORCID: \href{https://orcid.org/0009-0001-8673-3904}{0009-0001-8673-3904}). Department of Astronomy, University of California, Berkeley, CA 94720-3411, USA.

\textbf{Dhiraj Kuniyal} (ORCID: \href{https://orcid.org/0009-0000-5864-1060}{0009-0000-5864-1060}). Department of Physics and Astrophysics, University of Delhi, Delhi--110007, India.

\textbf{Mariia Marinichenko} (ORCID: \href{https://orcid.org/0009-0003-8861-636X}{0009-0003-8861-636X}). Leiden Observatory, Leiden University, PO Box 9513, NL-2300 RA Leiden, the Netherlands.

\textbf{Yash Mehta} (ORCID: \href{https://orcid.org/0009-0001-1604-4118}{0009-0001-1604-4118}). Department of Astronomy, University of California, Berkeley, CA 94720-3411, USA.

\textbf{Vasudev Mittal} (ORCID: \href{https://orcid.org/0000-0002-1708-6088}{0000-0002-1708-6088}). Sydney Institute for Astronomy, School of Physics A28, The University of Sydney, NSW 2006, Australia.

\textbf{Lily Myers} (ORCID: \href{https://orcid.org/0009-0006-8327-7890}{0009-0006-8327-7890}). Department of Astronomy, University of California, Berkeley, CA 94720-3411, USA.

\textbf{Devaraja Mallesha Naik} (ORCID: \href{https://orcid.org/0000-0003-2082-697X}{0000-0003-2082-697X}). Department of Mathematics, Government First Grade College Shikaripura, Kuvempu University, Shivamogga--577 427, Karnataka, India.

\textbf{José Antonio Nájera} (ORCID: \href{https://orcid.org/0000-0001-9738-7704}{0000-0001-9738-7704}). Institute of Cosmology \& Gravitation, University of Portsmouth, Dennis Sciama Building, Burnaby Road, Portsmouth PO1 3FX, UK.

\textbf{Pilar Ruiz-Lapuente} (ORCID: \href{https://orcid.org/0000-0001-9046-4420}{0000-0001-9046-4420}). Instituto de Física Fundamental, Consejo Superior de Investigaciones Científicas, c/. Serrano 121, E-28006, Madrid, Spain; Institut de Ciències del Cosmos (UB-IEEC), c/. Martí i Franqués 1, E-08028, Barcelona, Spain.

\textbf{Kathan Dave} (ORCID: \href{https://orcid.org/0009-0005-4908-7789}{0009-0005-4908-7789}). School of Engineering and Applied Sciences, Ahmedabad University, Ahmedabad, Gujarat 380009, India.

\textbf{Elaina Sadler} (ORCID: \href{https://orcid.org/0009-0002-8370-9551}{0009-0002-8370-9551}). Department of Astronomy, University of California, Berkeley, CA 94720-3411, USA.

\textbf{Gurnoor Sandhu} (ORCID: \href{https://orcid.org/0009-0008-8473-9983}{0009-0008-8473-9983}). Department of Astronomy and Astrophysics, University of California, Berkeley, CA, USA.

\textbf{Sarang Shah} (ORCID: \href{https://orcid.org/0000-0003-1959-8439}{0000-0003-1959-8439}). Inter-University Centre for Astronomy and Astrophysics, Ganeshkhind, Pune, India - 411007.

\textbf{Petar Suman} (ORCID: \href{https://orcid.org/0009-0009-2119-1110}{0009-0009-2119-1110}). Centre for Theoretical Cosmology, Department of Applied Mathematics and Theoretical Physics (DAMTP), University of Cambridge, Wilberforce Road, Cambridge, United Kingdom.

\textbf{Lilia Anguelova}. Institute for Nuclear Research and Nuclear Energy, Bulgarian Academy of Sciences, Sofia 1784, Bulgaria.

\textbf{Marika Asgari} (ORCID: \href{https://orcid.org/0000-0002-3064-083X}{0000-0002-3064-083X}). School of Mathematics, Statistics and Physics, Newcastle University, Herschel Building, NE1 7RU, Newcastle-upon-Tyne, UK.

\textbf{Stefano Casertano}. Space Telescope Science Institute, 3700 San Martin Drive, Baltimore, MD 21218, USA.

\textbf{Alexei V. Filippenko} (ORCID: \href{https://orcid.org/0000-0003-3460-0103}{0000-0003-3460-0103}). Department of Astronomy, University of California, Berkeley, CA 94720-3411, USA.

\textbf{Nihan Kat{\i}rc{\i}} (ORCID: \href{https://orcid.org/0000-0002-9492-3791}{0000-0002-9492-3791}). Department of Electrical and Electronics Engineering, Do\u{g}u\c{s} University, Istanbul, T\"{u}rkiye.

\textbf{Mar\'{i}a P\'{e}rez Garrote} (ORCID: \href{https://orcid.org/0009-0008-8108-7879}{0009-0008-8108-7879}). Departamento de F\'{i}sica Fundamental, Universidad de Salamanca, E-37008 Salamanca, Spain; Instituto Universitario de F\'{i}sica Fundamental y Matem\'{a}ticas IUFFyM, Universidad de Salamanca, E-37008 Salamanca, Spain.

\textbf{Leandros Perivolaropoulos} (ORCID: \href{https://orcid.org/0000-0001-9330-2371}{0000-0001-9330-2371}). Department of Physics, University of Ioannina, 45110 Ioannina, Greece.

\textbf{Pradyumn Kumar Sahoo} (ORCID: \href{https://orcid.org/0000-0003-2130-8832}{0000-0003-2130-8832}). Department of Mathematics, Birla Institute of Technology \& Science, Pilani, Hyderabad Campus, Jawahar Nagar, Kapra Mandal, Hyderabad 500 078, Telangana, India.

\textbf{Denitsa Staicova} (ORCID: \href{https://orcid.org/0000-0001-6139-3125}{0000-0001-6139-3125}). Institute for Nuclear Research and Nuclear Energy, Bulgarian Academy of Sciences, Bulgaria.

\textbf{Richard Stiskalek} (ORCID: \href{https://orcid.org/0000-0002-0986-314X}{0000-0002-0986-314X}). Astrophysics, University of Oxford, Denys Wilkinson Building, Keble Road, Oxford, OX1 3RH, UK; St John's College, University of Oxford, St Giles', Oxford, OX1 3JP, UK.

\textbf{Syed Ashraf Uddin} (ORCID: \href{https://orcid.org/0000-0002-9413-4186}{0000-0002-9413-4186}). Department of Physical Sciences, Independent University, Bashundhara R/A, Dhaka 1229, Bangladesh; Center for Astronomy, Space Science and Astrophysics, Independent University, Bashundhara R/A, Dhaka 1229, Bangladesh; American Public University System, 111 W. Congress St, Charles Town, WV 25414, USA.
\endgroup
\bigskip

\end{document}